\documentclass[fleqn,usenatbib]{mnras}

\usepackage[T1]{fontenc}

\usepackage{graphicx}	
\usepackage{amsmath}	
\usepackage{amssymb}	
\usepackage{mhchem}
\usepackage{xcolor}

\usepackage[normalem]{ulem}
\usepackage{pdflscape}

\newcommand{\Rdust}{$R_\mathrm{dust}$}
\newcommand{\Tcomp}{$T_\mathrm{comp}$}
\newcommand{\Rcomp}{$R_\mathrm{comp}$}
\newcommand{\msunyr}{M$_\odot$ yr$^{-1}$}
\newcommand{\fvol}{$f_\mathrm{vol}$}
\newcommand{\fic}{$f_\mathrm{ic}$}

\definecolor{cobalt}{rgb}{0.0, 0.28, 0.67}

\definecolor{rufous}{rgb}{0.66, 0.11, 0.03}
\newcommand{\red}[1]{\textcolor{rufous}{#1}}

\definecolor{royalblue(web)}{rgb}{0.25, 0.41, 0.88}
\newcommand{\greent}[1]{\textcolor{royalblue(web)}{#1}}

\title[Impact of companion UV photons on AGB outflows]{The impact of stellar companion UV photons on the chemistry of the circumstellar environments of AGB stars}

\author[Van de Sande \& Millar]{
M. Van de Sande$^{1,2}$\thanks{E-mail: m.vandesande@leeds.ac.uk} \&
T. J. Millar$^{3}$
\\
$^{1}$School of Physics and Astronomy, University of Leeds, Leeds LS2 9JT, UK\\
$^{2}$Institute of Astronomy, KU Leuven, Celestijnenlaan 200D, 3001 Leuven, Belgium\\
$^{3}$Astrophysics Research Centre, School of Mathematics and Physics, Queen's University Belfast, University Road, Belfast BT7 1NN, UK\\
}

\date{Accepted XXX. Received YYY; in original form ZZZ}

\pubyear{2021}

\begin{document}
\label{firstpage}
\pagerange{\pageref{firstpage}--\pageref{lastpage}}
\maketitle

\begin{abstract}
Spherical asymmetries are prevalent within the outflows of AGB stars. 
Since binary interaction with a stellar or planetary companion is thought to be the underlying mechanism behind large-scale structures, we included the effects of UV radiation originating from a stellar companion in our chemical kinetics model.
The one-dimensional model provides a first approximation of its effects on the chemistry throughout the outflow.
The presence of a close-by stellar companion can strongly influence the chemistry within the entire outflow.
Its impact depends on the intensity of the radiation (set by the stellar radius and blackbody temperature) and on the extinction the UV radiation experiences (set by the outflow density, density structure, and assumed radius of dust formation).
Parent species can be photodissociated by the companion, initiating a rich photon-driven chemistry in the inner parts of the outflow.
The outcome depends on the balance between two-body reactions and photoreactions. 
If two-body reactions dominate, chemical complexity within the outflow increases. 
This can make the abundance profiles of daughters appear like those of parents, with a larger inner abundance and a gaussian decline. If photoreactions dominate, the outflow can appear molecule-poor. 
We model three stellar companions.
The impact of a red dwarf companion is limited. 
Solar-like companions show the largest effect, followed by a white dwarf. 
A stellar companion can also lead to the formation of  unexpected species.
The outflow's molecular content, especially combined with abundance profiles, can indicate a stellar companion's presence.
Our results pave the way for further outflow-specific (three-dimensional) model development.
\end{abstract}


\begin{keywords}
Stars: AGB and post-AGB -- circumstellar matter -- astrochemistry -- molecular processes 
\end{keywords}



\section{Introduction}

The asymptotic giant branch (AGB) phase near the end of the lives of low-to-intermediate mass stars is characterised by strong mass loss.
AGB stars lose their outer layers by means of a stellar outflow or wind at a rate between $10^{-8}$ and $10^{-4}$ \msunyr.
The strong gradients in density and temperature present within the outflows make them rich astrochemical environments: close to 100 different molecules have been detected so far, as well as some 15 types of newly formed dust \citep{Decin2021}. 
Thanks to their outflows, AGB stars are important contributors to the chemical enrichment of the interstellar medium \cite[ISM,][]{Tielens2005}.

Recent observations have confirmed that spherical asymmetry is prevalent within AGB outflows.
Both small-scale asymmetries, such as density-enhanced clumps \cite[e.g.,][]{Khouri2016,Agundez2017,Leao2006}, and large-scale structures, such as spirals \cite[e.g.,][]{Mauron2006,Maercker2016} and disks \cite[e.g.,][]{Kervella2014,Homan2018}, have been widely observed.
Binary interaction with a stellar or planetary companion has been proposed to be the driving mechanism behind the large-scale asymmetries observed in the outflow \citep{Decin2015,Ramstedt2017,Moe2017,Decin2020}, potentially setting the stage for the asymmetrical structures observed around post-AGB stars and planetary nebulae \citep{DeMarco2009}. 

Further observational evidence of stellar companions comes from UV observations by the Galaxy Evolution Explorer (GALEX) mission. 
UV radiation could arise from sources intrinsic to the AGB star, such as chromospheres or pulsational shock waves \citep{Montez2017}, or from extrinsic sources, such as accretion disks or binary companions \citep{Sahai2008,Sahai2011}. 
Detections of far-UV emission could have an extrinsic origin, but a definite origin is not clear \citep{Ortiz2019}.
X-ray observations of AGB stars are correlated with far-UV emission, hinting towards X-ray emission from accretion disks around companions \citep{Ortiz2021}. 
Additionally, the detection of CI emission towards the known O-rich binary system omi Cet is thought to arise from irradiation by its white dwarf companion \citep{Saberi2018}, while the inner wind of the symbiotic system R Aqr shows evidence of photodissociation caused by its white dwarf \citep{Bujarrabal2021}.

In this paper, we consider the effect of stellar companion UV photons on the chemistry throughout the outflow in our one-dimensional gas-phase chemical model.
The presence of a stellar companion in the inner wind initiates a rich photochemistry in the dense inner outflow.
This type of chemistry is otherwise restricted to the tenuous outer regions of the outflow, where it is initiated by interstellar UV photons.
The model is based on previous work \citep{VandeSande2019a} and allows us to present a first analysis of any effect a stellar companion might have on the chemistry throughout the outflow, with the aim to identify chemical markers for its presence and guide future model development.
To that end, we assess the impact of an internal UV radiation field on a variety of C-rich and O-rich outflows.

The physics and chemistry of the model is described in Sect. \ref{sect:model}, along with the approximations and assumptions made. 
Our results for C-rich and O-rich outflows are presented in Sect. \ref{sect:results}.
Discussion and conclusions follow in Sects. \ref{sect:discussion} and \ref{sect:conclusions}, respectively.

\section{Chemical model}			\label{sect:model}

\begin{figure*}
 \includegraphics[width=1\textwidth]{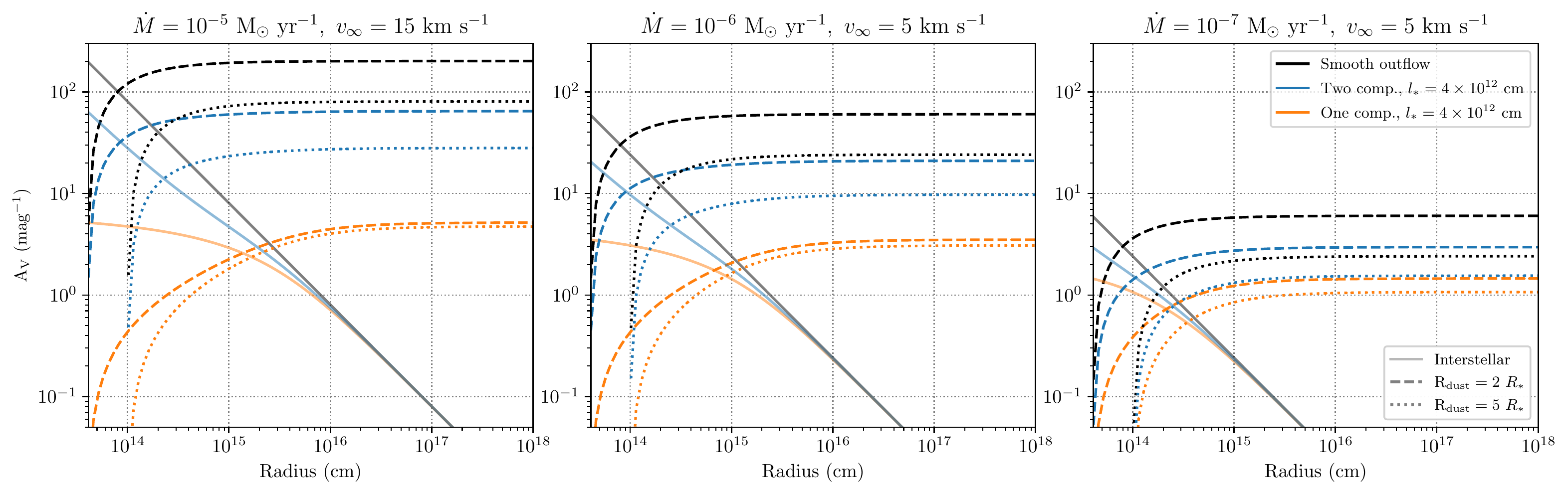}
 \caption{Visual extinction experienced by interstellar and internal UV photons for different outflow densities, density structures, and values of \Rdust. 
 Different line styles show the extinction experienced by different UV photons.
 {Solid} lines: $A_V$ experienced by interstellar UV photons. {Dashed/dotted} lines: $\Delta A_V$ experienced by internal UV photons for \Rdust\ = 2 $R_*$/5 $R_*$.
 Different colours show the extinction experienced by different density structures.
 Black: smooth outflow. 
 Blue: two-component outflow (\fic = 0.3, \fvol = 0.3, $l_* = 4 \times 10^{12}$ cm = 0.2 $R_*$). 
 Orange: one-component outflow (\fic = 0.0, \fvol = 0.3, $l_* = 4 \times 10^{12}$ cm = 0.2 $R_*$). 
  }
 \label{fig:extinction}
\end{figure*}

\subsection{Physics}			\label{subsect:model:physics}

The chemical kinetics model is that of \citet{VandeSande2019a}. 
It is based on the publicly available UMIST Database for Astrochemistry (UDfA) CSE model \citep{McElroy2013}\footnote{\url{http://www.udfa.net}}, adapted by \citet{VandeSande2018} to include a gas temperature profile that follows a power law,
\begin{equation}
T(r) = T_* \left( \frac{r}{R_*} \right)^{-\epsilon}.
\end{equation}
The model describes a spherically symmetric outflow with constant mass-loss rate and outflow velocity.
H$_2$ is assumed to be fully self-shielded; CO self-shielding is taken into account using a single-band approximation \citep{Morris1983}.

The rates of photoreactions, i.e. photodissociation and photoionisation, depend on the extinction experienced by UV radiation throughout the outflow. 
For interstellar UV photons, the visual extinction, $A_V$, is determined by the column density of dust radially from a point in the outflow to infinity.
Internal UV photons, either from the star or from a stellar companion, are both extinguished by dust and diluted geometrically.
The dust extinction experienced by internal photons, $\Delta A_V$, is determined by the outflow density set by $\dot{M}/v_\infty$, and the onset of dust extinction, \Rdust. 
Since we do not treat dust condensation in our model, our choice of \Rdust\ is meant to simulate a range of dust condensation radii. 
Here, we choose \Rdust\ = 2 and 5 $R_*$, which corresponds to gas temperatures of 1400 and 750 K, respectively, in our model.
These temperatures cover the range of possible dust condensation temperatures, from aluminium oxides and C-bearing dust at the higher end to silicates at the lower end \citep{Gail2013}. 
The model assumes that dust is already present and does not take its formation into account. 
The companion is assumed to be close to the AGB star and lie within the dust-free region before the onset of dust extinction, \Rdust\ (see also Sect. \ref{subsect:model:limits}).
Our models start at 1.025 $\times$ \Rdust.
Table \ref{table:model-params} lists the physical parameters of the grid of models.

The porosity formalism is used to approximate the effects of an inhomogeneous outflow.
This mathematical framework enables us to include the effects of local overdensities together with the influence of a clumpy outflow on the penetration of UV photons by modifying the optical depth, without making any assumptions on the locations of the clumps. 
The outflow is divided into a stochastic two-component medium composed of an overdense clumped component and a rarified interclump medium.
The specific clumpiness of the outflow and its influence on the density and optical depth is described by three parameters: the clump volume filling factor, \fvol, setting the fraction of the total volume of the outflow occupied by clumps; the interclump density contrast, \fic, setting the distribution of material between the components; and the clump size at the stellar surface, $l_*$.

Besides a smooth outflow, we model a two-component outflow, characterised by \fic = 0.3, \fvol = 0.3, $l_* = 4 \times 10^{12}$ cm = 0.2 $R_*$, and a one-component outflow, characterised by \fic = 0.0, \fvol = 0.3, $l_* = 4 \times 10^{12}$ cm = 0.2 $R_*$.
These three density structures allow us to vary over different behaviours of extinction experienced by interstellar and internal UV photons.
Although the one-component outflow with a void interclump medium is likely not realistic, it provides us with valuable insights in the heavily radiation-driven chemistry it induces.
Figure \ref{fig:extinction} shows the extinction experienced by interstellar and internal UV photons for different outflow densities, density structures, and values of \Rdust.

\begin{table}
	\caption{Physical parameters of the grid of chemical models.}
	\resizebox{1.0\columnwidth}{!}{%
	\centering
	\label{table:model-params}
	\begin{tabular}{ll} 
		\hline
    Outflow density, $\dot{M}$ - $v_\infty$        	&    $10^{-5}$  $\mathrm{M}_\odot\ \mathrm{yr}^{-1}$ - 15 km s$^{-1}$  \\
    									&    $10^{-6}$  $\mathrm{M}_\odot\ \mathrm{yr}^{-1}$ - 5 km s$^{-1}$  \\
    									&    $10^{-7}$  $\mathrm{M}_\odot\ \mathrm{yr}^{-1}$ - 5 km s$^{-1}$  \\
	\noalign{\smallskip}								
	Density structures		& Smooth outflow \\
							& Two-component clumpy outflow \\
							& \quad \fic = 0.3, \fvol = 0.3, $l_* = 4 \times 10^{12}$ cm \\
							& One-component clumpy outflow \\
							& \quad \fic = 0.0, \fvol = 0.3, $l_* = 4 \times 10^{12}$ cm \\
    \noalign{\smallskip}
    Stellar radius, $R_*$             & 2 $\times 10^{13}$ cm \\
    Stellar temperature, $T_*$        & 2330 K \\
    Exponent $T(r)$, $\epsilon$                    & 0.7 \\
    Onset of dust extinction, $R_\mathrm{dust}$		& 2, 5 $R_*$ \\
    Companion temp., \Tcomp, 					& 4000 K - $1.53\times 10^{10}$ cm - 0.041 \\
    \quad and radius, \Rcomp,					& 6000 K - $8.14\times 10^{10}$ cm - 633.8 \\
	\quad scaling factor for photorates		& 10 000 K - $6.96 \times 10^{8}$ cm - 9.291 \\
    \noalign{\smallskip}
    Start of the model		& $1.025 \times R_\mathrm{dust}$ \\
	\hline
	\end{tabular}
	}
\end{table}

\begin{table}
	\caption{Parent species for the C-rich and O-rich outflows, derived from observations as compiled by \citet{Agundez2020}. The initial abundances relative to H$_2$ are the mean of the observed ranges. 
	} 
    \centering
    \begin{tabular}{l r c  l r }
    \hline  
    \multicolumn{2}{c}{Carbon-rich} && \multicolumn{2}{c}{Oxygen-rich}  \\  
    \cline{1-2} \cline{4-5} 
    \noalign{\smallskip}
    Species & Abun. & & Species & Abun. \\
    \hline
    He		&  0.17				& & He		& 0.17  \\
    CO		& $8.00\times10^{-4}$	& & CO		& $3.00 \times 10^{-4}$  \\
    N$_2$		& $4.00 \times 10^{-5}$	& & H$_2$O	& $2.15 \times 10^{-4}$  \\
    CH$_4$	& $3.50 \times 10^{-6}$	& & N$_2$ 	& $4.00 \times 10^{-5}$  \\ 
    H$_2$O	& $2.55 \times 10^{-6}$	& & SiO 	& $2.71 \times 10^{-5}$  \\ 
    SiC$_2$	& $1.87 \times 10^{-5}$	& & H$_2$S 	& $1.75 \times 10^{-5}$  \\
    CS		& $1.06 \times 10^{-5}$	& & SO$_2$ 	& $3.72 \times 10^{-6}$  \\
    C$_2$H$_2$& $4.38 \times 10^{-5}$	& & SO 		& $3.06 \times 10^{-6}$  \\
    HCN		& $4.09 \times 10^{-5}$	& & SiS 		& $9.53 \times 10^{-7}$  \\
    SiS   		& $5.98 \times 10^{-6}$	& & NH$_3$ 	& $6.25 \times 10^{-7}$  \\ 
    SiO 		& $5.02 \times 10^{-6}$	& & CO$_2$ 	& $3.00 \times 10^{-7}$  \\   
    HCl		& $3.25 \times 10^{-7}$	& & HCN 	& $2.59 \times 10^{-7}$  \\  
    C$_2$H$_4$& $6.85 \times 10^{-8}$	& & PO 		& $7.75 \times 10^{-8}$  \\ 
    NH$_3$	& $6.00 \times 10^{-8}$	& & CS 		& $5.57 \times 10^{-8}$  \\
    HCP		& $2.50 \times 10^{-8}$	& & PN 		& $1.50 \times 10^{-8}$  \\
    HF    		& $1.70 \times 10^{-8}$	& &  HCl		& $1.00 \times 10^{-8}$  \\  
    H$_2$S	& $4.00 \times 10^{-9}$	& & 	HF	& $1.00 \times 10^{-8}$  \\    
    \hline 
    \end{tabular}%
    \label{table:model-parents}    
\end{table}

\subsection{Chemistry}			\label{subsect:model:chemistry}

The parent species, i.e. species present at the start of the model, and their initial abundances are listed in Table \ref{table:model-parents} for the O-rich and C-rich outflow.
They are taken from \citet{Agundez2020}, who compiled (ranges of) observed abundances in the inner regions of AGB outflows.

The companion's UV photon flux is approximated by blackbody radiation. 
While stellar atmosphere models can be used, we use this first-order approximation to describe the effect of a stellar companion in a more general way.
The blackbody temperature of the stellar companion, \Tcomp, and its radius, \Rcomp, are listed in Table \ref{table:model-params} and range from a red dwarf (\Tcomp\ = 4000 K, \Rdust\ =$1.53\times 10^{10}$ cm), to a solar-like star (\Tcomp\ = 6000 K, \Rdust\ = $8.14\times 10^{10}$ cm), to a white dwarf (\Tcomp\ = 10 000 K and \Rdust\ = $6.96 \times 10^{8}$ cm).
For simplicity, we assume the same power-law gas temperature profile for all models.

The same methods of \citet{VandeSande2019a} were used to expand the gas-phase reaction network, based on the UDfA \textsc{Rate12} network \citep{McElroy2013}, to include photodissociation and photoionisation reactions induced by companion UV photons. 
Cross sections are used to calculate the unshielded photorate coefficients and were taken in the main from the Leiden Observatory Database\footnote{\url{https://home.strw.leidenuniv.nl/~ewine/photo/}} \citep{Heays2017} or from other sources when available. 
If cross sections are not available we follow the approach taken in many studies of interstellar or protoplanetary disk chemistry, estimating the rate coefficient through scaling the unshielded interstellar rate by the ratio of the integrated fluxes of companion photons to interstellar photons over the $912-2150$ \AA\ range. 
The scaling factors calculated in this manner using companion fluxes at 50 $R_*$ are listed in Table \ref{table:model-params}.
This approach can, however, lead to large errors in the estimated rate coefficients, particularly at low blackbody temperatures which provide very few photons at energies typical of bond dissociation or ionisation. 
To gain an understanding of these uncertainties, we compared the photorates of species with known cross sections to their rates calculated via the scaling approximation for each of our blackbody fields.
For photodissociation, we find that the exact and scaled values are close, mostly to within a factor of a few for \Tcomp\ = 10 000 K and 6000 K albeit with some divergence, factors of $10-100$, in the latter case. 
As expected, the scaling is poorest for \Tcomp\ = 4000 K, where rates are overestimated with some differences of around $10^4$ although many are in reasonable agreement, within a factor of $10-100$.  
For photoionisation, scaling overestimates the exact rates by large factors, $\sim 10^4 - 10^6$ for 4000 K, about $100-1000$ for 6000 K, and is close to the exact values for 10 000 K.
These comparisons allow us to make specific scalings for the photorates of species with unknown cross sections, allowing for the fact that their adopted interstellar unshielded rates are essentially guesses.  
For photodissociation by all companions and for photoionisation by the 10 000 K UV field, we adopt the scaling approximation.
For photoionisation in the 6000 K and 4000 K cases, we adopt the interstellar rates in the former case (scaling = 1) and a scaling of (\Rcomp/$R_*$)$^2$ in the latter.
For more details, we refer to \citet{VandeSande2019a}.

\begin{figure*}
 \includegraphics[width=1\textwidth]{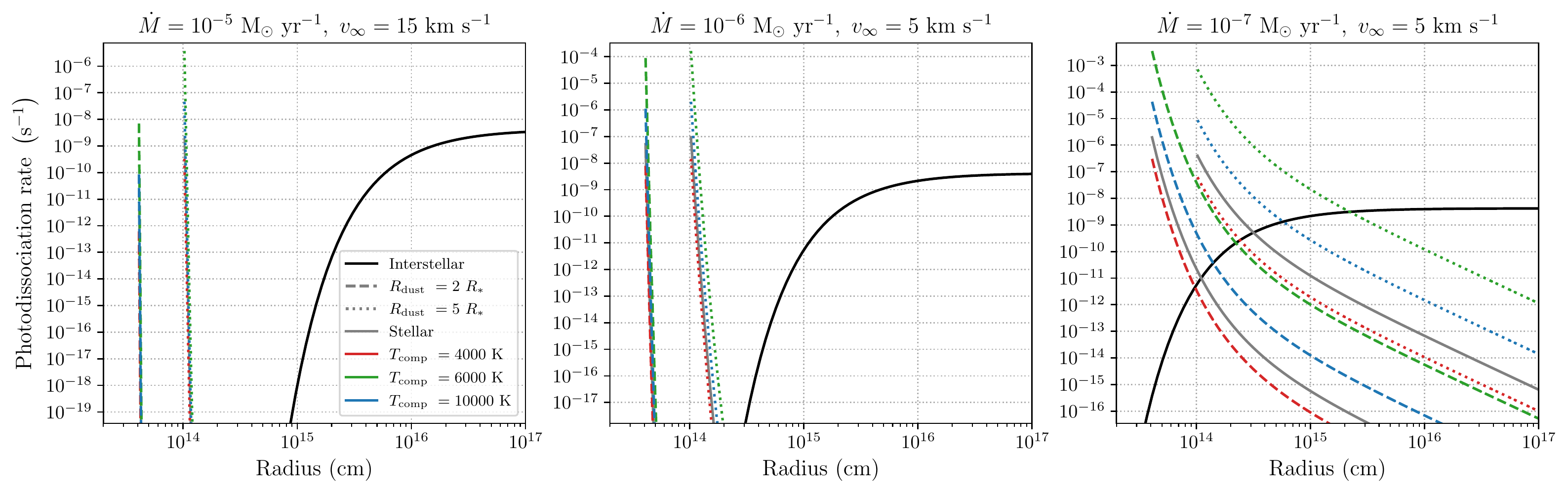}
 \caption{Photodissociation rates of SO in different outflow densities.
 Black solid line: rate due to interstellar UV photons.
 Different colours and line styles show the rate due to stellar or companion UV photons, with different values of \Rdust.
 Line styles: location of \Rdust\ (dashed 2 R$_*$,  dotted: 5 R$_*$).
 Grey: stellar UV photons ($T_*$ = 2330 K),
 red: red dwarf companion UV photons (\Tcomp = 4000 K, \Rcomp = $1.53\times 10^{10}$ cm),
 green: solar-like companion UV photons (\Tcomp = 6000 K, \Rcomp = $8.14\times 10^{10}$ cm).
 blue: white dwarf companion UV photons (\Tcomp = 10 000 K, \Rcomp = $6.96 \times 10^{8}$ cm).
 }
 \label{fig:rates-SO}
\end{figure*}

Figure \ref{fig:rates-SO} shows the photodissociation rate of SO caused by interstellar, stellar, and companion UV photons in a smooth outflow for different outflow densities, companions, and values of \Rdust.
The outflow density and \Rdust\ influence the onset of photodissociation, as they determine the extinction experienced by internal UV radiation.
The rate is determined by the intensity of the radiation, set by the blackbody temperature and radius adopted for the stellar companion.
The rate due to a red dwarf companion is similar to that of stellar UV photons. 
A solar-like companion shows the largest increase in photodissociation rate, about three orders of magnitude larger than that caused by stellar UV photons.
A white dwarf companion leads to an increase of more than an order of magnitude relative to stellar photons. The increase in rate is smaller than for a solar-like companion because of the compact nature of the white dwarf.

\subsection{Limitations of the model}			\label{subsect:model:limits}

Using the methods of \citet{VandeSande2019a} implies that the stellar companion is located at the centre of the star. 
However, the effect of misplacing the companion on the emitted UV radiation field by up to 5 $R_*$, the largest possible value of \Rdust, is negligible compared to the scale of the outflow, especially when considering the variations in extinction with outflow density and \Rdust\ (shown in Fig. \ref{fig:extinction}).

The lack of orbital motion in the model implies that the companion's radiation field is always present.
The fraction of the orbital period during which the companion can be hidden behind the star is largest in the orbital plane. 
It ranges from $20\%$ of the outflow at 10 R$_*$ to $17\%$ at infinity for a companion at 2 $R_*$, and from $10\%$ at 10 R$_*$ to $6\%$ at infinity for a companion at 5 $R_*$.
The fraction decreases as the angle with respect to the orbital plan increases. 
Considering the solid angle cast by the AGB star, the majority of the sphere is constantly irradiated by the companion (Appendix \ref{app:angles}).
Orbital parameters are generally not known for close AGB binaries. 
In symbiotic systems, composed of a white dwarf with a red giant companion, most orbital periods lie within $2-4$ yr \citep{Mikolajewska2012}.
Looking to the next evolutionary steps, carbon-enhanced metal-poor stars enriched in $s$-process elements (CEMP-$s$ stars) have orbital periods between 1 and 10 years \cite[e.g.,][]{Jorissen1998,Hansen2016} and  post-AGB binaries have periods between 100 and 3000 days \citep{Oomen2018}.
Considering the limited fraction of the period during which the companion can be occulted, the majority of these periods might be too short for the chemistry to reset during the companion occultation. 
At a particular radial distance, various parameters might influence this assumption, including the expansion velocity, the orbital separation between the AGB star and its companion, and the molecule under consideration. 
Additionally, the companion's radiation is diluted geometrically and extinguished by dust (Fig. \ref{fig:extinction}).
The region in which the companion can influence the chemistry is therefore confined to the inner regions of the outflow, where the effects of companion occultation on the chemistry are limited. 
Therefore, a constantly present companion radiation field is a reasonable first-order approximation for a close-by companion within the dust formation region.

Additionally, the chemical model does not include any hydrodynamics.
Binary interactions can lead to local density and temperature enhancements and complex kinematics, which will locally affect the chemistry \cite[e.g.,][]{Mastrodemos1999,Kim2012,Chen2017,ElMellah2020,Maes2021,Malfait2021}.
A comprehensive chemical model including such complex and three-dimensional structures is not yet possible.
By using the porosity formalism, we are able to approximate the effects of an inhomogeneous outflow on the chemistry.
The different density structures considered (Table \ref{table:model-params}) allow us to test the effects of different levels of extinction on the chemistry, a first step in studying the effects of binary interactions.

Here, we present the first analysis of any effect a stellar companion might have on the chemistry, with the aim to guide future chemical and hydrodynamical model development.
We calculate a grid of O-rich and C-rich models, varying the outflow density, density structure, companion temperature, and location of the onset of dust extinction.

\begin{figure*}
 \includegraphics[width=1\textwidth]{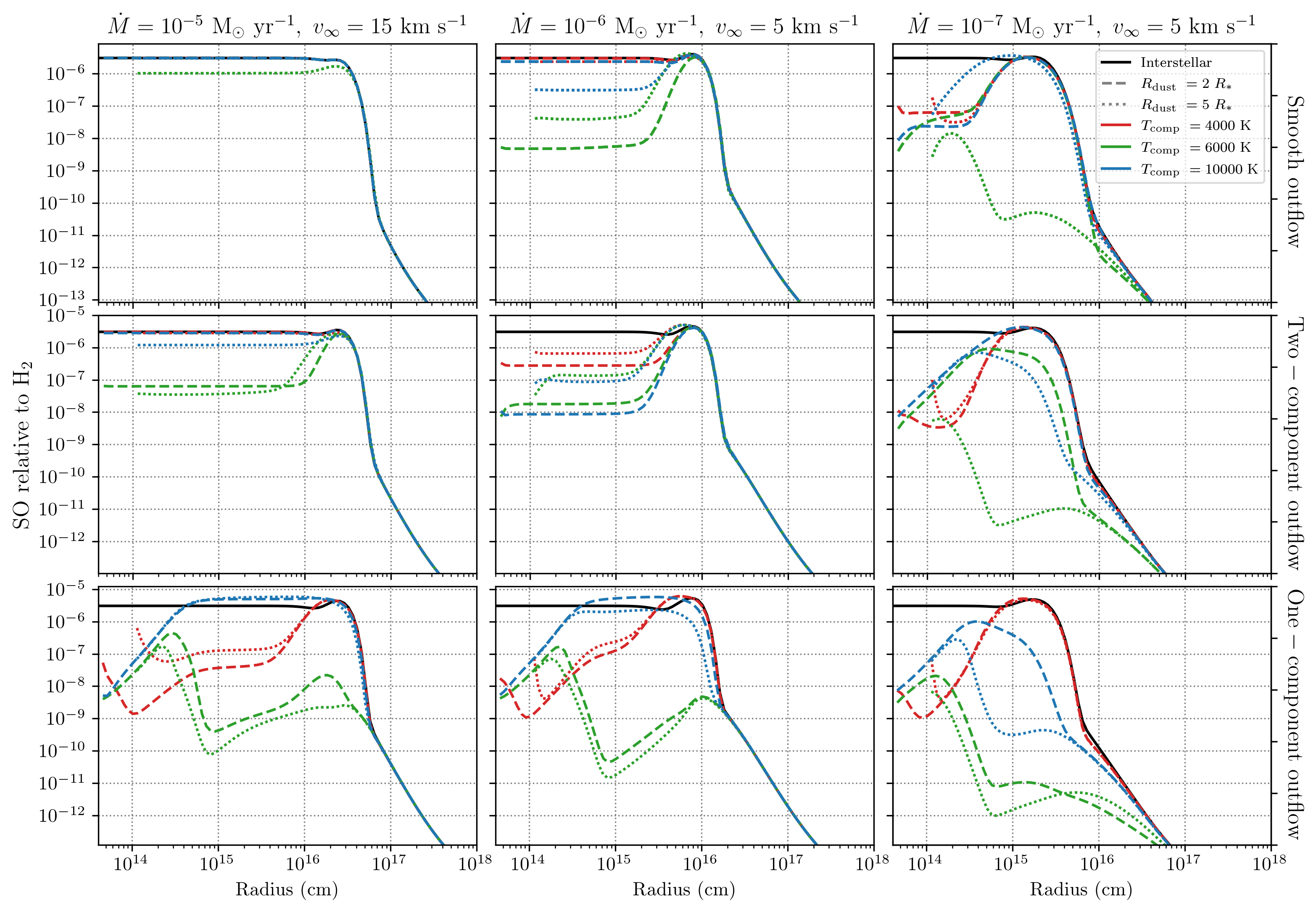}
 \caption{ Abundance of SO relative to H$_2$ in O-rich outflows with different density structures.
 Rows, from top to bottom: smooth outflow, two-component outflow, one-component outflow.
 Black, solid lines: interstellar photons only. 
 Line styles: location of \Rdust\ (dashed 2 R$_*$,  dotted: 5 R$_*$).
 Red: stellar + red dwarf companion, 
 green: stellar + solar-like companion,
 blue: stellar + white dwarf companion.
 }
 \label{fig:fracs-orich-SO}
\end{figure*}

\begin{figure*}
 \includegraphics[width=1\textwidth]{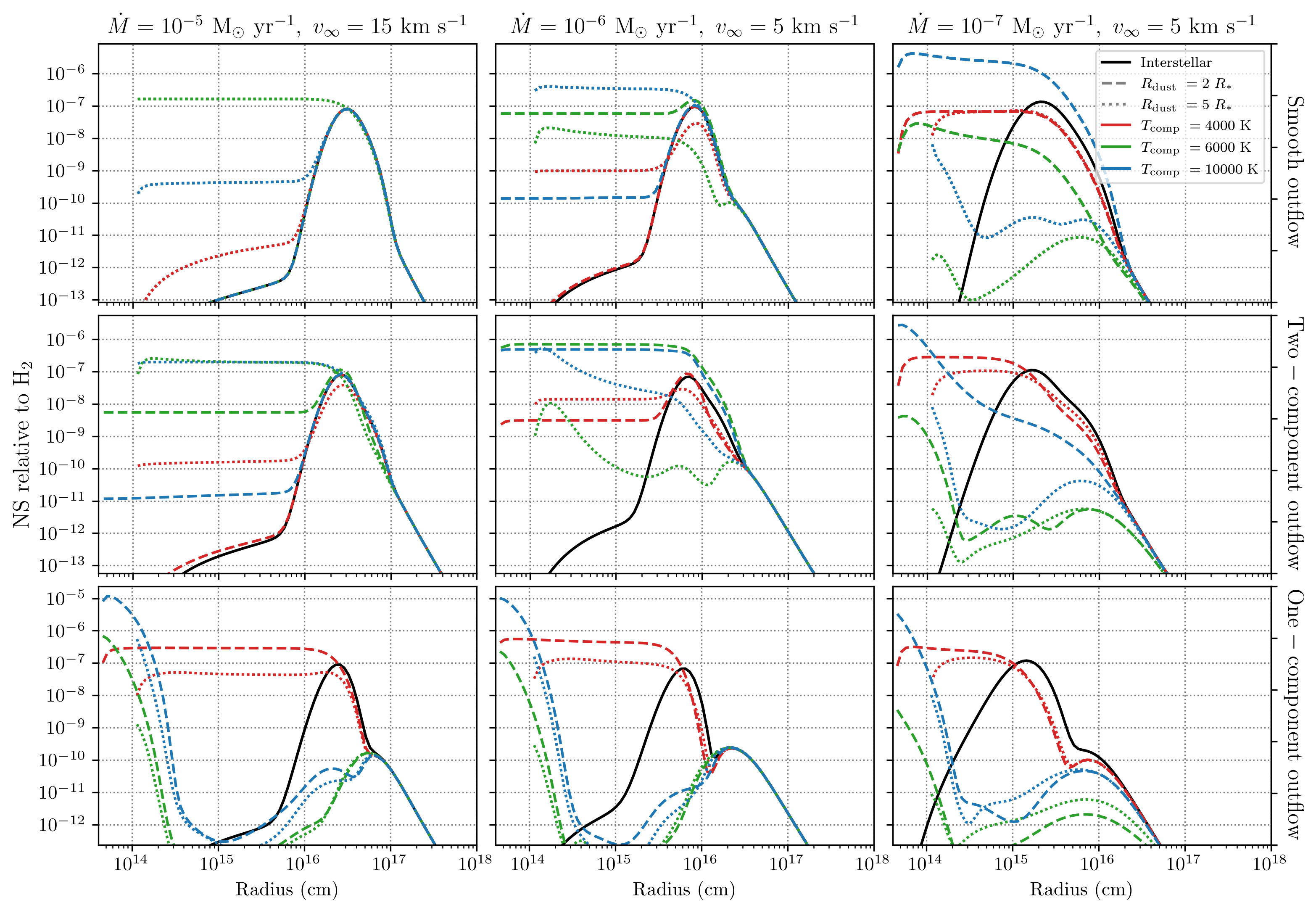}
 \caption{ Abundance of NS relative to H$_2$ in O-rich outflows with different density structures.
 Rows, from top to bottom: smooth outflow, two-component outflow, one-component outflow.
 Black, solid lines: interstellar photons only. 
 Line styles: location of \Rdust\ (dashed 2 R$_*$,  dotted: 5 R$_*$).
 Red: stellar + red dwarf companion, 
 green: stellar + solar-like companion,
 blue: stellar + white dwarf companion.
 }
 \label{fig:fracs-orich-NS}
\end{figure*}

\section{Results}				\label{sect:results}

The presence of a stellar companion in the inner wind can strongly influence the chemistry within the entire outflow. 
Companion photons start a rich photon-driven chemistry, typical for the outermost regions, already in the inner parts of the outflow.
The outcome of the inner photochemistry depends mainly on the extinction experienced by the internal UV photons, determined by the outflow density ($\dot{M}/v_\infty$), the density structure, and \Rdust, as well as on the intensity of the radiation, set by \Tcomp\ and \Rcomp.

Based on the extinction experienced by internal photons, we can differentiate between low UV outflows ($10 < \Delta A_V < 100$ mag) and high UV outflows ($\Delta A_V < 10$ mag), where $\Delta A_V$ is the maximum value of internal extinction at large radial distances.
Outflows with $\Delta A_V > 100$ mag experience very little impact.
In our grid, this is the smooth outflow with $\dot{M} = 10^{-5}$ \msunyr\ and \Rdust = 2 $R_*$.
Low UV outflows are all porous outflows with $\dot{M} = 10^{-5}$ \msunyr, and the smooth and two-component outflow with $\dot{M} = 10^{-6}$ \msunyr.
High UV outflows are the one-component outflow with $\dot{M} = 10^{-6}$ \msunyr\ and all outflows with $\dot{M} = 10^{-7}$ \msunyr.

In high UV outflows, photodissociation and photoionisation are faster than two-body reactions, inhibiting reformation of parents and chemistry among newly formed daughters. 
This can reduce the outflow to a mostly atomic and ionised state, which in smooth outflows is otherwise expected in the outermost regions where the outflow merges with the ISM.
In low UV outflows, chemical complexity increases as two-body reactions can occur before further photodissociation and photoionisation. 
Species that are otherwise produced only in the outer wind can have abundance profiles more similar to those of parent species, with a larger abundance in the inner region followed by a gaussian decline.

Red dwarf companions do not significantly influence the chemistry, with only low extinction outflows (low outflow density and porous density structure) showing an impact, at the same level as that of stellar UV photons (see Fig. \ref{fig:rates-SO}). 
We note that CO self-shielding was not properly taken into account in \citet{VandeSande2019a}, as discussed in \citet{VandeSande2018Err}.
The largest impact, for all outflow densities and structures, is seen for solar-like companions. 
A white dwarf companion has a smaller effect.
The specific outcome of the photochemistry in the inner wind is difficult to predict, as it not only depends on the intensity of UV radiation and the extinction it experiences, but also on the parent species and their assumed abundances.
However, general trends are present.

The results for the O-rich and C-rich outflows are presented in Sects. \ref{subsect:results:orich} and \ref{subsect:results:crich}, respectively. 
Abundance profiles of the more relevant species discussed in the text are included in the supplementary material, along with their calculated column densities.
In order to compare outflows with different \Rdust, and hence different starting radius, column densities are calculated from $8 \times 10^{14}$ cm (40 $R_*$) onwards. 
At this location, the chemistry has adapted to the governing physical conditions.

\begin{figure*}
 \includegraphics[width=1\textwidth]{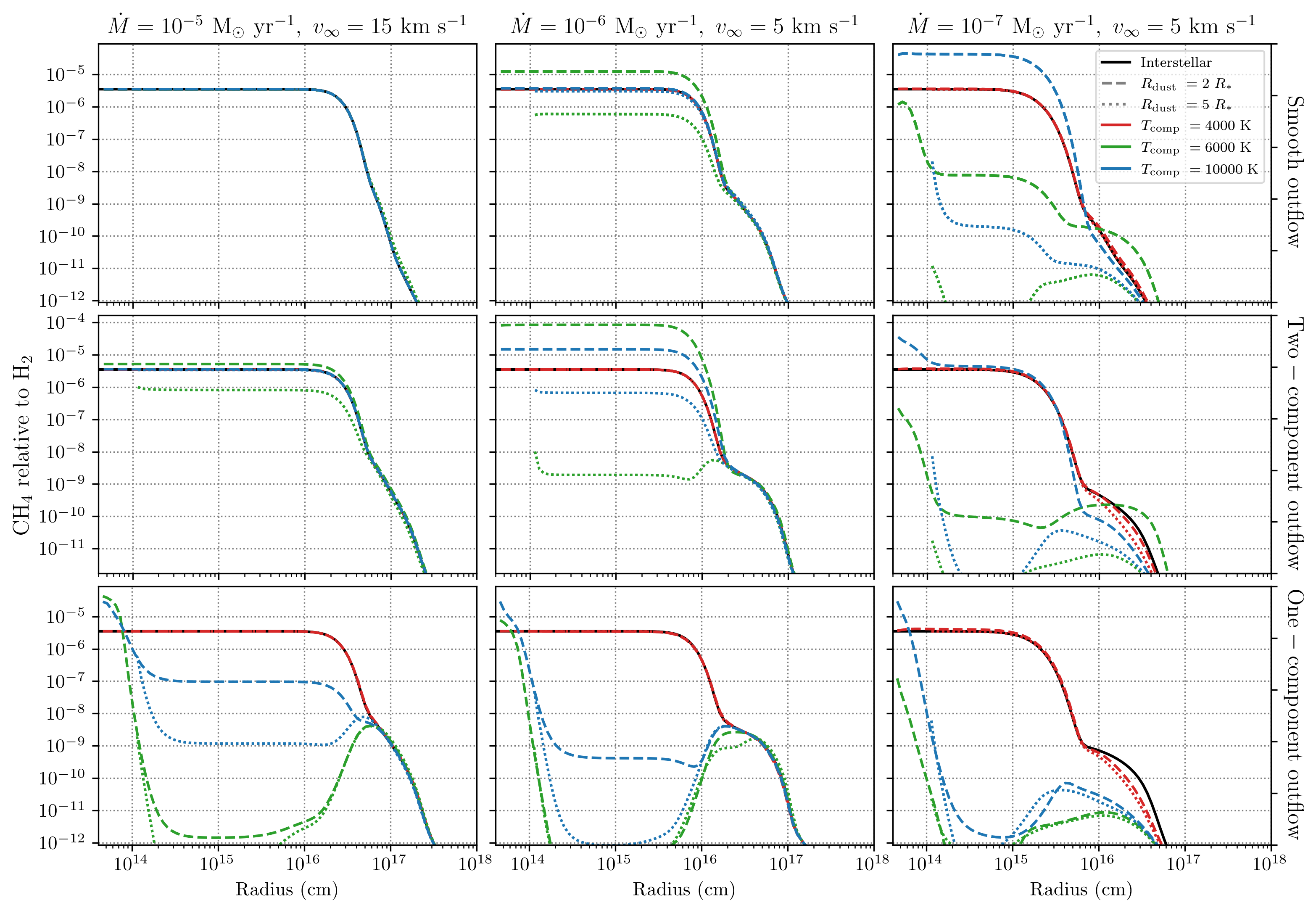}
 \caption{ Abundance of \ce{CH4} relative to H$_2$ in C-rich outflows with different density structures.
 Rows, from top to bottom: smooth outflow, two-component outflow, one-component outflow.
 Black, solid lines: interstellar photons only. 
 Line styles: location of \Rdust\ (dashed 2 R$_*$,  dotted: 5 R$_*$).
 Red: stellar + red dwarf companion, 
 green: stellar + solar-like companion,
 blue: stellar + white dwarf companion.
 }
 \label{fig:fracs-crich-CH4}
\end{figure*}

\begin{figure*}
 \includegraphics[width=1\textwidth]{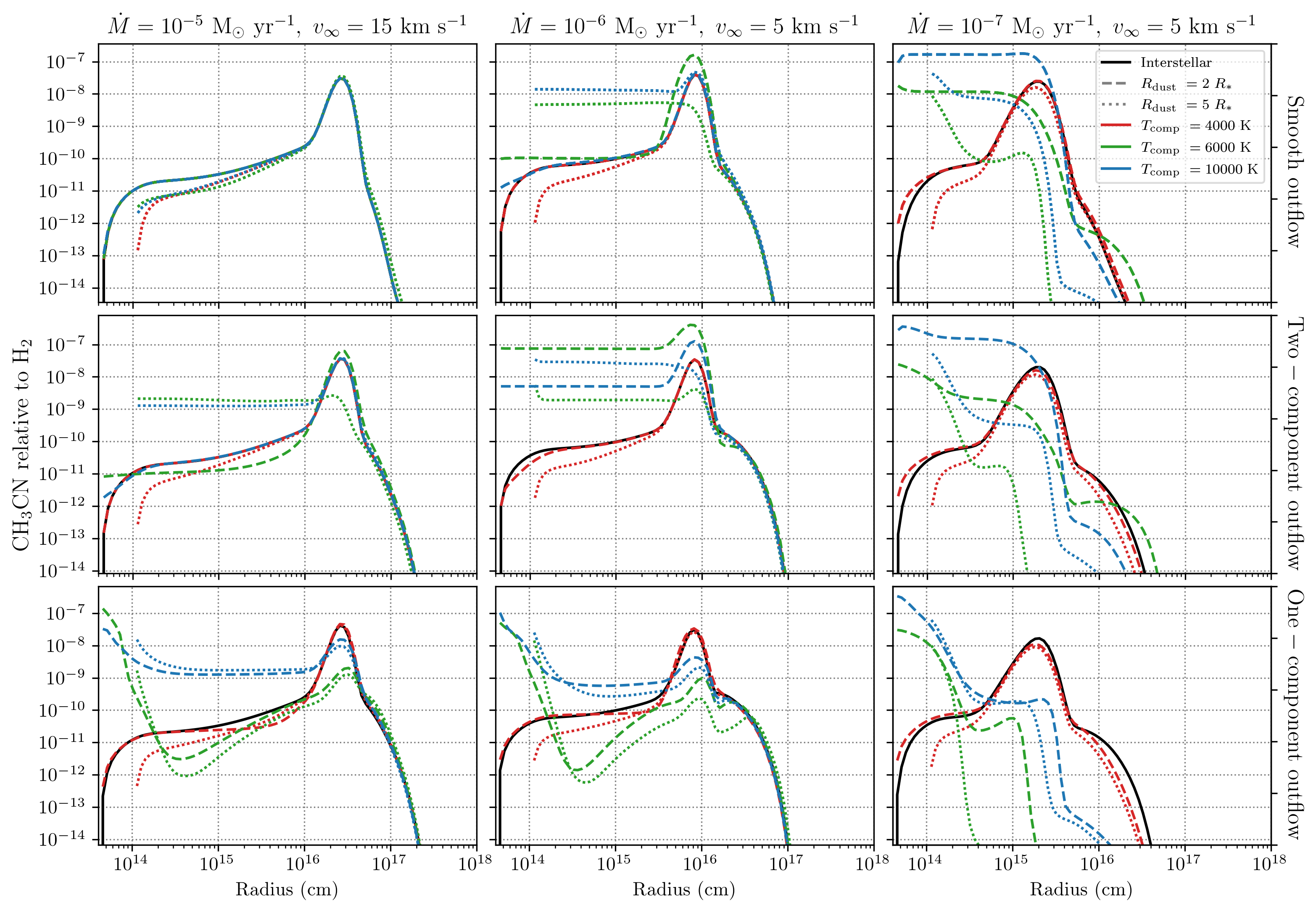}
 \caption{ Abundance of \ce{CH3CN} relative to H$_2$ in C-rich outflows with different density structures.
 Rows, from top to bottom: smooth outflow, two-component outflow, one-component outflow.
 Black, solid lines: interstellar photons only. 
 Line styles: location of \Rdust\ (dashed 2 R$_*$,  dotted: 5 R$_*$).
 Red: stellar + red dwarf companion, 
 green: stellar + solar-like companion,
 blue: stellar + white dwarf companion.
 }
 \label{fig:fracs-crich-CH3CN}
\end{figure*}

\begin{figure*}
 \includegraphics[width=1\textwidth]{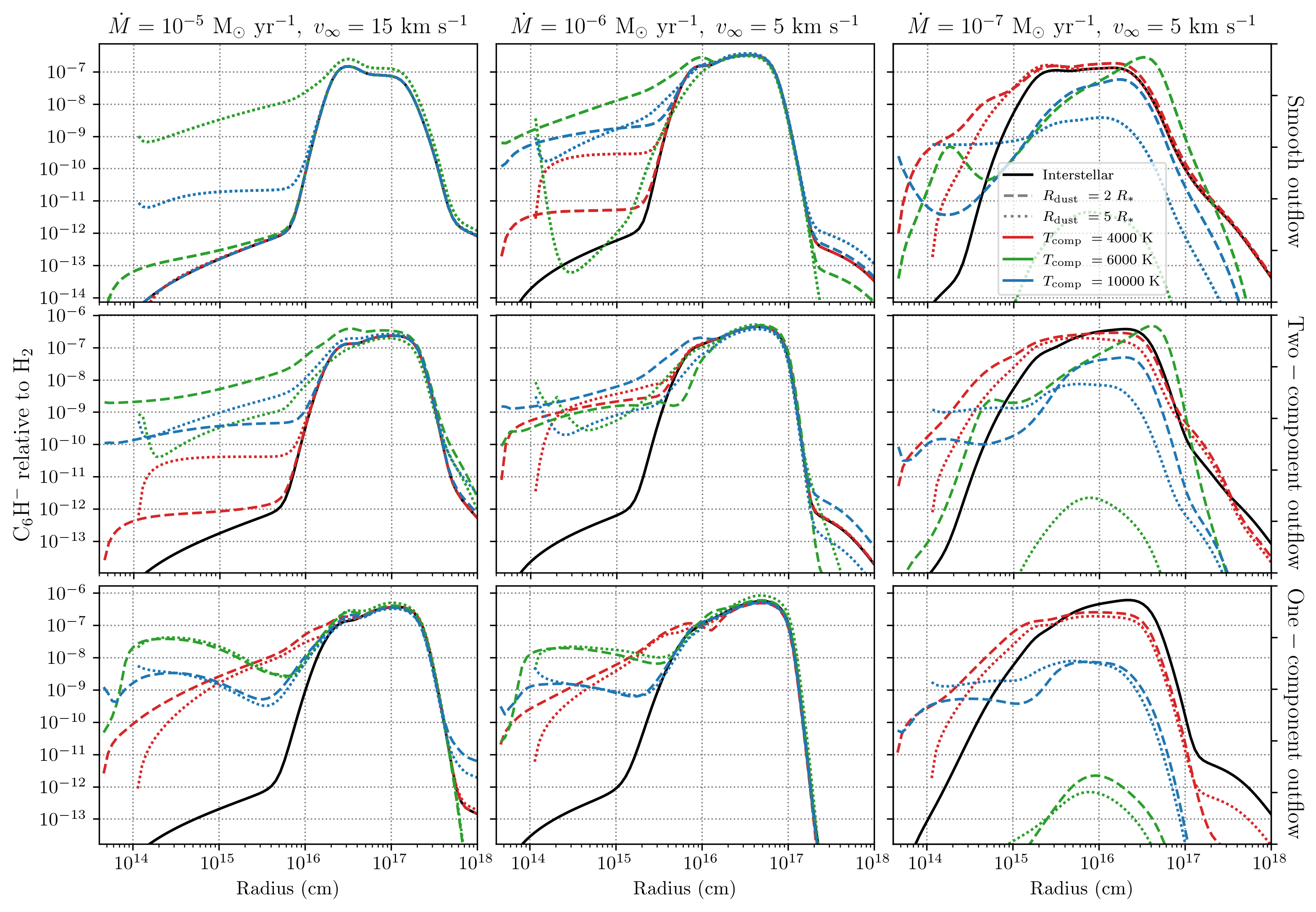}
 \caption{ Abundance of \ce{C6H-} relative to H$_2$ in C-rich outflows with different density structures.
 Rows, from top to bottom: smooth outflow, two-component outflow, one-component outflow.
 Black, solid lines: interstellar photons only. 
 Line styles: location of \Rdust\ (dashed 2 R$_*$,  dotted: 5 R$_*$).
 Red: stellar + red dwarf companion, 
 green: stellar + solar-like companion,
 blue: stellar + white dwarf companion.
 }
 \label{fig:fracs-crich-C6H-}
\end{figure*}

\subsection{O-rich outflows}			\label{subsect:results:orich}

The column densities of selected parent and daughter species compared to those obtained with interstellar UV photons only are listed in Table \ref{table:app-coldens-orich-parents}.
Figure \ref{fig:fracs-orich-SO} shows the abundance of the parent SO in the O-rich grid of outflows. 
While a red dwarf companion only influences the chemistry in high UV outflows, a solar-like or white dwarf companion destroys SO in the inner wind for both low and high UV outflows.
SO can be reformed via S + OH, with the latter formed by interstellar photodissociation of the parent \ce{H2O}.
In high UV outflows, OH itself is photodissociated, inhibiting the reformation of SO. 
The difference in UV radiation between the solar-like and white dwarf companion leads to a different radial behaviour of OH and subsequently SO: OH is more abundant in the presence of the white dwarf, leading to a larger radial extent of the increase in SO abundance.
The abundances of SO and \ce{H2O} are therefore closely linked, together with that of \ce{SO2}, as it is reformed via SO + OH.
In high UV outflows, SO is efficiently photodissociated for the solar-like companion and white dwarf companion, significantly reducing its envelope size.
While a red dwarf companion influences the shape of the SO abundance profile, its envelope size remains unaffected.

S is one of the final photodissociation products of S-bearing parents.
In low UV outflows, it is rapidly hydrogenated to HS and subsequently \ce{H2S}.
Its hydrogenation chain is disturbed in high UV outflows, halting at the formation of HS.
The (re)formation of all hydrogenated parents (\ce{H2S,\ NH3,\ H2O,\ HCN}) and their daughters is sensitive to the balance between two-body reactions and photoreactions.
While the ambient UV radiation field has the largest influence, the outflow density and temperature also play a role due to energy barriers within their hydrogenation chains. 
These radical-\ce{H2} reactions have rates (s$^{-1}$) that increase with gas density and temperature. 
Thus for a specific gas temperature, higher densities can drive hydrogenation faster than UV photons can dehydrogenate.

Figure \ref{fig:fracs-orich-NS} shows the abundance of NS.
Although the column density of the parent \ce{N2} is relatively unaffected in low UV and most high UV outflows, its destruction at the level of a few percent is the main source of N in the inner wind.
NS is produced by N + HS, formed mostly by photodissociation of the parent \ce{H2S}, along with the hydrogenation of S, liberated from the photodissociation of S-bearing parents.
In low UV outflows, the abundance profile of the daughter NS is altered from its typical shell-like form to a more parent-like shape, with a larger inner wind abundance.
Note that this is also the case in high UV outflows with a red dwarf companion, as its radiation field is not strong enough to photodissociate newly formed daughters.

\subsection{C-rich outflows}			\label{subsect:results:crich}

The column densities of selected parent and daughter species compared to those obtained with interstellar UV photons only are listed in Table \ref{table:app-coldens-crich-parents}.
Figure \ref{fig:fracs-crich-CH4} shows the abundance of the parent \ce{CH4} in the C-rich grid of outflows.
Efficient hydrogenation of C, released by the photodissociation of other parents, can lead to an increase in its abundance.
As for the O-rich outflows, the hydrogenation chain is highly sensitive to the balance between two-body reactions and photoreactions.
The column density of the parents HCN, \ce{H2S}, \ce{NH3}, and \ce{H2O} can also increase under favourable conditions.
The increase in photodissociation of the parent \ce{N2} liberates N, which can react with \ce{CH2} to form CN, followed by hydrogenation to HCN. 
The liberated N can also be hydrogenated to \ce{NH3}. 
Similarly, \ce{H2S} is formed via successive hydrogenation of S, liberated from the photodissociation of the parent CS, and \ce{H2O} via successive hydrogenation of O, liberated mostly from the photodissociation of the parent SiO.

Figure \ref{fig:fracs-crich-CH3CN} shows the abundance of \ce{CH3CN}.
In low UV outflows, this complex species can show a parent-like abundance profile, with larger abundances close to the star.
\ce{CH3CN} is formed by reaction of the parent HCN with \ce{CH3+}, formed by the photoionisation of \ce{CH3}, a photodissociation product of the parent \ce{CH4}.
The balance between photoreactions and two-body reactions is hence crucial for the formation of \ce{CH3}, as it should not be photodissociated itself nor should its (re)formation via hydrogenation of \ce{CH2} be impeded.
In high UV outflows, the increase in its inner wind abundance rapidly declines due to efficient photodissociation.
Note that for a red dwarf companion, the radiation field is not strong enough to efficiently photoionise \ce{CH3}.

Figure \ref{fig:fracs-crich-C6H-} shows the abundance of \ce{C6H-}.
Unlike for \ce{CH3CN}, a solar-like and white dwarf companion both yield approximately the same inner wind abundance, which is moreover relatively independent of outflow density and structure.
This is due to similar radiative electron attachment and photodetachment rates for \ce{C6H-}.
Other species, such as \ce{C8H-} and \ce{C3N-}, also show such a balance in radiative electron attachment and photodetachment rates and hence a large and outflow independent inner wind abundance.

\section{Discussion}				\label{sect:discussion}

The internal UV radiation of a stellar companion can have a large impact on the chemistry throughout the outflow.
The effect depends mainly on the extinction experienced by the internal UV photons, $\Delta A_V$, and the density structure of the outflow. 
The intensity of the radiation, set by \Tcomp\ and \Rcomp, plays a secondary role since the intensity is exponentially dependent on extinction. 
We find that a red dwarf companion does not significantly influence the chemistry in low UV outflows, with an effect similar to that of stellar UV photons.

Outflows with $\Delta A_V > 100$ at large radial distances are not significantly affected.
In low UV outflows, with $10 < \Delta A_V < 100$ mag, the outcome of the chemistry critically depends on the balance between photodissociation and two-body reactions.
Parents are only partly photodissociated, so that atoms and radicals are liberated while the parents themselves are still present and can participate in the rich photochemistry of the inner wind. 
The subsequent two-body reactions determine the reformation of parents and production of daughters in the inner region. 
This transforms the shell-like abundance profile of some daughter species to a more parent-like profile, with an increased abundance in the inner wind.
Energy barriers of certain reactions, most notably hydrogenation reactions, play an important role in (re)production pathways.

High UV outflows, with $\Delta A_V <10$ at large radial distances, are most affected by internal UV radiation. 
The low extinction experienced by internal photons leads to a strong decline in the abundance of parent species.
Photoreactions are faster than two-body reactions in high UV outflows, reducing the chemistry to a mostly atomic and ionised state from the inner region onwards.
As a result, the outflow appears to be molecule-poor: both parents and daughters are readily photodissociated and photoionised.
Because of the limited impact of a red dwarf companion, these outflows do not appear molecule-poor.

The origin of chemical complexity in the inner wind is discussed in Sect. \ref{subsect:discussion:complexity}.
Whether chemistry can be used as a tool to detect stellar companions is discussed in Sect. \ref{subsect:discussion:starplanet}.
We address the formation of unexpected C-bearing species in O-rich outflows, and vice versa, in Sect. \ref{subsect:discussion:unexpected}.
The modelling results are compared to observations in Sect. \ref{subsect:discussion:obs}.

\subsection{Inner wind chemical complexity}			\label{subsect:discussion:complexity}

Chemical complexity can increase in the inner wind of low UV outflows, as two-body reactions among newly formed daughter species are faster than photoreactions.
This can lead to daughter species showing abundance profiles typical of parent species, with a large inner wind abundance followed by a gaussian decline.
Although photodissociation of the parent \ce{N2} does not significantly impact its column density, it leads to the formation of NS in O-rich and C-rich outflows. 
Other N-bearing species can also be abundantly produced, such as SiN. 
Its main formation pathway is via the parent \ce{NH3} reacting with \ce{Si+} to form \ce{SiNH2+}, which can dissociatively recombine with electrons to form \ce{SiN}.
In both O-rich and C-rich outflows, the photodissociation of SiS is the main source of Si and consecutively of \ce{Si+}.

The increase in chemical complexity and diversity in the inner wind is larger for C-rich outflows thanks to the reactivity of carbon.
Photodissociation of the parent \ce{C2H2} leads to the formation of the radicals \ce{C2H} and \ce{C2}.
The continued presence of \ce{C2H2} in low UV outflows allows it to participate in the chemistry, starting off the reaction chains forming \ce{C_{2n}H2} and \ce{C_{2n}H} ($n = 1,2,...$), respectively, aiding to the formation of \ce{HC_{2n+1}N} ($n = 1,2,...$, (supplementary material)).
Reactions of \ce{C2H2} with photodissociation products Si and \ce{Si+} form \ce{SiC2}, \ce{SiC2H} and \ce{SiC2H^+}, starting off a Si-C chemistry.
The formation of \ce{CH3CN} (Fig. \ref{fig:fracs-crich-CH3CN}) requires the presence of the parent HCN.
It is therefore necessary that parents \ce{C2H2}, \ce{CH4}, and \ce{HCN} are only partly photodissociated for complexity to increase, highlighting again the balance between two-body reactions and photoreactions.

\ce{CH3} is a gateway species to the inclusion of atoms other than C in daughter species.
In O-rich outflows, it is mainly formed via the hydrogenation of C, mostly liberated from CS. 
Reactions of \ce{CH3} with Si or S leads to \ce{SiCH2} or \ce{H2CS}, respectively. 
\ce{SiCH2} can be photodissociated into HCSi, followed by SiC.
In C-rich outflows, \ce{CH3} is more abundant thanks to the presence of the parent \ce{CH4} and its reactions with Si, \ce{Si+}, S and N lead to \ce{SiCH2}, \ce{SiCH2+}, \ce{H2CS} and \ce{H2CN}, respectively.
Because of the high \ce{H2 + S} and \ce{H2 + Si} reaction barriers, \ce{H2CS}, \ce{SiCH2} and \ce{SiCH2+} are more efficiently formed.

\subsection{Chemistry as a tool to detect stellar companions}			\label{subsect:discussion:starplanet}

Observations and non-observations of specific molecules could help identify a stellar companion. 
However, the model predictions are highly dependent on the extinction experienced by internal UV photons (determined by the outflow density and structure), the temperature and radius of the companion, as well as the assumed parent species and their relative abundances.
Certain trends can be distinguished, keeping in mind that specific outflows will require specific chemical models.

In general, a red dwarf companion does not significantly impact the chemistry throughout the outflow, with an effect comparable to that of stellar UV photons (Fig. \ref{fig:rates-SO}). 
A solar-like companion has a larger influence than a 10 000 K white dwarf due to their difference in size.
High UV outflows (low outflow density and/or highly porous density structure) with a solar-like or white dwarf companion are likely to appear molecule-poor.
CO is often the only detectable molecule.
The envelope size of the other parents and of daughters is severely reduced.
The presence of atoms and atomic ions, especially combined with a lack of molecules, appears to be a clear indicator of the presence of a (hot) stellar companion.

In low UV, O-rich outflows, the presence in the inner wind of N-bearing species such as NS and SiN, as well as SiC, point towards the presence of a solar-like or white dwarf companion.
Retrieving the abundance profile can also indicate its presence: while the SO column density remains large, its abundance profile shows a clear shell-like shape. 
SO$_2$ behaves similarly.
In low UV, C-rich outflows, a large inner wind abundance of various complex species (e.g., \ce{CH3CN,\ HC3N}, \ce{C6H-}) can indicate the presence of a solar-like or white dwarf companion.
The parents SiO, SiS, and \ce{NH3} are severely reduced in abundance. 
The behaviour of the other parents and daughters, in particular of hydrogenated species, is however highly dependent on the exact input parameters.

\begin{figure*}
 \includegraphics[width=1\textwidth]{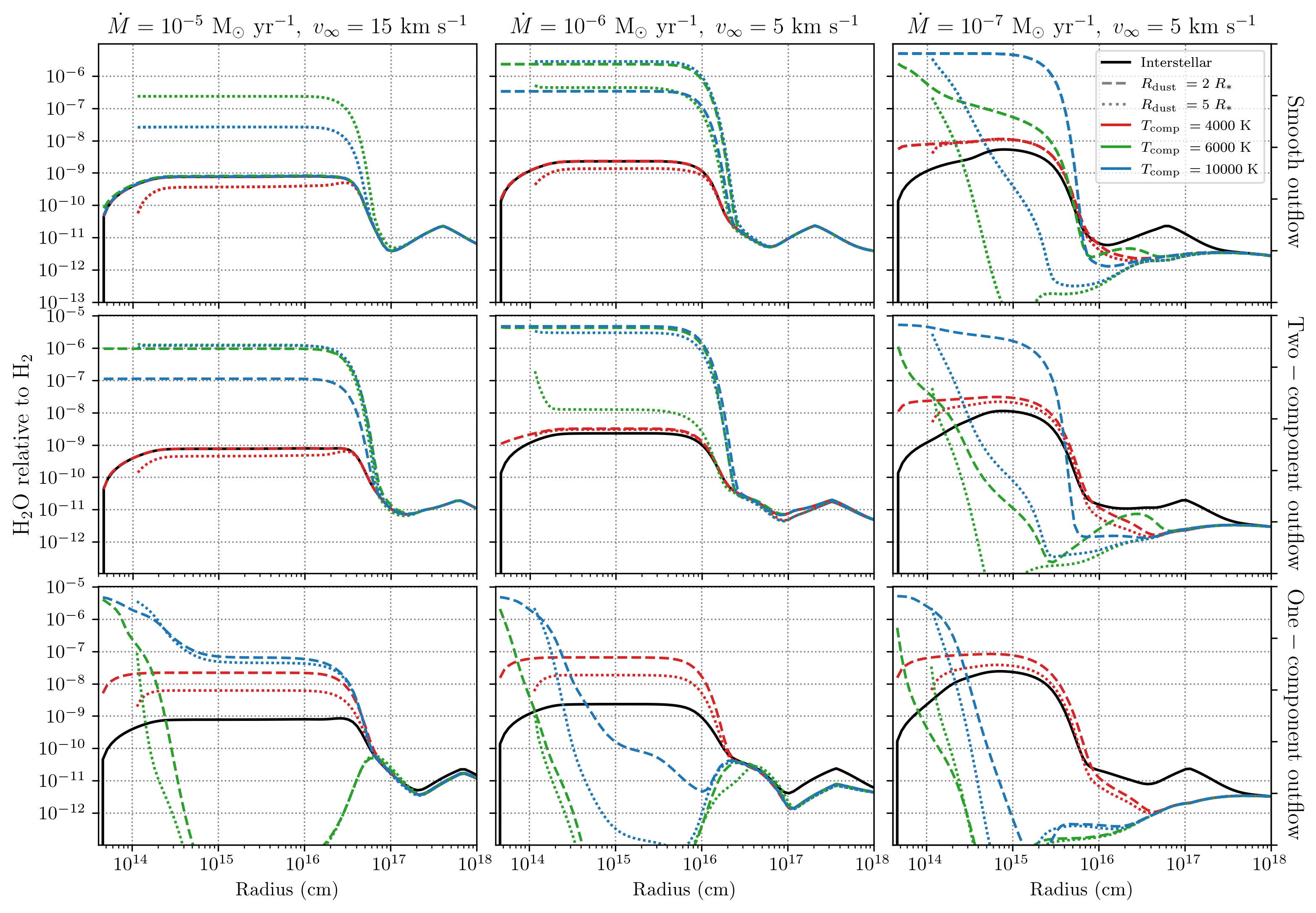}
 \caption{ Abundance of \ce{H2O} relative to H$_2$ in C-rich outflows with different density structures.
 The unexpected species \ce{H2O} and \ce{NH3} are not included as parent species.
 Rows, from top to bottom: smooth outflow, two-component outflow, one-component outflow.
 Black, solid lines: interstellar photons only. 
 Line styles: location of \Rdust\ (dashed 2 R$_*$,  dotted: 5 R$_*$).
 Red: stellar + red dwarf companion, 
 green: stellar + solar-like companion,
 blue: stellar + white dwarf companion.
 }
 \label{fig:fracs-crich-H2O}
\end{figure*}

\begin{figure*}
 \includegraphics[width=1\textwidth]{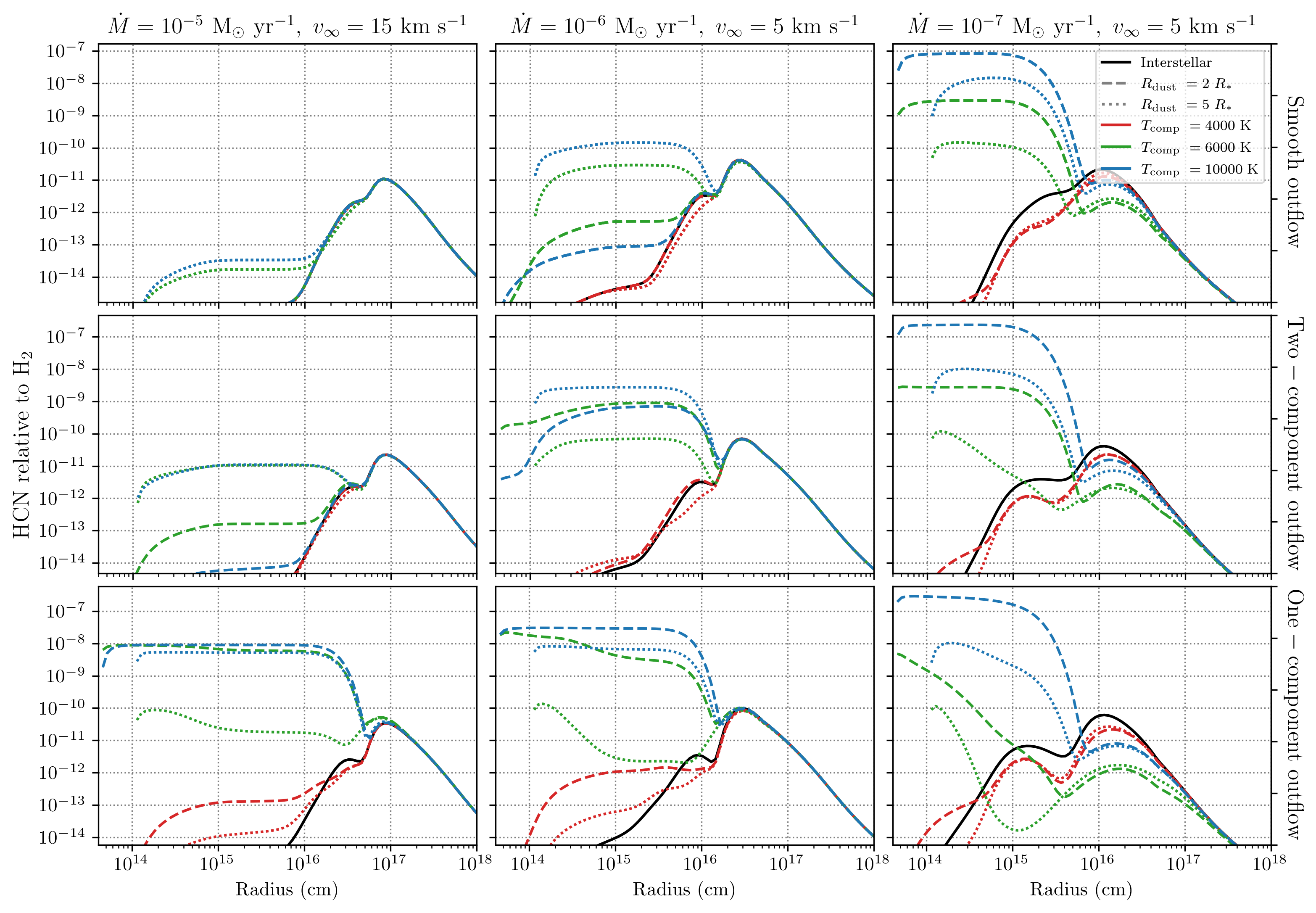}
 \caption{ Abundance of HCN relative to H$_2$ in O-rich outflows with different density structures.
 The unexpected species CS, HCN, and \ce{NH3} are not included as parent species.
 Rows, from top to bottom: smooth outflow, two-component outflow, one-component outflow.
 Black, solid lines: interstellar photons only. 
 Line styles: location of \Rdust\ (dashed 2 R$_*$,  dotted: 5 R$_*$).
 Red: stellar + red dwarf companion, 
 green: stellar + solar-like companion,
 blue: stellar + white dwarf companion.
 }
 \label{fig:fracs-orich-HCN}
\end{figure*}

\subsection{Formation of unexpected species}			\label{subsect:discussion:unexpected}

Several so-called unexpected species have been detected in the inner winds of AGB stars.
Under the assumption of thermodynamic equilibrium (TE) in the inner wind, these species are expected to have a smaller abundance than observed or be absent.
In O-rich outflows, the unexpected species are C-bearing species, such as HCN, CS, and CN \citep{Omont1993,Bujarrabal1994,Justtanont1996}.
In C-rich outflows, they are O-bearing species, such as \ce{H2O} \citep{Decin2010a,Neufeld2011}, but also complex species such as \ce{CH3CN} \citep{Agundez2015}.
Non-TE inner wind chemical models that take the effect of shocks, caused by stellar pulsations, into account are relatively successful in reproducing the abundances of the unexpected species \cite[e.g.,][]{Cherchneff2006}. 
However, they fail to reproduce certain species, such as \ce{NH3} in the high density O-rich outflow of IK Tau \citep{Gobrecht2016}.
A clumpy density distribution could help explain the presence of these unexpected species.
Non-TE models do not include photodissociation, while a porous outflow can allow interstellar UV photons to reach the inner wind.
This allows C to be liberated in the inner wind in O-rich outflows, and similarly O in C-rich outflows.
\citet{Agundez2010} included clumpiness by allowing a certain fraction of UV photons to reach the inner winds. 
\citet{VandeSande2018} introduced the statistical porosity formalism for a more general approach.
However, while both models find that the abundance of \ce{NH3} can increase in both O-rich and C-rich outflows, it does not reach the observed values, and the effects on the other unexpected species remains limited \citep{Agundez2010,VandeSande2018Err}.

To investigate whether the presence of a stellar companion can yield enough UV radiation in the inner wind to produce unexpected species, we recalculated our grid of models with adapted sets of parent species.
\ce{H2O} and \ce{NH3} are omitted from the C-rich parent species; and HCN, CS, and \ce{NH3} from the O-rich parent species.

Figure \ref{fig:fracs-crich-H2O} shows the abundance of \ce{H2O} in the grid of C-rich outflows obtained using the adapted set of parent species.
\ce{H2O} can be abundantly formed in the inner winds of most C-rich outflows.
O is released from the photodissociation of mostly the parent SiO, which is then successively hydrogenated to \ce{H2O}.
Solar-like or white dwarf companions yield the largest inner wind abundances. 
A maximum abundance of roughly $5 \times 10^{-6}$ relative to \ce{H2} is achieved in smooth and two-component outflows.
For one-component outflows, the large inner wind abundance rapidly decreases. 
A red dwarf companion does not result in a larger inner wind abundance, except for in the one-component outflows where a maximum abundance of about $10^{-7}$ relative to \ce{H2} can be reached.
The abundances obtained in low UV outflows are in agreement with observations, which lie around $10^{-7} - 10^{-5}$ relative to \ce{H2} \citep{Decin2010a,Lombaert2016}.

The abundance of \ce{NH3} is shown in the supplementary material. 
A red dwarf companion does not lead to an increase in the inner wind abundance of \ce{NH3} in any of the outflows.
A solar-like companion can increase its inner wind abundance up to some $10^{-9}$ relative to \ce{H2}, a white dwarf companion up to a few times $10^{-6}$ relative to \ce{H2}. 
\ce{NH3} has been observed in the high density outflow of IRC+10216 with an abundance of $1.7 \times 10^{-7}$ relative to \ce{H2}. 
Our high density models reach an inner wind abundance of only $10^{-9}$ in the two-component case, suggesting that non-TE effects play a large role in the formation of \ce{NH3}.

Figure \ref{fig:fracs-orich-HCN} shows the abundance of HCN in the grid of O-rich outflows obtained using the adapted set of parent species.
A red dwarf companion does not result in a larger inner wind abundance in any of the outflows.
The presence of a solar-like or white dwarf companion can dramatically increase the inner HCN abundance, with lower outflow densities and more porous density structures yielding larger abundances.
C is liberated from photodissociation of CO. 
This does not impact the CO abundance because of its large abundance and its (re)formation via photodissociation of \ce{CO2}.
The abundances obtained are consistent with observations of R Dor, which has an inner wind abundance of $5 \times 10^{-7}$ relative to \ce{H2} in an outflow with $\dot{M} \approx 10^{-7}$ \msunyr \citep{VandeSandeRDor}. 
They are an order of magnitude too low for IK Tau, with $\dot{M} = 5 \times 10^{-6}$ \msunyr \citep{Decin2010b}, and two orders of magnitude too low for TX Cam, with $\dot{M} = 3 \times 10^{-6}$ \msunyr \citep{Bujarrabal1994}.
Similarly, our results are consistent with the HCN abundances derived by \citet{Schoier2013} for lower density outflows and not for higher density outflows.
These differences could be mitigated by a more porous outflow or with some HCN as a parent, produced by non-TE chemistry.

The abundances of \ce{NH3} and CS are shown in the supplementary material.
The results obtained for \ce{NH3} are similar to those obtained in the C-rich outflows. 
It has been observed in the outflow of IK Tau with an abundance of $6 \times 10^{-7}$ relative to \ce{H2} \citep{Decin2010b}, with other outflows showing a similar abundance of about $10^{-6}$ \citep{Wong2018}.
Such abundances are again only achieved in the two-component case for a solar-like or white dwarf companion.
CS is not efficiently producetd in the inner wind, with an increase in inner wind abundance of only an order of magnitude, reaching up to $10^{-9}$ relative to \ce{H2}.
This is far below the observed abundances around IK Tau and TX Cam of approximately $10^{-7}$ and $10^{-6}$, respectively \citep{Bujarrabal1994,Decin2010b,Kim2010,Lindqvist1988,Olofsson1991}.
A combination of non-TE chemistry producing (some) CS in the inner wind appears to be necessary.
Including CS as a parent with its original abundance (Table \ref{table:model-parents}) increases the inner HCN abundance by an order of magnitude.

\begin{figure*}
 \includegraphics[width=1\textwidth]{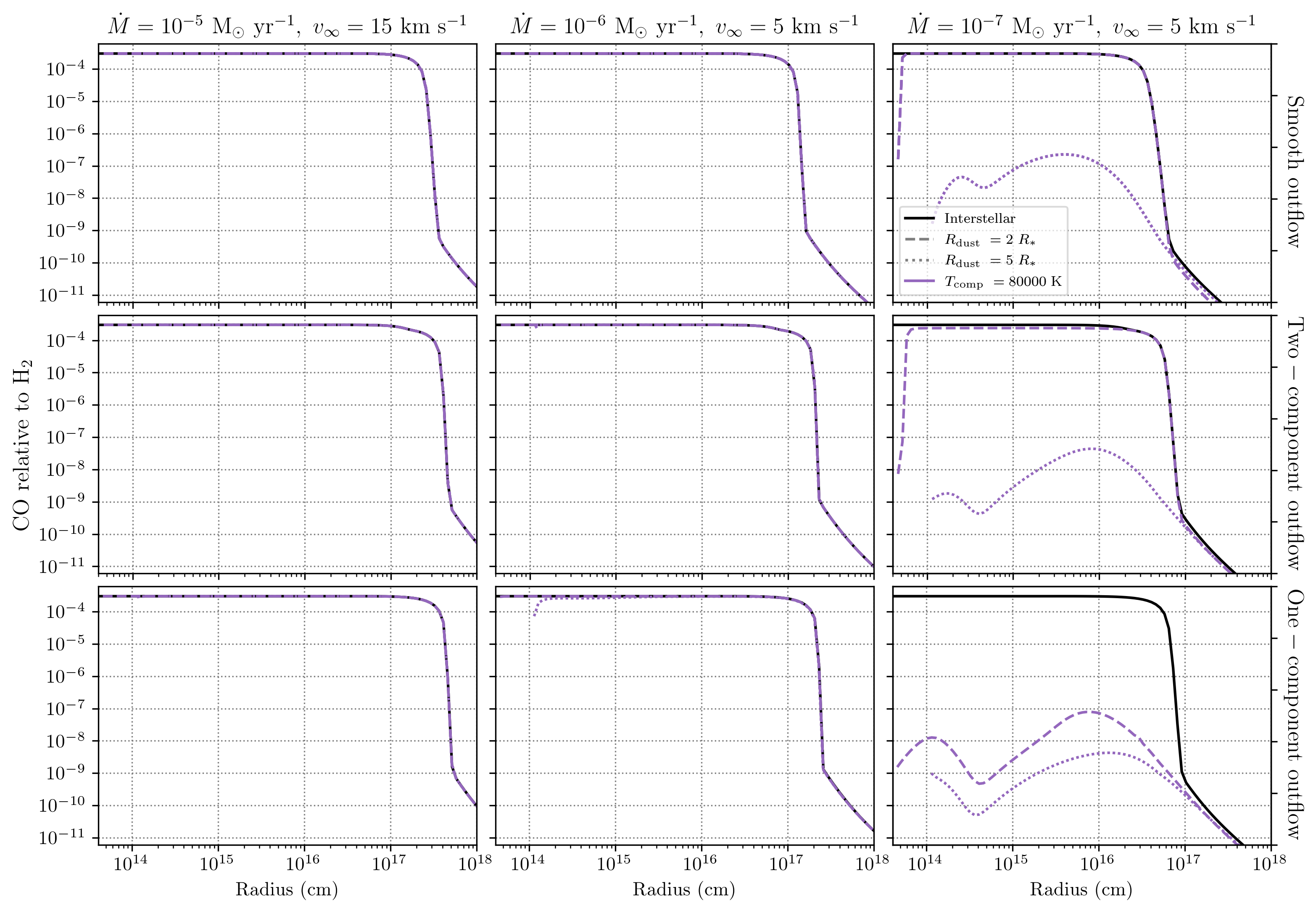}
 \caption{ Abundance of CO relative to H$_2$ in O-rich outflows with different density structures and a 80 000 K white dwarf companion.
 Black, solid lines: interstellar photons only. 
 Line styles: location of \Rdust\ (dashed 2 R$_*$,  dotted: 5 R$_*$).
 Purple: stellar + white dwarf companion with \Tcomp\ = 80 000 K.
 }
 \label{fig:fracs-orich-CO-80000K}
\end{figure*}

\subsection{Comparison to observations}			\label{subsect:discussion:obs}

CI emission has been detected in a shell-like structure around C-rich AGB stars, where it appears to be formed by interstellar UV photodissociation of CO \citep{Keene1993,vanderVeen1998,Olofsson2015,Knapp2000}.
\citet{Saberi2018} detected CI emission in the Mira AB system, where the primary Mira A (omi Cet) is an O-rich AGB star and the secondary Mira B is likely to be a white dwarf with a temperature around 20 000 K \citep{Sokoloski2010}.
The outflow of Mira A has a mass-loss rate of a few times $10^{-7}$ \msunyr\ and contains a bubble-like structure, formed by material blowing from the primary to the secondary \citep{Ramstedt2014}. 
\citet{Saberi2018} suggest that the CI emission arises from a more compact region near Mira B and find a CI column density of $1.1 \times 10^{19}$ cm$^{-2}$. 
We do not find such large column densities for C in our O-rich models with $\dot{M} = 10^{-7}$ \msunyr and a white dwarf companion (Table \ref{table:app-coldens-orich-parents}).
The Mira AB system has a large separation of approximately $6.9 \times 10^{14}$ cm \citep{Ramstedt2014}, which corresponds to $\sim35\ R_*$ in our model.
Our models assume a close-by companion within the dust formation zone of the AGB star.
The influence of the orbital motion and differences in dust extinction may underlie the discrepancy with the CI observations.

\citet{Bujarrabal2021} investigated the intricate structure within inner wind of the symbiotic system R Aqr, composed of an AGB star and a 80 000 K white dwarf.
They found that photodissociation caused by the white dwarf destroys most molecules, including CO, except for in the densest regions. 
They also found evidence for time-dependence effects associated with the white dwarf's orbital motion. 
The separation between the two stars ($1.6 \times 10^{14}$ cm = 8 $R_*$ in our model) reasonably lies within our model assumption of a close-by companion.
However, R Aqr's orbit is relatively long (about 42 years) and has a large eccentricity. 
These factors make our model not representative, as orbital motion is not included.
Nevertheless, we do not find that CO is destroyed in the outflows assumed in this paper.
When increasing the temperature of the white dwarf to 80 000 K, CO is efficiently destroyed in low mass-loss rate outflows, with larger decreases in abundance seen for higher porosities.
This is shown Figure \ref{fig:fracs-orich-CO-80000K}.
Outflows with $\dot{M} = 10^{-6}$ \msunyr only show a significant destruction for outflows with a highly porous density structure. 
As noted by \citet{Bujarrabal2021}, modelling the complex system of R Aqr - and by extension all binary models - requires a tailored and hydrochemical approach.

The shape of the SO abundance profile in O-rich outflows has been observed to depend on the mass-loss rate, with higher density outflows showing a shell-like shape with a lower initial abundance, while lower density outflows show the typical gaussian profile expected from parent species \citep{Danilovich2016,Danilovich2020}.
The \ce{SO2} abundance profile might follow this behaviour \citep{Danilovich2020}.
Higher density outflows are observed to have \ce{H2S} present within their outflows, showing a gaussian-like profile \citep{Danilovich2017}.
Our O-rich results show that the presence of a companion changes the abundance profile of the parents SO and SO$_2$ to a shell-like shape in higher density outflows (Figs \ref{fig:fracs-orich-SO} and supplementary material).
In lower density outflows, both species are efficiently photodissociated. 
\ce{H2S} retains its gaussian profile for higher density outflows with a slightly porous density structure and is efficiently destroyed in lower density outflows (supplementary material).
In the case of IK Tau, which has a mass-loss rate of $5 \times 10^{-6}$ \msunyr\ \citep{Danilovich2016}, the shell-like shape of SO combined with the presence of \ce{H2S} indicates against the presence of a stellar companion.
Therefore, a broad overview of the molecular content of the outflow is crucial. 

\citet{VelillaPrieto2017} detected NS in the outflow of the O-rich IK Tau, characterised by $\dot{M} \sim 6 \times 10^{-6}$ \msunyr\ \citep{Decin2010a}.
They retrieved an abundance of about $10^{-8}$ relative to \ce{H2}.
This corresponds to the results of our higher density outflow models (Fig. \ref{fig:fracs-orich-NS}). 
\citet{Decin2018} detected NS in IK Tau, but not in the low mass-loss rate outflow of R Dor, which is thought to harbour an equatorial density enhancement caused by binary interaction \citep{Homan2018}.
We find that NS is rapidly destroyed in low density outflows with a higher porosity and a solar-like or white dwarf companion. 
However, \citet{VelillaPrieto2017} also detected \ce{H2CO} around IK Tau with an abundance of $\sim 10^{-7} - 10^{-8}$ relative to \ce{H2}.
Our models do not predict an increase in inner wind abundance of formaldehyde (supplementary material), suggesting non-TE effects might be at play.

SiN has been detected in the inner winds of the C-rich IRC+10216 \citep{Turner1992} and the S-type star W Aql \citep{DeBeck2020}, which has a mass-loss rate of $3 \times 10^{-6}$ \msunyr\ and shows evidence of eccentric binary interaction \citep{Ramstedt2017}.
The SiN abundance in W Aql is some $4 \times 10^{-8}$ relative to \ce{H2}, about five times larger than its abundance in IRC+10216 \citep{DeBeck2020}.
In both the O-rich and C-rich model results, we find that the presence of a companion leads to the efficient production of SiN in higher density outflows, in agreement with the observed abundances (supplementary material).

\citet{Agundez2015} retrieved the spatial distribution of \ce{CH3CN} in the high mass-loss rate outflow of the C-rich star IRC+10216.
The outflow of IRC+10216 contains a spiral structure, thought to be induced by binary interaction \citep{Mauron2000,Leao2006,Decin2015}.
This complex species shows a profile different to other daughter species. 
They found that methyl cyanide is formed close to the star, within 2\arcsec or $\sim 3.5 \times 10^{15}$ cm at 120 pc \citep{Groenewegen2012}. 
This is not predicted by standard chemical models nor their model with a lower extinction throughout the entire outflow.
We find that the presence of a solar-like or white dwarf companion in a porous, high density outflow can increase the inner wind abundance of \ce{CH3CN}.
As with the formation of unexpected species (Sect. \ref{subsect:discussion:unexpected}), an internal source of UV radiation is necessary to initiate complex chemistry close to the star.

The anions \ce{C4H-}, \ce{C6H-} and \ce{C8H-} have been observed in the outflow of IRC+10216 \citep{McCarthy2006,Cernicharo2007,Remijan2007}.
Their column densities have been retrieved, along with those of their neutral counterparts.
The ratio of the \ce{C4H-} and \ce{C4H} column density is $2.4 \times 10^{-5}$ \citep{Cernicharo2007}, that of \ce{C6H-} and \ce{C6H} is 0.17 \citep{McCarthy2006,Cernicharo2007}, and that of \ce{C8H-} and \ce{C8H} is 0.32 \citep{Remijan2007}.
In outflows with $\dot{M} = 10^{-5}$ \msunyr, the presence of a stellar companion can increase the inner wind abundance of \ce{C6H-} and \ce{C8H-} and decrease that of \ce{C4H-} (Fig. \ref{fig:fracs-crich-C6H-} and supplementary material).
The inner wind abundances of \ce{C4H}, \ce{C6H} and \ce{C8H} can be larger, showing a more parent-like abundance profile.
The column density of these molecules increases for most outflow densities and structures, except for \ce{C4H-}.
Therefore, the ratios are relatively stable for higher density outflows and correspond well to those retrieved for IRC+10216. 
Consequently, the abundance profiles shapes are crucial to determine and quantify the presence of a stellar companion.

\section{Conclusions}			\label{sect:conclusions}

We presented the first calculations of the impact of stellar companion UV photons on the chemistry throughout an AGB outflow.
The chemical model is a one-dimensional approximation based on our previous work on stellar UV photons.
We find that the effect of a stellar companion depends mainly on the extinction experienced by its UV radiation, $\Delta A_V$, set by the outflow density, the onset of dust extinction, and the density structure of the outflow, as well as on the intensity of its radiation, set by the stellar radius and blackbody temperature.

Outflows with $\Delta A_V > 100$ mag at large radial distances experience very little impact.
In low UV outflows, with $10 < \Delta A_V < 100$ mag, chemical complexity increases.
Parents are only partly photodissociated, which initiates a rich photochemistry of the inner wind. 
Two-body reactions can occur before further photodissociation and photoionisation, leading to the reformation of parents and production of daughters in the inner region. 
Daughter species can now show abundance profiles with a larger abundance in the inner region followed by a gaussian decline, i.e. more similar to those of parent species.
In high UV outflows, with $\Delta A_V < 10$ mag, photodissociation and photoionisation are faster than two-body reactions.
This reduces the outflow to a mostly atomic and ionised state and makes them apparently molecule-poor. 

The chemical composition of the outflow can be used as a tool to detect stellar companions.
However, the model is not only highly dependent on the extinction experienced by internal UV photons and the size and temperature of the companion, but also on the specific parent species and their initial abundances.
Generally, we find that red dwarf companions do not significantly influence the chemistry, with an effect similar to that of stellar UV photons. 
Solar-like companions have a larger impact on the chemistry than the white dwarf companion considered here, because of the latter's compact size. 
In low UV outflows, more complex or uncommon species can be formed in the inner wind, such as \ce{CH3CN} and \ce{C6H-} in C-rich outflows and NS in O-rich outflows.
The shape of the abundance profile of daughter species is another powerful indicator.
The apparent lack of molecules in high UV outflows can point to the presence of a solar-like or white dwarf companion, especially when atoms and their ions are detected close to the star.

The additional source of UV radiation in the inner wind can lead to the formation of unexpected species in O-rich and C-rich outflows.
While our models can reproduce the observed abundances of \ce{H2O} in C-rich outflows and HCN in O-rich outflows, non-thermodynamical equilibrium effects appear to play a role in the formation of CS in O-rich outflows and \ce{NH3} in both C-rich and O-rich outflows.

Several approximations go into the chemical model.
Our results are the first to quantify the impact of a stellar companion's UV field and show that further model development is necessary when modelling specific, asymmetric outflows.
Future (three-dimensional) chemical models should take the orbital motion of the companion and its impact on the morphology into account.
Moreover, modelling the outflows of specific binary systems will require specific chemical models.

\section*{Acknowledgements}

We thank the referee for their careful reading of the work and valuable input.
MVdS acknowledges support from the Research Foundation Flanders (FWO) through grant 12X6419N and the European Union’s Horizon 2020 research and innovation programme under the Marie Skłodowska-Curie grant agreement No 882991.
TJM gratefully acknowledges the receipt of a Leverhulme Emeritus Fellowship and the STFC for support under grant reference ST/P000312/1 and ST/T000198/1.
We also thank Hanne Baeyens and Jolien Covens for their bachelor thesis work, which contributed to the discussion.

\section*{Data Availability}
 
The data underlying this article will be shared on reasonable request to the corresponding author.



\bibliographystyle{mnras}
\bibliography{chemistry} 

\begin{thebibliography}{}
\makeatletter
\relax
\def\mn@urlcharsother{\let\do\@makeother \do\$\do\&\do\#\do\^\do\_\do\%\do\~}
\def\mn@doi{\begingroup\mn@urlcharsother \@ifnextchar [ {\mn@doi@}
  {\mn@doi@[]}}
\def\mn@doi@[#1]#2{\def\@tempa{#1}\ifx\@tempa\@empty \href
  {http://dx.doi.org/#2} {doi:#2}\else \href {http://dx.doi.org/#2} {#1}\fi
  \endgroup}
\def\mn@eprint#1#2{\mn@eprint@#1:#2::\@nil}
\def\mn@eprint@arXiv#1{\href {http://arxiv.org/abs/#1} {{\tt arXiv:#1}}}
\def\mn@eprint@dblp#1{\href {http://dblp.uni-trier.de/rec/bibtex/#1.xml}
  {dblp:#1}}
\def\mn@eprint@#1:#2:#3:#4\@nil{\def\@tempa {#1}\def\@tempb {#2}\def\@tempc
  {#3}\ifx \@tempc \@empty \let \@tempc \@tempb \let \@tempb \@tempa \fi \ifx
  \@tempb \@empty \def\@tempb {arXiv}\fi \@ifundefined
  {mn@eprint@\@tempb}{\@tempb:\@tempc}{\expandafter \expandafter \csname
  mn@eprint@\@tempb\endcsname \expandafter{\@tempc}}}

\bibitem[\protect\citeauthoryear{{Ag{\'u}ndez}, {Cernicharo}  \&
  {Gu{\'e}lin}}{{Ag{\'u}ndez} et~al.}{2010}]{Agundez2010}
{Ag{\'u}ndez} M.,  {Cernicharo} J.,   {Gu{\'e}lin} M.,  2010, \mn@doi [\apjl]
  {10.1088/2041-8205/724/2/L133}, \href
  {http://adsabs.harvard.edu/abs/2010ApJ...724L.133A} {724, L133}

\bibitem[\protect\citeauthoryear{{Ag{\'u}ndez}, {Cernicharo},
  {Quintana-Lacaci}, {Velilla Prieto}, {Castro-Carrizo}, {Marcelino}  \&
  {Gu{\'e}lin}}{{Ag{\'u}ndez} et~al.}{2015}]{Agundez2015}
{Ag{\'u}ndez} M.,  {Cernicharo} J.,  {Quintana-Lacaci} G.,  {Velilla Prieto}
  L.,  {Castro-Carrizo} A.,  {Marcelino} N.,   {Gu{\'e}lin} M.,  2015, \mn@doi
  [\apj] {10.1088/0004-637X/814/2/143}, \href
  {http://adsabs.harvard.edu/abs/2015ApJ...814..143A} {814, 143}

\bibitem[\protect\citeauthoryear{{Ag{\'u}ndez} et~al.,}{{Ag{\'u}ndez}
  et~al.}{2017}]{Agundez2017}
{Ag{\'u}ndez} M.,  et~al., 2017, \mn@doi [\aap] {10.1051/0004-6361/201630274},
  \href {http://adsabs.harvard.edu/abs/2017A%26A...601A...4A} {601, A4}

\bibitem[\protect\citeauthoryear{{Ag{\'u}ndez}, {Mart{\'\i}nez}, {de Andres},
  {Cernicharo}  \& {Mart{\'\i}n-Gago}}{{Ag{\'u}ndez}
  et~al.}{2020}]{Agundez2020}
{Ag{\'u}ndez} M.,  {Mart{\'\i}nez} J.~I.,  {de Andres} P.~L.,  {Cernicharo} J.,
    {Mart{\'\i}n-Gago} J.~A.,  2020, \mn@doi [\aap]
  {10.1051/0004-6361/202037496}, \href
  {https://ui.adsabs.harvard.edu/abs/2020A&A...637A..59A} {637, A59}

\bibitem[\protect\citeauthoryear{{Bujarrabal}, {Fuente}  \&
  {Omont}}{{Bujarrabal} et~al.}{1994}]{Bujarrabal1994}
{Bujarrabal} V.,  {Fuente} A.,   {Omont} A.,  1994, \aap, \href
  {http://adsabs.harvard.edu/abs/1994A%26A...285..247B} {285, 247}

\bibitem[\protect\citeauthoryear{{Bujarrabal}, {Ag{\'u}ndez},
  {G{\'o}mez-Garrido}, {Kim}, {Santander-Garc{\'\i}a}, {Alcolea},
  {Castro-Carrizo}  \& {Miko{\l}ajewska}}{{Bujarrabal}
  et~al.}{2021}]{Bujarrabal2021}
{Bujarrabal} V.,  {Ag{\'u}ndez} M.,  {G{\'o}mez-Garrido} M.,  {Kim} H.,
  {Santander-Garc{\'\i}a} M.,  {Alcolea} J.,  {Castro-Carrizo} A.,
  {Miko{\l}ajewska} J.,  2021, \mn@doi [\aap] {10.1051/0004-6361/202141002},
  \href {https://ui.adsabs.harvard.edu/abs/2021A&A...651A...4B} {651, A4}

\bibitem[\protect\citeauthoryear{{Cernicharo}, {Gu{\'e}lin}, {Ag{\'u}ndez},
  {Kawaguchi}, {McCarthy}  \& {Thaddeus}}{{Cernicharo}
  et~al.}{2007}]{Cernicharo2007}
{Cernicharo} J.,  {Gu{\'e}lin} M.,  {Ag{\'u}ndez} M.,  {Kawaguchi} K.,
  {McCarthy} M.,   {Thaddeus} P.,  2007, \mn@doi [\aap]
  {10.1051/0004-6361:20077415}, \href
  {https://ui.adsabs.harvard.edu/abs/2007A&A...467L..37C} {467, L37}

\bibitem[\protect\citeauthoryear{{Chen}, {Frank}, {Blackman}, {Nordhaus}  \&
  {Carroll-Nellenback}}{{Chen} et~al.}{2017}]{Chen2017}
{Chen} Z.,  {Frank} A.,  {Blackman} E.~G.,  {Nordhaus} J.,
  {Carroll-Nellenback} J.,  2017, \mn@doi [\mnras] {10.1093/mnras/stx680},
  \href {https://ui.adsabs.harvard.edu/abs/2017MNRAS.468.4465C} {468, 4465}

\bibitem[\protect\citeauthoryear{{Cherchneff}}{{Cherchneff}}{2006}]{Cherchneff2006}
{Cherchneff} I.,  2006, \mn@doi [\aap] {10.1051/0004-6361:20064827}, \href
  {http://adsabs.harvard.edu/abs/2006A%26A...456.1001C} {456, 1001}

\bibitem[\protect\citeauthoryear{{Danilovich}, {De Beck}, {Black}, {Olofsson}
  \& {Justtanont}}{{Danilovich} et~al.}{2016}]{Danilovich2016}
{Danilovich} T.,  {De Beck} E.,  {Black} J.~H.,  {Olofsson} H.,   {Justtanont}
  K.,  2016, \mn@doi [\aap] {10.1051/0004-6361/201527943}, \href
  {http://adsabs.harvard.edu/abs/2016A%26A...588A.119D} {588, A119}

\bibitem[\protect\citeauthoryear{{Danilovich}, {Van de Sande}, {De Beck},
  {Decin}, {Olofsson}, {Ramstedt}  \& {Millar}}{{Danilovich}
  et~al.}{2017}]{Danilovich2017}
{Danilovich} T.,  {Van de Sande} M.,  {De Beck} E.,  {Decin} L.,  {Olofsson}
  H.,  {Ramstedt} S.,   {Millar} T.~J.,  2017, \mn@doi [\aap]
  {10.1051/0004-6361/201731203}, \href
  {https://ui.adsabs.harvard.edu/abs/2017A&A...606A.124D} {606, A124}

\bibitem[\protect\citeauthoryear{{Danilovich}, {Richards}, {Decin}, {Van de
  Sande}  \& {Gottlieb}}{{Danilovich} et~al.}{2020}]{Danilovich2020}
{Danilovich} T.,  {Richards} A.~M.~S.,  {Decin} L.,  {Van de Sande} M.,
  {Gottlieb} C.~A.,  2020, \mn@doi [\mnras] {10.1093/mnras/staa693}, \href
  {https://ui.adsabs.harvard.edu/abs/2020MNRAS.494.1323D} {494, 1323}

\bibitem[\protect\citeauthoryear{{De Beck} \& {Olofsson}}{{De Beck} \&
  {Olofsson}}{2020}]{DeBeck2020}
{De Beck} E.,  {Olofsson} H.,  2020, \mn@doi [\aap]
  {10.1051/0004-6361/202038335}, \href
  {https://ui.adsabs.harvard.edu/abs/2020A&A...642A..20D} {642, A20}

\bibitem[\protect\citeauthoryear{{De Marco}}{{De Marco}}{2009}]{DeMarco2009}
{De Marco} O.,  2009, \mn@doi [\pasp] {10.1086/597765}, \href
  {https://ui.adsabs.harvard.edu/abs/2009PASP..121..316D} {121, 316}

\bibitem[\protect\citeauthoryear{{Decin}}{{Decin}}{2021}]{Decin2021}
{Decin} L.,  2021, \mn@doi [Annual Review of Astronomy and Astrophysics]
  {10.1146/annurev-astro-090120-033712}, 59, 337

\bibitem[\protect\citeauthoryear{{Decin} et~al.,}{{Decin}
  et~al.}{2010a}]{Decin2010b}
{Decin} L.,  et~al., 2010a, \mn@doi [\nat] {10.1038/nature09344}, \href
  {http://adsabs.harvard.edu/abs/2010Natur.467...64D} {467, 64}

\bibitem[\protect\citeauthoryear{{Decin} et~al.,}{{Decin}
  et~al.}{2010b}]{Decin2010a}
{Decin} L.,  et~al., 2010b, \mn@doi [\aap] {10.1051/0004-6361/201014136}, \href
  {http://adsabs.harvard.edu/abs/2010A%26A...516A..69D} {516, A69}

\bibitem[\protect\citeauthoryear{{Decin}, {Richards}, {Neufeld}, {Steffen},
  {Melnick}  \& {Lombaert}}{{Decin} et~al.}{2015}]{Decin2015}
{Decin} L.,  {Richards} A.~M.~S.,  {Neufeld} D.,  {Steffen} W.,  {Melnick} G.,
   {Lombaert} R.,  2015, \mn@doi [\aap] {10.1051/0004-6361/201424593}, \href
  {https://ui.adsabs.harvard.edu/abs/2015A&A...574A...5D} {574, A5}

\bibitem[\protect\citeauthoryear{{Decin}, {Richards}, {Danilovich}, {Homan}  \&
  {Nuth}}{{Decin} et~al.}{2018}]{Decin2018}
{Decin} L.,  {Richards} A.~M.~S.,  {Danilovich} T.,  {Homan} W.,   {Nuth}
  J.~A.,  2018, \mn@doi [\aap] {10.1051/0004-6361/201732216}, \href
  {https://ui.adsabs.harvard.edu/abs/2018A&A...615A..28D} {615, A28}

\bibitem[\protect\citeauthoryear{{Decin} et~al.,}{{Decin}
  et~al.}{2020}]{Decin2020}
{Decin} L.,  et~al., 2020, \mn@doi [Science] {10.1126/science.abb1229}, \href
  {https://ui.adsabs.harvard.edu/abs/2020Sci...369.1497D} {369, 1497}

\bibitem[\protect\citeauthoryear{{El Mellah}, {Bolte}, {Decin}, {Homan}  \&
  {Keppens}}{{El Mellah} et~al.}{2020}]{ElMellah2020}
{El Mellah} I.,  {Bolte} J.,  {Decin} L.,  {Homan} W.,   {Keppens} R.,  2020,
  \mn@doi [\aap] {10.1051/0004-6361/202037492}, \href
  {https://ui.adsabs.harvard.edu/abs/2020A&A...637A..91E} {637, A91}

\bibitem[\protect\citeauthoryear{{Gail} \& {Sedlmayr}}{{Gail} \&
  {Sedlmayr}}{2013}]{Gail2013}
{Gail} H.-P.,  {Sedlmayr} E.,  2013, {Physics and Chemistry of Circumstellar
  Dust Shells}.
Cambridge University Press

\bibitem[\protect\citeauthoryear{{Gobrecht}, {Cherchneff}, {Sarangi}, {Plane}
  \& {Bromley}}{{Gobrecht} et~al.}{2016}]{Gobrecht2016}
{Gobrecht} D.,  {Cherchneff} I.,  {Sarangi} A.,  {Plane} J.~M.~C.,   {Bromley}
  S.~T.,  2016, \mn@doi [\aap] {10.1051/0004-6361/201425363}, \href
  {http://adsabs.harvard.edu/abs/2016A%26A...585A...6G} {585, A6}

\bibitem[\protect\citeauthoryear{{Groenewegen} et~al.,}{{Groenewegen}
  et~al.}{2012}]{Groenewegen2012}
{Groenewegen} M.~A.~T.,  et~al., 2012, \mn@doi [\aap]
  {10.1051/0004-6361/201219604}, \href
  {https://ui.adsabs.harvard.edu/abs/2012A&A...543L...8G} {543, L8}

\bibitem[\protect\citeauthoryear{{Hansen}, {Andersen}, {Nordstr{\"o}m},
  {Beers}, {Placco}, {Yoon}  \& {Buchhave}}{{Hansen} et~al.}{2016}]{Hansen2016}
{Hansen} T.~T.,  {Andersen} J.,  {Nordstr{\"o}m} B.,  {Beers} T.~C.,  {Placco}
  V.~M.,  {Yoon} J.,   {Buchhave} L.~A.,  2016, \mn@doi [\aap]
  {10.1051/0004-6361/201527409}, \href
  {https://ui.adsabs.harvard.edu/abs/2016A&A...588A...3H} {588, A3}

\bibitem[\protect\citeauthoryear{{Heays}, {Bosman}  \& {van Dishoeck}}{{Heays}
  et~al.}{2017}]{Heays2017}
{Heays} A.~N.,  {Bosman} A.~D.,   {van Dishoeck} E.~F.,  2017, \mn@doi [\aap]
  {10.1051/0004-6361/201628742}, \href
  {http://cdsads.u-strasbg.fr/abs/2017A%26A...602A.105H} {602, A105}

\bibitem[\protect\citeauthoryear{{Homan}, {Danilovich}, {Decin}, {de Koter},
  {Nuth}  \& {Van de Sande}}{{Homan} et~al.}{2018}]{Homan2018}
{Homan} W.,  {Danilovich} T.,  {Decin} L.,  {de Koter} A.,  {Nuth} J.,   {Van
  de Sande} M.,  2018, \mn@doi [\aap] {10.1051/0004-6361/201732246}, \href
  {https://ui.adsabs.harvard.edu/abs/2018A&A...614A.113H} {614, A113}

\bibitem[\protect\citeauthoryear{{Jorissen}, {Van Eck}, {Mayor}  \&
  {Udry}}{{Jorissen} et~al.}{1998}]{Jorissen1998}
{Jorissen} A.,  {Van Eck} S.,  {Mayor} M.,   {Udry} S.,  1998, \aap, \href
  {https://ui.adsabs.harvard.edu/abs/1998A&A...332..877J} {332, 877}

\bibitem[\protect\citeauthoryear{{Justtanont} et~al.,}{{Justtanont}
  et~al.}{1996}]{Justtanont1996}
{Justtanont} K.,  et~al., 1996, \aap, \href
  {http://adsabs.harvard.edu/abs/1996A%26A...315L.217J} {315, L217}

\bibitem[\protect\citeauthoryear{{Keene}, {Young}, {Phillips}, {Buettgenbach}
  \& {Carlstrom}}{{Keene} et~al.}{1993}]{Keene1993}
{Keene} J.,  {Young} K.,  {Phillips} T.~G.,  {Buettgenbach} T.~H.,
  {Carlstrom} J.~E.,  1993, \mn@doi [\apjl] {10.1086/187050}, \href
  {https://ui.adsabs.harvard.edu/abs/1993ApJ...415L.131K} {415, L131}

\bibitem[\protect\citeauthoryear{{Kervella} et~al.,}{{Kervella}
  et~al.}{2014}]{Kervella2014}
{Kervella} P.,  et~al., 2014, \mn@doi [\aap] {10.1051/0004-6361/201323273},
  \href {http://adsabs.harvard.edu/abs/2014A%26A...564A..88K} {564, A88}

\bibitem[\protect\citeauthoryear{{Khouri} et~al.,}{{Khouri}
  et~al.}{2016}]{Khouri2016}
{Khouri} T.,  et~al., 2016, \mn@doi [\aap] {10.1051/0004-6361/201628435}, \href
  {http://adsabs.harvard.edu/abs/2016A%26A...591A..70K} {591, A70}

\bibitem[\protect\citeauthoryear{{Kim} \& {Taam}}{{Kim} \&
  {Taam}}{2012}]{Kim2012}
{Kim} H.,  {Taam} R.~E.,  2012, \mn@doi [\apj] {10.1088/0004-637X/759/1/59},
  \href {https://ui.adsabs.harvard.edu/abs/2012ApJ...759...59K} {759, 59}

\bibitem[\protect\citeauthoryear{{Kim}, {Wyrowski}, {Menten}  \& {Decin}}{{Kim}
  et~al.}{2010}]{Kim2010}
{Kim} H.,  {Wyrowski} F.,  {Menten} K.~M.,   {Decin} L.,  2010, \mn@doi [\aap]
  {10.1051/0004-6361/201014094}, \href
  {http://adsabs.harvard.edu/abs/2010A%26A...516A..68K} {516, A68}

\bibitem[\protect\citeauthoryear{{Knapp}, {Crosas}, {Young}  \&
  {Ivezi{\'c}}}{{Knapp} et~al.}{2000}]{Knapp2000}
{Knapp} G.~R.,  {Crosas} M.,  {Young} K.,   {Ivezi{\'c}} {\v{Z}}.,  2000,
  \mn@doi [\apj] {10.1086/308731}, \href
  {https://ui.adsabs.harvard.edu/abs/2000ApJ...534..324K} {534, 324}

\bibitem[\protect\citeauthoryear{{Le{\~a}o}, {de Laverny}, {M{\'e}karnia}, {de
  Medeiros}  \& {Vandame}}{{Le{\~a}o} et~al.}{2006}]{Leao2006}
{Le{\~a}o} I.~C.,  {de Laverny} P.,  {M{\'e}karnia} D.,  {de Medeiros} J.~R.,
  {Vandame} B.,  2006, \mn@doi [\aap] {10.1051/0004-6361:20054577}, \href
  {https://ui.adsabs.harvard.edu/abs/2006A&A...455..187L} {455, 187}

\bibitem[\protect\citeauthoryear{{Lindqvist}, {Nyman}, {Olofsson}  \&
  {Winnberg}}{{Lindqvist} et~al.}{1988}]{Lindqvist1988}
{Lindqvist} M.,  {Nyman} L.-A.,  {Olofsson} H.,   {Winnberg} A.,  1988, \aap,
  \href {http://adsabs.harvard.edu/abs/1988A%26A...205L..15L} {205, L15}

\bibitem[\protect\citeauthoryear{{Lombaert} et~al.,}{{Lombaert}
  et~al.}{2016}]{Lombaert2016}
{Lombaert} R.,  et~al., 2016, \mn@doi [\aap] {10.1051/0004-6361/201527049},
  \href {http://adsabs.harvard.edu/abs/2016A%26A...588A.124L} {588, A124}

\bibitem[\protect\citeauthoryear{{Maercker}, {Danilovich}, {Olofsson}, {De
  Beck}, {Justtanont}, {Lombaert}  \& {Royer}}{{Maercker}
  et~al.}{2016}]{Maercker2016}
{Maercker} M.,  {Danilovich} T.,  {Olofsson} H.,  {De Beck} E.,  {Justtanont}
  K.,  {Lombaert} R.,   {Royer} P.,  2016, \mn@doi [\aap]
  {10.1051/0004-6361/201628310}, \href
  {http://adsabs.harvard.edu/abs/2016A%26A...591A..44M} {591, A44}

\bibitem[\protect\citeauthoryear{{Maes}, {Homan}, {Malfait}, {Siess}, {Bolte},
  {De Ceuster}  \& {Decin}}{{Maes} et~al.}{2021}]{Maes2021}
{Maes} S.,  {Homan} W.,  {Malfait} J.,  {Siess} L.,  {Bolte} J.,  {De Ceuster}
  F.,   {Decin} L.,  2021, \mn@doi [\aap] {10.1051/0004-6361/202140823}, \href
  {https://ui.adsabs.harvard.edu/abs/2021A&A...653A..25M} {653, A25}

\bibitem[\protect\citeauthoryear{{Malfait}, {Homan}, {Maes}, {Bolte}, {Siess},
  {De Ceuster}  \& {Decin}}{{Malfait} et~al.}{2021}]{Malfait2021}
{Malfait} J.,  {Homan} W.,  {Maes} S.,  {Bolte} J.,  {Siess} L.,  {De Ceuster}
  F.,   {Decin} L.,  2021, \mn@doi [\aap] {10.1051/0004-6361/202141161}, \href
  {https://ui.adsabs.harvard.edu/abs/2021A&A...652A..51M} {652, A51}

\bibitem[\protect\citeauthoryear{Mastrodemos \& Morris}{Mastrodemos \&
  Morris}{1999}]{Mastrodemos1999}
Mastrodemos N.,  Morris M.,  1999, \mn@doi [The Astrophysical Journal]
  {10.1086/307717}, 523, 357

\bibitem[\protect\citeauthoryear{{Mauron} \& {Huggins}}{{Mauron} \&
  {Huggins}}{2000}]{Mauron2000}
{Mauron} N.,  {Huggins} P.~J.,  2000, \aap, \href
  {http://adsabs.harvard.edu/abs/2000A%26A...359..707M} {359, 707}

\bibitem[\protect\citeauthoryear{{Mauron} \& {Huggins}}{{Mauron} \&
  {Huggins}}{2006}]{Mauron2006}
{Mauron} N.,  {Huggins} P.~J.,  2006, \mn@doi [\aap]
  {10.1051/0004-6361:20054739}, \href
  {http://adsabs.harvard.edu/abs/2006A%26A...452..257M} {452, 257}

\bibitem[\protect\citeauthoryear{{McCarthy}, {Gottlieb}, {Gupta}  \&
  {Thaddeus}}{{McCarthy} et~al.}{2006}]{McCarthy2006}
{McCarthy} M.~C.,  {Gottlieb} C.~A.,  {Gupta} H.,   {Thaddeus} P.,  2006,
  \mn@doi [\apjl] {10.1086/510238}, \href
  {https://ui.adsabs.harvard.edu/abs/2006ApJ...652L.141M} {652, L141}

\bibitem[\protect\citeauthoryear{{McElroy}, {Walsh}, {Markwick}, {Cordiner},
  {Smith}  \& {Millar}}{{McElroy} et~al.}{2013}]{McElroy2013}
{McElroy} D.,  {Walsh} C.,  {Markwick} A.~J.,  {Cordiner} M.~A.,  {Smith} K.,
  {Millar} T.~J.,  2013, \mn@doi [\aap] {10.1051/0004-6361/201220465}, \href
  {http://adsabs.harvard.edu/abs/2013A%26A...550A..36M} {550, A36}

\bibitem[\protect\citeauthoryear{{Miko{\l}ajewska}}{{Miko{\l}ajewska}}{2012}]{Mikolajewska2012}
{Miko{\l}ajewska} J.,  2012, \mn@doi [Baltic Astronomy]
  {10.1515/astro-2017-0352}, \href
  {https://ui.adsabs.harvard.edu/abs/2012BaltA..21....5M} {21, 5}

\bibitem[\protect\citeauthoryear{{Moe} \& {Di Stefano}}{{Moe} \& {Di
  Stefano}}{2017}]{Moe2017}
{Moe} M.,  {Di Stefano} R.,  2017, \mn@doi [\apjs] {10.3847/1538-4365/aa6fb6},
  \href {https://ui.adsabs.harvard.edu/abs/2017ApJS..230...15M} {230, 15}

\bibitem[\protect\citeauthoryear{{Montez}, {Ramstedt}, {Kastner}, {Vlemmings}
  \& {Sanchez}}{{Montez} et~al.}{2017}]{Montez2017}
{Montez} Rodolfo J.,  {Ramstedt} S.,  {Kastner} J.~H.,  {Vlemmings} W.,
  {Sanchez} E.,  2017, \mn@doi [\apj] {10.3847/1538-4357/aa704d}, \href
  {https://ui.adsabs.harvard.edu/abs/2017ApJ...841...33M} {841, 33}

\bibitem[\protect\citeauthoryear{{Morris} \& {Jura}}{{Morris} \&
  {Jura}}{1983}]{Morris1983}
{Morris} M.,  {Jura} M.,  1983, \mn@doi [\apj] {10.1086/160622}, \href
  {http://adsabs.harvard.edu/abs/1983ApJ...264..546M} {264, 546}

\bibitem[\protect\citeauthoryear{{Neufeld} et~al.,}{{Neufeld}
  et~al.}{2011}]{Neufeld2011}
{Neufeld} D.~A.,  et~al., 2011, \mn@doi [\apjl] {10.1088/2041-8205/727/2/L29},
  \href {http://adsabs.harvard.edu/abs/2011ApJ...727L..29N} {727, L29}

\bibitem[\protect\citeauthoryear{{Olofsson}, {Lindqvist}, {Winnberg}, {Nyman}
  \& {Nguyen-Q-Rieu}}{{Olofsson} et~al.}{1991}]{Olofsson1991}
{Olofsson} H.,  {Lindqvist} M.,  {Winnberg} A.,  {Nyman} L.-A.,
  {Nguyen-Q-Rieu} 1991, \aap, \href
  {http://adsabs.harvard.edu/abs/1991A%26A...245..611O} {245, 611}

\bibitem[\protect\citeauthoryear{{Olofsson}, {Bergman}  \&
  {Lindqvist}}{{Olofsson} et~al.}{2015}]{Olofsson2015}
{Olofsson} H.,  {Bergman} P.,   {Lindqvist} M.,  2015, \mn@doi [\aap]
  {10.1051/0004-6361/201526741}, \href
  {https://ui.adsabs.harvard.edu/abs/2015A&A...582A.102O} {582, A102}

\bibitem[\protect\citeauthoryear{{Omont}, {Lucas}, {Morris}  \&
  {Guilloteau}}{{Omont} et~al.}{1993}]{Omont1993}
{Omont} A.,  {Lucas} R.,  {Morris} M.,   {Guilloteau} S.,  1993, \aap, \href
  {http://adsabs.harvard.edu/abs/1993A%26A...267..490O} {267, 490}

\bibitem[\protect\citeauthoryear{{Oomen}, {Van Winckel}, {Pols}, {Nelemans},
  {Escorza}, {Manick}, {Kamath}  \& {Waelkens}}{{Oomen}
  et~al.}{2018}]{Oomen2018}
{Oomen} G.-M.,  {Van Winckel} H.,  {Pols} O.,  {Nelemans} G.,  {Escorza} A.,
  {Manick} R.,  {Kamath} D.,   {Waelkens} C.,  2018, \mn@doi [\aap]
  {10.1051/0004-6361/201833816}, \href
  {https://ui.adsabs.harvard.edu/abs/2018A&A...620A..85O} {620, A85}

\bibitem[\protect\citeauthoryear{{Ortiz} \& {Guerrero}}{{Ortiz} \&
  {Guerrero}}{2021}]{Ortiz2021}
{Ortiz} R.,  {Guerrero} M.~A.,  2021, \mn@doi [\apj]
  {10.3847/1538-4357/abefd7}, \href
  {https://ui.adsabs.harvard.edu/abs/2021ApJ...912...93O} {912, 93}

\bibitem[\protect\citeauthoryear{{Ortiz}, {Guerrero}  \& {Costa}}{{Ortiz}
  et~al.}{2019}]{Ortiz2019}
{Ortiz} R.,  {Guerrero} M.~A.,   {Costa} R. D.~D.,  2019, \mn@doi [\mnras]
  {10.1093/mnras/sty3076}, \href
  {https://ui.adsabs.harvard.edu/abs/2019MNRAS.482.4697O} {482, 4697}

\bibitem[\protect\citeauthoryear{{Ramstedt} et~al.,}{{Ramstedt}
  et~al.}{2014}]{Ramstedt2014}
{Ramstedt} S.,  et~al., 2014, \mn@doi [\aap] {10.1051/0004-6361/201425029},
  \href {https://ui.adsabs.harvard.edu/abs/2014A&A...570L..14R} {570, L14}

\bibitem[\protect\citeauthoryear{{Ramstedt} et~al.,}{{Ramstedt}
  et~al.}{2017}]{Ramstedt2017}
{Ramstedt} S.,  et~al., 2017, \mn@doi [\aap] {10.1051/0004-6361/201730934},
  \href {https://ui.adsabs.harvard.edu/abs/2017A&A...605A.126R} {605, A126}

\bibitem[\protect\citeauthoryear{{Remijan}, {Hollis}, {Lovas}, {Cordiner},
  {Millar}, {Markwick-Kemper}  \& {Jewell}}{{Remijan}
  et~al.}{2007}]{Remijan2007}
{Remijan} A.~J.,  {Hollis} J.~M.,  {Lovas} F.~J.,  {Cordiner} M.~A.,  {Millar}
  T.~J.,  {Markwick-Kemper} A.~J.,   {Jewell} P.~R.,  2007, \mn@doi [\apjl]
  {10.1086/520704}, \href
  {https://ui.adsabs.harvard.edu/abs/2007ApJ...664L..47R} {664, L47}

\bibitem[\protect\citeauthoryear{{Saberi}, {Vlemmings}, {De Beck}, {Montez}  \&
  {Ramstedt}}{{Saberi} et~al.}{2018}]{Saberi2018}
{Saberi} M.,  {Vlemmings} W.~H.~T.,  {De Beck} E.,  {Montez} R.,   {Ramstedt}
  S.,  2018, \mn@doi [\aap] {10.1051/0004-6361/201833080}, \href
  {https://ui.adsabs.harvard.edu/abs/2018A&A...612L..11S} {612, L11}

\bibitem[\protect\citeauthoryear{{Sahai}, {Findeisen}, {Gil de Paz}  \&
  {S{\'a}nchez Contreras}}{{Sahai} et~al.}{2008}]{Sahai2008}
{Sahai} R.,  {Findeisen} K.,  {Gil de Paz} A.,   {S{\'a}nchez Contreras} C.,
  2008, \mn@doi [\apj] {10.1086/592559}, \href
  {https://ui.adsabs.harvard.edu/abs/2008ApJ...689.1274S} {689, 1274}

\bibitem[\protect\citeauthoryear{{Sahai}, {Neill}, {Gil de Paz}  \&
  {S{\'a}nchez Contreras}}{{Sahai} et~al.}{2011}]{Sahai2011}
{Sahai} R.,  {Neill} J.~D.,  {Gil de Paz} A.,   {S{\'a}nchez Contreras} C.,
  2011, \mn@doi [\apjl] {10.1088/2041-8205/740/2/L39}, \href
  {https://ui.adsabs.harvard.edu/abs/2011ApJ...740L..39S} {740, L39}

\bibitem[\protect\citeauthoryear{{Sch{\"o}ier}, {Ramstedt}, {Olofsson},
  {Lindqvist}, {Bieging}  \& {Marvel}}{{Sch{\"o}ier}
  et~al.}{2013}]{Schoier2013}
{Sch{\"o}ier} F.~L.,  {Ramstedt} S.,  {Olofsson} H.,  {Lindqvist} M.,
  {Bieging} J.~H.,   {Marvel} K.~B.,  2013, \mn@doi [\aap]
  {10.1051/0004-6361/201220400}, \href
  {http://cdsads.u-strasbg.fr/abs/2013A%26A...550A..78S} {550, A78}

\bibitem[\protect\citeauthoryear{{Sokoloski} \& {Bildsten}}{{Sokoloski} \&
  {Bildsten}}{2010}]{Sokoloski2010}
{Sokoloski} J.~L.,  {Bildsten} L.,  2010, \mn@doi [\apj]
  {10.1088/0004-637X/723/2/1188}, \href
  {https://ui.adsabs.harvard.edu/abs/2010ApJ...723.1188S} {723, 1188}

\bibitem[\protect\citeauthoryear{{Tielens}}{{Tielens}}{2005}]{Tielens2005}
{Tielens} A.~G.~G.~M.,  2005, {The Physics and Chemistry of the Interstellar
  Medium}.
Cambridge University Press

\bibitem[\protect\citeauthoryear{{Turner}}{{Turner}}{1992}]{Turner1992}
{Turner} B.~E.,  1992, \mn@doi [\apjl] {10.1086/186324}, \href
  {https://ui.adsabs.harvard.edu/abs/1992ApJ...388L..35T} {388, L35}

\bibitem[\protect\citeauthoryear{{Van de Sande} \& {Millar}}{{Van de Sande} \&
  {Millar}}{2019}]{VandeSande2019a}
{Van de Sande} M.,  {Millar} T.~J.,  2019, \mn@doi [\apj]
  {10.3847/1538-4357/ab03d4}, \href
  {https://ui.adsabs.harvard.edu/abs/2019ApJ...873...36V} {873, 36}

\bibitem[\protect\citeauthoryear{{Van de Sande}, {Decin}, {Lombaert}, {Khouri},
  {de Koter}, {Wyrowski}, {De Nutte}  \& {Homan}}{{Van de Sande}
  et~al.}{2018a}]{VandeSandeRDor}
{Van de Sande} M.,  {Decin} L.,  {Lombaert} R.,  {Khouri} T.,  {de Koter} A.,
  {Wyrowski} F.,  {De Nutte} R.,   {Homan} W.,  2018a, \aap, 609, A63

\bibitem[\protect\citeauthoryear{{Van de Sande}, {Sundqvist}, {Millar},
  {Keller}, {Homan}, {de Koter}, {Decin}  \& {De Ceuster}}{{Van de Sande}
  et~al.}{2018b}]{VandeSande2018}
{Van de Sande} M.,  {Sundqvist} J.~O.,  {Millar} T.~J.,  {Keller} D.,  {Homan}
  W.,  {de Koter} A.,  {Decin} L.,   {De Ceuster} F.,  2018b, \mn@doi [\aap]
  {10.1051/0004-6361/201732276}, \href
  {https://ui.adsabs.harvard.edu/\#abs/2018A&A...616A.106V} {616, A106}

\bibitem[\protect\citeauthoryear{{Van de Sande}, {Sundqvist}, {Millar},
  {Keller}, {Homan}, {de Koter}, {Decin}  \& {De Ceuster}}{{Van de Sande}
  et~al.}{2020}]{VandeSande2018Err}
{Van de Sande} M.,  {Sundqvist} J.~O.,  {Millar} T.~J.,  {Keller} D.,  {Homan}
  W.,  {de Koter} A.,  {Decin} L.,   {De Ceuster} F.,  2020, \mn@doi [\aap]
  {10.1051/0004-6361/201732276e}, \href
  {https://ui.adsabs.harvard.edu/abs/2020A&A...634C...1V} {634, C1}

\bibitem[\protect\citeauthoryear{{Velilla Prieto} et~al.,}{{Velilla Prieto}
  et~al.}{2017}]{VelillaPrieto2017}
{Velilla Prieto} L.,  et~al., 2017, \mn@doi [\aap]
  {10.1051/0004-6361/201628776}, \href
  {http://adsabs.harvard.edu/abs/2017A%26A...597A..25V} {597, A25}

\bibitem[\protect\citeauthoryear{{Wong}, {Menten}, {Kami{\'n}ski}, {Wyrowski},
  {Lacy}  \& {Greathouse}}{{Wong} et~al.}{2018}]{Wong2018}
{Wong} K.~T.,  {Menten} K.~M.,  {Kami{\'n}ski} T.,  {Wyrowski} F.,  {Lacy}
  J.~H.,   {Greathouse} T.~K.,  2018, \mn@doi [\aap]
  {10.1051/0004-6361/201731873}, \href
  {https://ui.adsabs.harvard.edu/abs/2018A%26A...612A..48W} {612, A48}

\bibitem[\protect\citeauthoryear{{van der Veen}, {Huggins}  \& {Matthews}}{{van
  der Veen} et~al.}{1998}]{vanderVeen1998}
{van der Veen} W.~E.~C.~J.,  {Huggins} P.~J.,   {Matthews} H.~E.,  1998,
  \mn@doi [\apj] {10.1086/306191}, \href
  {https://ui.adsabs.harvard.edu/abs/1998ApJ...505..749V} {505, 749}

\makeatother
\end{thebibliography}



\appendix

\section{Occulted region within the outflow}		\label{app:angles}

\begin{figure}
 \includegraphics[width=0.95\columnwidth]{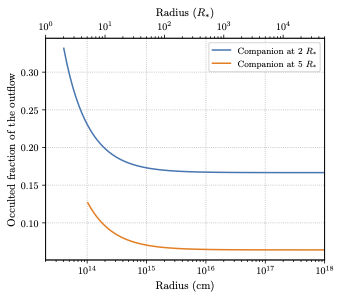}
 \caption{Occulted fraction of the orbit within the orbital plane as a function of distance to the star for a companion at 2 $R_*$ (blue) and at 5 R$_*$ (orange).
 }
 \label{fig:angles:occulted}
\end{figure}

The fraction of the period during which the companion can be occulted by the star is largest in the orbital plane.  
Following Fig. \ref{fig:angles:cone}, we have that $\mathrm{sin}(\phi) = {R_*}/{R_\mathrm{comp}}$ and $\mathrm{sin}(\theta) = {R_*}/{R}$.
The angle $\alpha$ with respect to the orbital plane is given by
\begin{equation}
\alpha = \mathrm{arcsin}\ \frac{R_*}{R} + \mathrm{arcsin}\ \frac{R_*}{R_\mathrm{comp}}.
\end{equation}
The hidden fraction of the period is then $2\alpha / 2\pi$.
The approximation of the companion by a point source is reasonable since an AGB star is $\sim300$ times larger than a solar-like star (\Tcomp\ = 6000 K) and $\sim 3000$ times larger than a red dwarf (\Tcomp\ = 4000 K).
The variation of the fraction $\alpha / \pi$ throughout the outflow is shown in Fig. \ref{fig:angles:occulted} for a companion at 2 $R_*$ and at 5 $R_*$.

The majority of the sphere is constantly irradiated by the companion.
For a companion at $R_\mathrm{comp}$, the solid angle cast by the AGB star is equal to $\pi (R_*/R_\mathrm{comp})^2$.
The fraction of the total sphere in which the companion is hidden by the star is then $(R_*/2R_\mathrm{comp})^2$.
For a companion at 2 $R_*$, only 6\% of the outflow is occulted by the stellar companion, decreasing to 1\% for a companion at 5 $R_*$.

\begin{figure}
 \includegraphics[width=1\columnwidth]{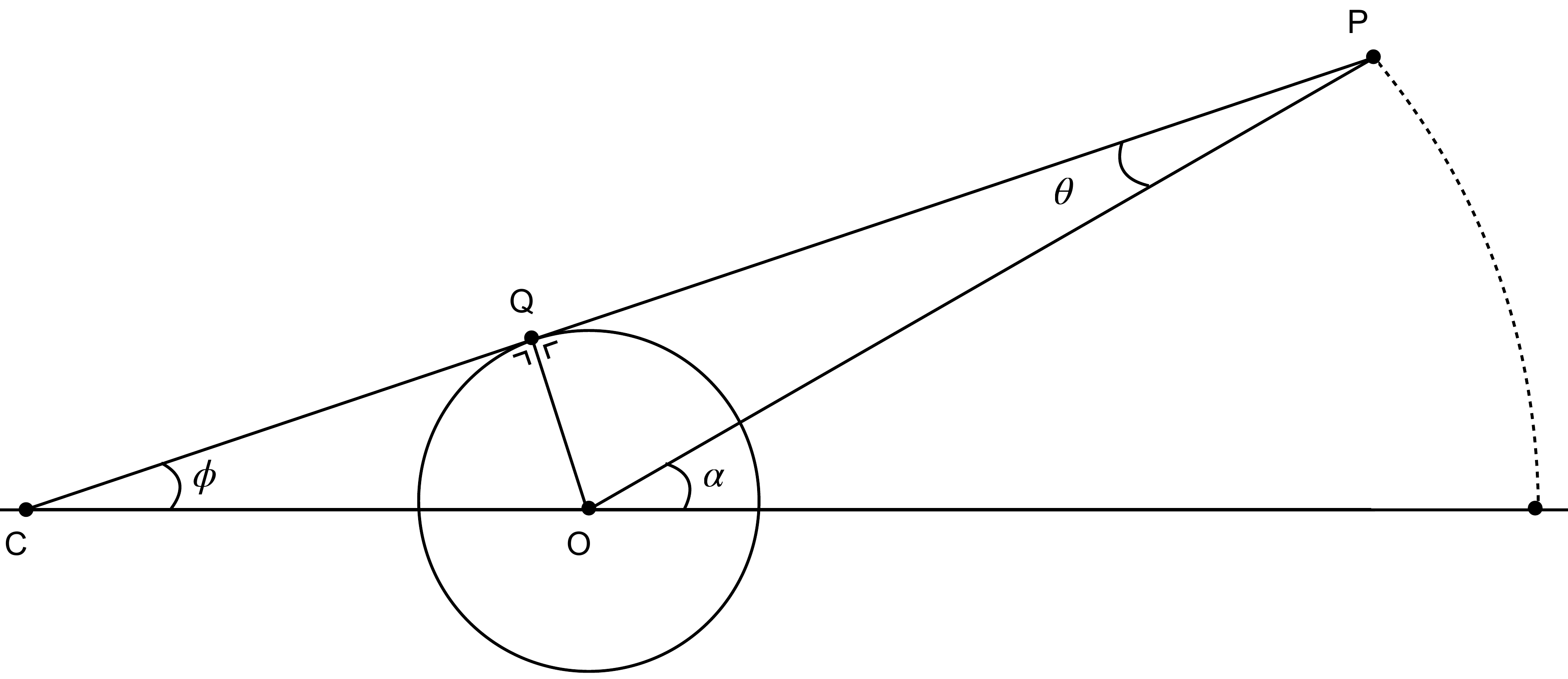}
 \caption{
Illustration of the AGB star, centred on $O$, and the companion, centred on $C$, as seen from point $P$ above the orbital plane at an angle $\alpha$.
 }
 \label{fig:angles:cone}
\end{figure}

\section{Column densities of all parent species}

Tables \ref{table:app-coldens-orich-parents} and \ref{table:app-coldens-crich-parents} list the column densities for selected parent and daughter species obtained in the grid of O-rich and C-rich outflows, respectively.
Column densities are calculated from $8 \times 10^{14}$ cm = 40 R$_*$ onwards to enable a comparison between models with a different \Rdust.

\setcounter{table}{0}
\renewcommand{\thetable}{B.\arabic{table}}

\begin{table*}
\caption{Column densities of selected parents and daughters in O-rich outflows. 
For each outflow density, the column density $[$cm$^{-2}]$ obtained including interstellar UV photons only (IS) is listed, followed by the ratios of the column density obtained including stellar and companion UV photons over that obtained including interstellar UV photons only, rounded to two significant figures. 
RD: red dwarf companion (\Tcomp\ = 4000 K, \Rcomp\ = $1.53\times 10^{10}$ cm), SL: solar-like companion (\Tcomp\ = 6000 K, \Rcomp\ = $8.14\times 10^{10}$ cm), WD: white dwarf companion (\Tcomp\ = 10 000 K, \Rcomp\ = $6.96 \times 10^{8}$ cm).
The different rows per molecule correspond to the smooth outflow (Sm.), two-component outflow (2C; \fic = 0.3, \fvol = 0.3, $l_* = 4 \times 10^{12}$ cm) and one-component outflow (1C; \fvol = 0.3, $l_* = 4 \times 10^{12}$ cm) for different onsets of dust extinction, \Rdust.
Increases/decreases larger than 10\% are marked in blue/red.
Changes larger than a factor five are marked in bold.
}
\begin{tabular}{l l l c c c c c c c c c c c c}
\hline  
& & $R_\mathrm{dust}$ & \multicolumn{4}{c}{$\dot{M}= 10^{-5}\ \mathrm{M_\odot/yr},\ v_\infty = 15\ \mathrm{km/s}$} & \multicolumn{4}{c}{$\dot{M}= 10^{-6}\ \mathrm{M_\odot/yr},\ v_\infty = 5\ \mathrm{km/s}$} & \multicolumn{4}{c}{$\dot{M}= 10^{-7}\ \mathrm{M_\odot/yr},\ v_\infty = 5\ \mathrm{km/s}$}  \\  
\cline{3-3} \cline{4-7} \cline{8-11} \cline{12-15} 
\noalign{\smallskip}
& &  & IS & RD & SL & WD & IS & RD & SL & WD & IS & RD & SL & WD \\
 \cline{4-7} \cline{8-11} \cline{12-15} 
\noalign{\smallskip}
 & Sm. & 2 $R_*$ & 3.64e+17 & 1.0 & 1.0 & 1.0 & 1.05e+17 & 1.0 & 1.0 & 1.0 & 8.94e+15 & 1.0 & \red{0.88} & \red{0.9} \\
 & & 5 $R_*$ & 3.64e+17 & 1.01 & 1.01 & 1.01 & 1.05e+17 & 1.01 & 1.01 & 1.01 & 8.94e+15 & 1.01 & \red{0.82} & \red{0.8} \\
\ce{N2} & 2C & 2 $R_*$ & 3.64e+17 & 1.0 & 1.0 & 1.0 & 1.05e+17 & 1.0 & 0.99 & 0.98 & 8.92e+15 & 1.0 & \red{0.79} & \red{0.75} \\
 & & 5 $R_*$ & 3.64e+17 & 1.01 & 1.01 & 1.01 & 1.05e+17 & 1.01 & 0.97 & 0.99 & 8.92e+15 & 1.01 & \red{0.63} & \red{0.72} \\
 & 1C & 2 $R_*$ & 3.64e+17 & 1.0 & 0.97 & 0.97 & 1.05e+17 & 1.0 & 0.93 & 0.97 & 8.90e+15 & 1.0 & \red{0.33} & \red{0.77} \\
 & & 5 $R_*$ & 3.64e+17 & 1.01 & 0.95 & 0.97 & 1.05e+17 & 1.01 & \red{0.84} & 0.97 & 8.90e+15 & 1.01 & \red{0.42} & \red{0.54} \\
\noalign{\smallskip}
 & Sm. & 2 $R_*$ & 1.92e+18 & 1.0 & 1.0 & 1.0 & 5.34e+17 & 1.0 & 1.08 & 1.01 & 3.32e+16 & 1.04 & 1.06 & 1.07 \\
 & & 5 $R_*$ & 1.92e+18 & 1.01 & 1.04 & 1.01 & 5.34e+17 & 1.02 & 1.09 & 1.09 & 3.32e+16 & 1.06 & \red{\textbf{0.0}} & \red{0.75} \\
\ce{H2O} & 2C & 2 $R_*$ & 1.92e+18 & 1.0 & 1.06 & 1.01 & 5.32e+17 & 1.03 & 1.08 & 1.1 & 3.26e+16 & 1.04 & \red{\textbf{0.17}} & 1.01 \\
 & & 5 $R_*$ & 1.92e+18 & 1.01 & 1.09 & 1.06 & 5.32e+17 & 1.04 & 1.08 & \greent{1.1} & 3.26e+16 & 1.05 & \red{\textbf{0.0}} & \red{\textbf{0.02}} \\
 & 1C & 2 $R_*$ & 1.92e+18 & 1.05 & \red{\textbf{0.0}} & 0.95 & 5.29e+17 & 1.05 & \red{\textbf{0.0}} & \red{0.68} & 3.22e+16 & 1.03 & \red{\textbf{0.0}} & \red{\textbf{0.02}} \\
 & & 5 $R_*$ & 1.92e+18 & 1.06 & \red{\textbf{0.0}} & \red{0.73} & 5.29e+17 & 1.06 & \red{\textbf{0.0}} & \red{\textbf{0.2}} & 3.22e+16 & 1.04 & \red{\textbf{0.0}} & \red{\textbf{0.0}} \\
\noalign{\smallskip}
 & Sm. & 2 $R_*$ & 2.43e+17 & 1.0 & 1.0 & 1.0 & 6.81e+16 & 1.0 & \red{0.79} & 0.95 & 4.38e+15 & 1.02 & 0.95 & \red{0.86} \\
 & & 5 $R_*$ & 2.43e+17 & 1.01 & 0.97 & 1.01 & 6.81e+16 & 1.01 & \red{0.81} & \red{0.72} & 4.38e+15 & 1.03 & \red{\textbf{0.01}} & 0.98 \\
\ce{SiO} & 2C & 2 $R_*$ & 2.43e+17 & 1.0 & \red{0.87} & 0.99 & 6.81e+16 & 1.0 & \red{0.8} & \red{0.67} & 4.51e+15 & 1.01 & \red{0.88} & 0.96 \\
 & & 5 $R_*$ & 2.43e+17 & 1.01 & \red{0.77} & \red{0.83} & 6.81e+16 & 1.01 & \red{0.82} & \red{0.7} & 4.51e+15 & 1.03 & \red{\textbf{0.0}} & \red{0.54} \\
 & 1C & 2 $R_*$ & 2.43e+17 & 1.0 & \red{0.54} & 0.9 & 6.82e+16 & 0.99 & \red{\textbf{0.01}} & 1.0 & 4.73e+15 & 1.01 & \red{\textbf{0.0}} & \red{0.6} \\
 & & 5 $R_*$ & 2.43e+17 & 1.01 & \red{\textbf{0.02}} & 1.01 & 6.82e+16 & 1.01 & \red{\textbf{0.0}} & 0.99 & 4.73e+15 & 1.03 & \red{\textbf{0.0}} & \red{\textbf{0.0}} \\
\noalign{\smallskip}
 & Sm. & 2 $R_*$ & 1.54e+17 & 1.0 & 1.0 & 1.0 & 4.10e+16 & 1.0 & \greent{1.11} & 1.06 & 1.32e+15 & \red{\textbf{0.0}} & \red{\textbf{0.0}} & \red{\textbf{0.0}} \\
 & & 5 $R_*$ & 1.54e+17 & 1.01 & \red{\textbf{0.0}} & 0.94 & 4.10e+16 & \red{\textbf{0.18}} & \red{\textbf{0.0}} & \red{\textbf{0.0}} & 1.32e+15 & \red{\textbf{0.0}} & \red{\textbf{0.0}} & \red{\textbf{0.0}} \\
\ce{H2S} & 2C & 2 $R_*$ & 1.54e+17 & 1.0 & 1.09 & 1.01 & 4.05e+16 & 0.98 & \greent{1.11} & 1.08 & 1.19e+15 & \red{\textbf{0.0}} & \red{\textbf{0.0}} & \red{\textbf{0.0}} \\
 & & 5 $R_*$ & 1.54e+17 & \red{0.45} & \red{\textbf{0.0}} & \red{\textbf{0.01}} & 4.05e+16 & \red{\textbf{0.01}} & \red{\textbf{0.0}} & \red{\textbf{0.0}} & 1.19e+15 & \red{\textbf{0.0}} & \red{\textbf{0.0}} & \red{\textbf{0.0}} \\
 & 1C & 2 $R_*$ & 1.53e+17 & \red{\textbf{0.0}} & \red{\textbf{0.0}} & \red{\textbf{0.0}} & 3.99e+16 & \red{\textbf{0.0}} & \red{\textbf{0.0}} & \red{\textbf{0.0}} & 1.06e+15 & \red{\textbf{0.0}} & \red{\textbf{0.0}} & \red{\textbf{0.0}} \\
 & & 5 $R_*$ & 1.53e+17 & \red{\textbf{0.0}} & \red{\textbf{0.0}} & \red{\textbf{0.0}} & 3.99e+16 & \red{\textbf{0.0}} & \red{\textbf{0.0}} & \red{\textbf{0.0}} & 1.06e+15 & \red{\textbf{0.0}} & \red{\textbf{0.0}} & \red{\textbf{0.0}} \\
\noalign{\smallskip}
 & Sm. & 2 $R_*$ & 3.31e+16 & 1.0 & 1.0 & 1.0 & 9.41e+15 & 0.99 & \red{\textbf{0.02}} & \red{0.89} & 5.48e+14 & \red{0.24} & \red{0.2} & \red{0.21} \\
 & & 5 $R_*$ & 3.31e+16 & 1.01 & \red{0.53} & 1.0 & 9.41e+15 & \red{0.88} & \red{\textbf{0.04}} & \red{\textbf{0.17}} & 5.48e+14 & \red{0.22} & \red{\textbf{0.0}} & \red{0.37} \\
\ce{SO2} & 2C & 2 $R_*$ & 3.33e+16 & 1.0 & \red{\textbf{0.06}} & 0.96 & 9.89e+15 & \red{0.62} & \red{\textbf{0.06}} & \red{\textbf{0.05}} & 7.04e+14 & \red{0.41} & \red{\textbf{0.03}} & \red{0.4} \\
 & & 5 $R_*$ & 3.33e+16 & 0.97 & \red{\textbf{0.01}} & \red{0.54} & 9.89e+15 & \red{0.41} & \red{\textbf{0.07}} & \red{\textbf{0.07}} & 7.04e+14 & \red{0.37} & \red{\textbf{0.0}} & \red{\textbf{0.01}} \\
 & 1C & 2 $R_*$ & 3.36e+16 & \red{\textbf{0.02}} & \red{\textbf{0.0}} & \red{0.51} & 1.05e+16 & \red{\textbf{0.12}} & \red{\textbf{0.0}} & \red{0.77} & 8.69e+14 & \red{0.46} & \red{\textbf{0.0}} & \red{\textbf{0.01}} \\
 & & 5 $R_*$ & 3.36e+16 & \red{\textbf{0.04}} & \red{\textbf{0.0}} & \red{0.75} & 1.05e+16 & \red{\textbf{0.12}} & \red{\textbf{0.0}} & \red{\textbf{0.11}} & 8.69e+14 & \red{0.38} & \red{\textbf{0.0}} & \red{\textbf{0.0}} \\
\noalign{\smallskip}
 & Sm. & 2 $R_*$ & 2.74e+16 & 1.0 & 1.0 & 1.0 & 7.87e+15 & 0.98 & \red{\textbf{0.08}} & \red{0.79} & 5.87e+14 & 0.93 & 0.92 & \red{0.9} \\
 & & 5 $R_*$ & 2.74e+16 & 1.01 & \red{0.35} & 0.99 & 7.87e+15 & \red{0.8} & \red{\textbf{0.2}} & \red{0.25} & 5.87e+14 & 0.93 & \red{\textbf{0.0}} & 1.03 \\
\ce{SO} & 2C & 2 $R_*$ & 2.75e+16 & 0.99 & \red{\textbf{0.04}} & 0.92 & 7.89e+15 & \red{0.22} & \red{\textbf{0.14}} & \red{\textbf{0.12}} & 6.87e+14 & 1.08 & \red{\textbf{0.17}} & 1.1 \\
 & & 5 $R_*$ & 2.75e+16 & 0.93 & \red{\textbf{0.05}} & \red{0.41} & 7.89e+15 & \red{0.45} & \red{0.3} & \red{0.29} & 6.87e+14 & 1.03 & \red{\textbf{0.0}} & \red{\textbf{0.06}} \\
 & 1C & 2 $R_*$ & 2.76e+16 & \red{\textbf{0.09}} & \red{\textbf{0.0}} & \greent{1.6} & 7.88e+15 & \red{0.43} & \red{\textbf{0.0}} & \greent{1.72} & 8.05e+14 & \greent{1.11} & \red{\textbf{0.0}} & \red{\textbf{0.06}} \\
 & & 5 $R_*$ & 2.76e+16 & \red{\textbf{0.12}} & \red{\textbf{0.0}} & \greent{1.77} & 7.88e+15 & \red{0.46} & \red{\textbf{0.0}} & \red{0.67} & 8.05e+14 & 1.0 & \red{\textbf{0.0}} & \red{\textbf{0.0}} \\
\noalign{\smallskip}
 & Sm. & 2 $R_*$ & 8.55e+15 & 1.0 & 1.0 & 1.0 & 2.38e+15 & 1.0 & \red{0.61} & 0.99 & 1.56e+14 & \red{0.46} & \red{\textbf{0.0}} & \red{\textbf{0.03}} \\
 & & 5 $R_*$ & 8.55e+15 & 1.01 & 0.97 & 1.01 & 2.38e+15 & 1.0 & \red{\textbf{0.0}} & \red{0.76} & 1.56e+14 & \red{0.46} & \red{\textbf{0.0}} & \red{\textbf{0.0}} \\
\ce{SiS} & 2C & 2 $R_*$ & 8.54e+15 & 1.0 & \red{0.85} & 1.0 & 2.37e+15 & 0.99 & \red{\textbf{0.0}} & \red{0.76} & 1.49e+14 & \red{0.34} & \red{\textbf{0.0}} & \red{\textbf{0.0}} \\
 & & 5 $R_*$ & 8.54e+15 & 1.01 & \red{\textbf{0.17}} & 0.94 & 2.37e+15 & 0.92 & \red{\textbf{0.0}} & \red{\textbf{0.06}} & 1.49e+14 & \red{0.36} & \red{\textbf{0.0}} & \red{\textbf{0.0}} \\
 & 1C & 2 $R_*$ & 8.52e+15 & \red{0.86} & \red{\textbf{0.0}} & \red{\textbf{0.0}} & 2.35e+15 & \red{0.61} & \red{\textbf{0.0}} & \red{\textbf{0.0}} & 1.38e+14 & \red{\textbf{0.17}} & \red{\textbf{0.0}} & \red{\textbf{0.0}} \\
 & & 5 $R_*$ & 8.52e+15 & 0.94 & \red{\textbf{0.0}} & \red{\textbf{0.0}} & 2.35e+15 & \red{0.79} & \red{\textbf{0.0}} & \red{\textbf{0.0}} & 1.38e+14 & \red{0.22} & \red{\textbf{0.0}} & \red{\textbf{0.0}} \\
\noalign{\smallskip}
 & Sm. & 2 $R_*$ & 5.53e+15 & 1.0 & 1.0 & 1.0 & 1.50e+15 & 1.0 & \red{0.81} & 1.01 & 7.11e+13 & \red{0.67} & \red{\textbf{0.0}} & \red{0.85} \\
 & & 5 $R_*$ & 5.53e+15 & 1.01 & \red{0.7} & 1.01 & 1.50e+15 & 1.01 & \red{\textbf{0.0}} & \red{0.42} & 7.11e+13 & \red{0.52} & \red{\textbf{0.0}} & \red{\textbf{0.0}} \\
\ce{NH3} & 2C & 2 $R_*$ & 5.52e+15 & 1.0 & 0.97 & 1.0 & 1.49e+15 & 0.99 & 0.95 & \greent{1.7} & 6.77e+13 & \red{0.22} & \red{\textbf{0.0}} & \red{\textbf{0.0}} \\
 & & 5 $R_*$ & 5.52e+15 & 1.01 & \red{\textbf{0.0}} & \red{0.84} & 1.49e+15 & 0.98 & \red{\textbf{0.0}} & \red{\textbf{0.0}} & 6.77e+13 & \red{0.21} & \red{\textbf{0.0}} & \red{\textbf{0.0}} \\
 & 1C & 2 $R_*$ & 5.51e+15 & \red{0.48} & \red{\textbf{0.0}} & \red{\textbf{0.0}} & 1.47e+15 & \red{\textbf{0.05}} & \red{\textbf{0.0}} & \red{\textbf{0.0}} & 6.43e+13 & \red{\textbf{0.0}} & \red{\textbf{0.0}} & \red{\textbf{0.0}} \\
 & & 5 $R_*$ & 5.51e+15 & \red{0.78} & \red{\textbf{0.0}} & \red{\textbf{0.0}} & 1.47e+15 & \red{0.43} & \red{\textbf{0.0}} & \red{\textbf{0.0}} & 6.43e+13 & \red{\textbf{0.07}} & \red{\textbf{0.0}} & \red{\textbf{0.0}} \\
\noalign{\smallskip}
 & Sm. & 2 $R_*$ & 2.70e+15 & 1.0 & 1.0 & 1.0 & 7.63e+14 & 1.0 & \red{\textbf{0.02}} & \red{\textbf{0.07}} & 7.33e+13 & \red{0.87} & \red{0.44} & \red{0.36} \\
 & & 5 $R_*$ & 2.70e+15 & 1.01 & \red{0.49} & 0.94 & 7.63e+14 & 1.01 & \red{\textbf{0.05}} & \red{\textbf{0.02}} & 7.33e+13 & 0.98 & \greent{1.34} & \greent{1.55} \\
\ce{CO2} & 2C & 2 $R_*$ & 2.69e+15 & 1.0 & \red{\textbf{0.02}} & \red{0.25} & 7.73e+14 & \red{0.65} & \red{\textbf{0.03}} & \red{\textbf{0.03}} & 9.24e+13 & \red{0.68} & \greent{3.32} & \red{0.53} \\
 & & 5 $R_*$ & 2.69e+15 & 1.01 & \red{\textbf{0.07}} & \red{\textbf{0.08}} & 7.73e+14 & 0.98 & \red{0.37} & \red{\textbf{0.05}} & 9.24e+13 & 0.91 & \red{\textbf{0.06}} & \greent{1.23} \\
 & 1C & 2 $R_*$ & 2.69e+15 & \red{\textbf{0.04}} & \greent{3.03} & \red{0.52} & 7.97e+14 & \red{\textbf{0.06}} & \greent{1.23} & \greent{2.1} & 1.12e+14 & \red{0.57} & \red{\textbf{0.09}} & \greent{1.18} \\
 & & 5 $R_*$ & 2.69e+15 & 0.95 & \red{0.78} & \greent{1.98} & 7.97e+14 & \red{0.83} & \red{0.43} & \greent{2.88} & 1.12e+14 & \red{0.87} & \red{\textbf{0.0}} & \red{0.33} \\
\hline 
\end{tabular}
\label{table:app-coldens-orich-parents}    
\end{table*}

\begin{table*}
\label{tab:continued}
\begin{tabular}{l l l c c c c c c c c c c c c}
\hline  
& & $R_\mathrm{dust}$ & \multicolumn{4}{c}{$\dot{M}= 10^{-5}\ \mathrm{M_\odot/yr},\ v_\infty = 15\ \mathrm{km/s}$} & \multicolumn{4}{c}{$\dot{M}= 10^{-6}\ \mathrm{M_\odot/yr},\ v_\infty = 5\ \mathrm{km/s}$} & \multicolumn{4}{c}{$\dot{M}= 10^{-7}\ \mathrm{M_\odot/yr},\ v_\infty = 5\ \mathrm{km/s}$}  \\  
\cline{3-3} \cline{4-7} \cline{8-11} \cline{12-15} 
\noalign{\smallskip}
& &  & IS & RD & SL & WD & IS & RD & SL & WD & IS & RD & SL & WD \\
 \cline{4-7} \cline{8-11} \cline{12-15} 
\noalign{\smallskip}
 & Sm. & 2 $R_*$ & 2.30e+15 & 1.0 & 1.0 & 1.0 & 6.29e+14 & 1.0 & 1.0 & 1.0 & 3.33e+13 & 1.02 & 1.02 & \greent{1.52} \\
 & & 5 $R_*$ & 2.30e+15 & 1.01 & 1.01 & 1.01 & 6.29e+14 & 1.02 & 1.03 & 1.04 & 3.33e+13 & 1.04 & \red{0.4} & 1.06 \\
\ce{HCN} & 2C & 2 $R_*$ & 2.30e+15 & 1.0 & 1.0 & 1.0 & 6.24e+14 & 1.0 & 1.09 & 1.01 & 3.27e+13 & 1.04 & \red{0.83} & \greent{2.12} \\
 & & 5 $R_*$ & 2.30e+15 & 1.01 & 1.04 & 1.02 & 6.24e+14 & 1.02 & 1.01 & \greent{1.14} & 3.27e+13 & 1.05 & \red{\textbf{0.01}} & \red{0.69} \\
 & 1C & 2 $R_*$ & 2.29e+15 & 1.01 & \red{0.84} & \greent{1.23} & 6.19e+14 & 1.02 & \red{\textbf{0.19}} & \greent{1.28} & 3.23e+13 & 1.06 & \red{\textbf{0.0}} & \greent{1.49} \\
 & & 5 $R_*$ & 2.29e+15 & 1.01 & \red{0.21} & \greent{1.12} & 6.19e+14 & 1.02 & \red{\textbf{0.01}} & \red{0.85} & 3.23e+13 & 1.06 & \red{\textbf{0.0}} & \red{\textbf{0.19}} \\
\noalign{\smallskip}
 & Sm. & 2 $R_*$ & 4.95e+14 & 1.0 & 1.0 & 1.0 & 1.28e+14 & 1.0 & \red{0.83} & 0.98 & 4.42e+12 & 1.01 & \red{\textbf{0.05}} & \red{0.43} \\
 & & 5 $R_*$ & 4.95e+14 & 1.01 & 0.99 & 1.01 & 1.28e+14 & 1.02 & \red{\textbf{0.15}} & \red{0.78} & 4.42e+12 & 1.02 & \red{\textbf{0.0}} & \red{\textbf{0.08}} \\
\ce{CS} & 2C & 2 $R_*$ & 4.89e+14 & 1.0 & 0.93 & 0.99 & 1.21e+14 & 0.97 & \red{0.29} & \red{0.44} & 3.10e+12 & 1.03 & \red{\textbf{0.02}} & \red{\textbf{0.16}} \\
 & & 5 $R_*$ & 4.89e+14 & 1.01 & \red{0.43} & 0.93 & 1.21e+14 & 1.03 & \red{\textbf{0.01}} & \red{\textbf{0.09}} & 3.10e+12 & 1.06 & \red{\textbf{0.0}} & \red{\textbf{0.02}} \\
 & 1C & 2 $R_*$ & 4.82e+14 & \red{0.85} & \red{\textbf{0.0}} & \red{\textbf{0.0}} & 1.10e+14 & \red{0.72} & \red{\textbf{0.01}} & \red{\textbf{0.01}} & 1.43e+12 & \greent{1.12} & \red{\textbf{0.01}} & \red{\textbf{0.07}} \\
 & & 5 $R_*$ & 4.82e+14 & 0.95 & \red{\textbf{0.0}} & \red{\textbf{0.0}} & 1.10e+14 & 0.97 & \red{\textbf{0.01}} & \red{\textbf{0.01}} & 1.43e+12 & \greent{1.25} & \red{\textbf{0.01}} & \red{\textbf{0.02}} \\
\noalign{\smallskip}
 & Sm. & 2 $R_*$ & 1.42e+13 & 1.0 & 1.0 & 1.0 & 1.69e+13 & 1.01 & \greent{\textbf{9.7}} & \greent{1.11} & 1.54e+13 & \red{0.83} & \red{\textbf{0.07}} & \greent{\textbf{22.32}} \\
 & & 5 $R_*$ & 1.42e+13 & 1.01 & \greent{\textbf{105.06}} & \greent{1.28} & 1.69e+13 & \red{0.47} & \greent{1.67} & \greent{\textbf{49.95}} & 1.54e+13 & 0.9 & \red{\textbf{0.0}} & \red{\textbf{0.0}} \\
\ce{NS} & 2C & 2 $R_*$ & 1.57e+13 & 1.0 & \greent{4.48} & 1.04 & 1.63e+13 & \greent{1.58} & \greent{\textbf{104.52}} & \greent{\textbf{73.2}} & 1.54e+13 & \greent{1.63} & \red{\textbf{0.0}} & \red{\textbf{0.04}} \\
 & & 5 $R_*$ & 1.57e+13 & \red{0.59} & \greent{\textbf{114.64}} & \greent{\textbf{113.7}} & 1.63e+13 & \greent{2.48} & \red{\textbf{0.02}} & \greent{4.84} & 1.54e+13 & 0.97 & \red{\textbf{0.0}} & \red{\textbf{0.0}} \\
 & 1C & 2 $R_*$ & 1.86e+13 & \greent{\textbf{137.22}} & \red{\textbf{0.0}} & \red{\textbf{0.0}} & 1.73e+13 & \greent{\textbf{55.68}} & \red{\textbf{0.0}} & \red{\textbf{0.0}} & 1.66e+13 & \red{0.75} & \red{\textbf{0.0}} & \red{\textbf{0.0}} \\
 & & 5 $R_*$ & 1.86e+13 & \greent{\textbf{21.93}} & \red{\textbf{0.0}} & \red{\textbf{0.0}} & 1.73e+13 & \greent{\textbf{14.07}} & \red{\textbf{0.0}} & \red{\textbf{0.0}} & 1.66e+13 & \red{0.61} & \red{\textbf{0.0}} & \red{\textbf{0.0}} \\
\noalign{\smallskip}
 & Sm. & 2 $R_*$ & 4.97e+12 & 1.0 & 1.0 & 1.0 & 4.12e+12 & 1.0 & \greent{\textbf{32.91}} & \greent{1.58} & 2.51e+12 & \red{0.82} & \red{\textbf{0.0}} & \greent{\textbf{73.39}} \\
 & & 5 $R_*$ & 4.97e+12 & 1.0 & \greent{1.57} & 1.02 & 4.12e+12 & 1.0 & \red{\textbf{0.03}} & \greent{\textbf{24.09}} & 2.51e+12 & \red{0.61} & \red{\textbf{0.0}} & \red{\textbf{0.0}} \\
\ce{SiN} & 2C & 2 $R_*$ & 4.56e+12 & 1.0 & \greent{\textbf{23.57}} & \greent{1.19} & 3.31e+12 & 0.98 & \greent{\textbf{66.67}} & \greent{\textbf{352.37}} & 2.19e+12 & \red{0.29} & \red{\textbf{0.0}} & \red{\textbf{0.19}} \\
 & & 5 $R_*$ & 4.56e+12 & 1.0 & \red{0.82} & \greent{\textbf{20.14}} & 3.31e+12 & 1.01 & \red{\textbf{0.01}} & \greent{\textbf{13.13}} & 2.19e+12 & \red{0.27} & \red{\textbf{0.0}} & \red{\textbf{0.0}} \\
 & 1C & 2 $R_*$ & 4.34e+12 & \red{0.48} & \red{\textbf{0.01}} & \red{\textbf{0.0}} & 2.54e+12 & \red{\textbf{0.07}} & \red{\textbf{0.02}} & \red{\textbf{0.01}} & 2.01e+12 & \red{\textbf{0.0}} & \red{\textbf{0.0}} & \red{\textbf{0.0}} \\
 & & 5 $R_*$ & 4.34e+12 & \red{0.75} & \red{\textbf{0.02}} & \red{\textbf{0.0}} & 2.54e+12 & \red{0.53} & \red{\textbf{0.02}} & \red{\textbf{0.01}} & 2.01e+12 & \red{\textbf{0.09}} & \red{\textbf{0.0}} & \red{\textbf{0.0}} \\
\noalign{\smallskip}
 & Sm. & 2 $R_*$ & 1.62e+16 & 1.0 & 1.0 & 1.0 & 1.19e+16 & 1.0 & \greent{1.11} & 1.02 & 7.72e+15 & 0.92 & \red{0.88} & 0.94 \\
 & & 5 $R_*$ & 1.62e+16 & 1.0 & 1.03 & 1.0 & 1.19e+16 & 0.99 & \red{0.87} & 0.94 & 7.72e+15 & 0.92 & \red{\textbf{0.0}} & \red{0.59} \\
\ce{OH} & 2C & 2 $R_*$ & 1.48e+16 & 1.0 & 1.09 & 1.01 & 1.02e+16 & 1.08 & 1.05 & 1.08 & 6.39e+15 & \red{0.89} & \red{\textbf{0.1}} & \red{0.78} \\
 & & 5 $R_*$ & 1.48e+16 & 1.01 & 0.98 & 1.04 & 1.02e+16 & \red{0.82} & \red{0.8} & \red{0.82} & 6.39e+15 & \red{0.88} & \red{\textbf{0.0}} & \red{\textbf{0.02}} \\
 & 1C & 2 $R_*$ & 1.29e+16 & \red{0.84} & \red{\textbf{0.0}} & \red{0.69} & 7.65e+15 & \red{0.71} & \red{\textbf{0.0}} & \red{0.3} & 4.66e+15 & \red{0.81} & \red{\textbf{0.0}} & \red{\textbf{0.02}} \\
 & & 5 $R_*$ & 1.29e+16 & \red{0.84} & \red{\textbf{0.0}} & \red{0.44} & 7.65e+15 & \red{0.71} & \red{\textbf{0.0}} & \red{\textbf{0.06}} & 4.66e+15 & \red{0.78} & \red{\textbf{0.0}} & \red{\textbf{0.0}} \\
\noalign{\smallskip}
 & Sm. & 2 $R_*$ & 1.15e+15 & 1.0 & 1.0 & 1.0 & 8.54e+14 & 1.01 & \greent{2.44} & \greent{1.13} & 4.94e+14 & \red{\textbf{0.01}} & \red{\textbf{0.0}} & \red{\textbf{0.0}} \\
 & & 5 $R_*$ & 1.15e+15 & \greent{1.37} & \red{\textbf{0.01}} & \greent{4.51} & 8.54e+14 & \red{0.21} & \red{\textbf{0.0}} & \red{\textbf{0.02}} & 4.94e+14 & \red{\textbf{0.01}} & \red{\textbf{0.0}} & \red{\textbf{0.0}} \\
\ce{HS} & 2C & 2 $R_*$ & 1.06e+15 & 1.04 & 1.05 & \greent{1.96} & 7.84e+14 & \red{0.89} & 0.98 & 0.96 & 3.74e+14 & \red{\textbf{0.01}} & \red{\textbf{0.0}} & \red{\textbf{0.0}} \\
 & & 5 $R_*$ & 1.06e+15 & \greent{2.4} & \red{\textbf{0.0}} & \red{\textbf{0.02}} & 7.84e+14 & \red{\textbf{0.03}} & \red{\textbf{0.0}} & \red{\textbf{0.0}} & 3.74e+14 & \red{\textbf{0.0}} & \red{\textbf{0.0}} & \red{\textbf{0.0}} \\
 & 1C & 2 $R_*$ & 9.46e+14 & \red{\textbf{0.02}} & \red{\textbf{0.0}} & \red{\textbf{0.0}} & 6.62e+14 & \red{\textbf{0.02}} & \red{\textbf{0.0}} & \red{\textbf{0.0}} & 2.29e+14 & \red{\textbf{0.0}} & \red{\textbf{0.0}} & \red{\textbf{0.0}} \\
 & & 5 $R_*$ & 9.46e+14 & \red{\textbf{0.02}} & \red{\textbf{0.0}} & \red{\textbf{0.0}} & 6.62e+14 & \red{\textbf{0.03}} & \red{\textbf{0.0}} & \red{\textbf{0.0}} & 2.29e+14 & \red{\textbf{0.0}} & \red{\textbf{0.0}} & \red{\textbf{0.0}} \\
\noalign{\smallskip}
 & Sm. & 2 $R_*$ & 3.13e+13 & 1.0 & 1.0 & 1.0 & 2.64e+13 & 1.0 & 1.06 & 1.01 & 1.58e+13 & 1.09 & 1.03 & \greent{1.62} \\
 & & 5 $R_*$ & 3.13e+13 & 1.0 & 1.0 & 1.0 & 2.64e+13 & 0.98 & 1.09 & 1.06 & 1.58e+13 & 1.01 & \red{0.25} & 0.92 \\
\ce{CN} & 2C & 2 $R_*$ & 2.92e+13 & 1.0 & 1.05 & 1.0 & 2.42e+13 & 1.02 & \greent{1.2} & 1.1 & 1.41e+13 & 1.08 & \red{0.46} & \greent{1.97} \\
 & & 5 $R_*$ & 2.92e+13 & 0.99 & \greent{1.11} & 1.02 & 2.42e+13 & 0.98 & 0.97 & \greent{1.14} & 1.41e+13 & 1.0 & \red{\textbf{0.01}} & \red{0.36} \\
 & 1C & 2 $R_*$ & 2.66e+13 & 1.04 & \red{0.5} & \greent{1.12} & 2.13e+13 & 1.06 & \red{\textbf{0.09}} & \red{0.63} & 1.21e+13 & 1.07 & \red{\textbf{0.0}} & \red{0.54} \\
 & & 5 $R_*$ & 2.66e+13 & 1.01 & \red{\textbf{0.16}} & \red{0.8} & 2.13e+13 & 1.0 & \red{\textbf{0.03}} & \red{0.27} & 1.21e+13 & 0.97 & \red{\textbf{0.0}} & \red{\textbf{0.07}} \\
\noalign{\smallskip}
 & Sm. & 2 $R_*$ & 2.51e+13 & 1.0 & 1.0 & 1.0 & 1.78e+13 & 1.0 & \red{0.84} & 1.01 & 1.22e+13 & \red{0.87} & \red{\textbf{0.0}} & 0.99 \\
 & & 5 $R_*$ & 2.51e+13 & 1.0 & \red{0.73} & 1.0 & 1.78e+13 & 1.01 & \red{\textbf{0.0}} & \red{0.43} & 1.22e+13 & 0.98 & \red{\textbf{0.0}} & \red{\textbf{0.0}} \\
\ce{NH} & 2C & 2 $R_*$ & 2.24e+13 & 1.0 & 1.0 & 1.0 & 1.54e+13 & 1.03 & \red{0.8} & \greent{1.76} & 1.00e+13 & \red{0.43} & \red{\textbf{0.0}} & \red{\textbf{0.0}} \\
 & & 5 $R_*$ & 2.24e+13 & 1.01 & \red{\textbf{0.0}} & \red{0.87} & 1.54e+13 & 0.94 & \red{\textbf{0.0}} & \red{\textbf{0.0}} & 1.00e+13 & \red{0.85} & \red{\textbf{0.0}} & \red{\textbf{0.0}} \\
 & 1C & 2 $R_*$ & 1.83e+13 & \red{0.53} & \red{\textbf{0.0}} & \red{\textbf{0.0}} & 1.13e+13 & \red{\textbf{0.05}} & \red{\textbf{0.0}} & \red{\textbf{0.0}} & 6.83e+12 & \red{\textbf{0.01}} & \red{\textbf{0.0}} & \red{\textbf{0.0}} \\
 & & 5 $R_*$ & 1.83e+13 & 0.92 & \red{\textbf{0.0}} & \red{\textbf{0.0}} & 1.13e+13 & \red{0.52} & \red{\textbf{0.0}} & \red{\textbf{0.0}} & 6.83e+12 & \red{0.42} & \red{\textbf{0.0}} & \red{\textbf{0.0}} \\
\noalign{\smallskip}
 & Sm. & 2 $R_*$ & 2.76e+18 & 1.0 & 1.0 & 1.0 & 8.24e+17 & 1.0 & 1.0 & 1.0 & 8.01e+16 & 1.0 & 1.0 & 1.0 \\
 & & 5 $R_*$ & 2.76e+18 & 1.01 & 1.01 & 1.01 & 8.24e+17 & 1.01 & 1.01 & 1.01 & 8.01e+16 & 1.01 & 1.01 & 1.01 \\
\ce{CO} & 2C & 2 $R_*$ & 2.76e+18 & 1.0 & 1.0 & 1.0 & 8.25e+17 & 1.0 & 1.0 & 1.0 & 8.05e+16 & 1.0 & 1.0 & 1.0 \\
 & & 5 $R_*$ & 2.76e+18 & 1.01 & 1.01 & 1.01 & 8.25e+17 & 1.01 & 1.01 & 1.01 & 8.05e+16 & 1.01 & 1.01 & 1.01 \\
 & 1C & 2 $R_*$ & 2.77e+18 & 1.0 & 1.0 & 1.0 & 8.27e+17 & 1.0 & 1.0 & 1.0 & 8.14e+16 & 1.0 & 1.0 & 1.0 \\
 & & 5 $R_*$ & 2.77e+18 & 1.01 & 1.01 & 1.01 & 8.27e+17 & 1.01 & 1.01 & 1.01 & 8.14e+16 & 1.01 & 1.01 & 1.01 \\
\noalign{\smallskip}
 & Sm. & 2 $R_*$ & 8.72e+14 & 1.0 & 1.0 & 1.0 & 3.88e+14 & 1.0 & 1.0 & 1.0 & 3.20e+14 & 1.01 & 1.01 & 1.01 \\
 & & 5 $R_*$ & 8.72e+14 & 1.0 & 1.0 & 1.0 & 3.88e+14 & 1.0 & 1.08 & 1.0 & 3.20e+14 & 1.0 & 1.05 & 1.02 \\
\ce{C} & 2C & 2 $R_*$ & 7.81e+14 & 1.0 & 1.0 & 1.0 & 3.54e+14 & 1.0 & 1.0 & 1.0 & 2.88e+14 & 1.01 & 1.03 & 1.07 \\
 & & 5 $R_*$ & 7.81e+14 & 1.0 & 1.05 & 1.0 & 3.54e+14 & 1.01 & 1.02 & 1.08 & 2.88e+14 & 1.0 & \greent{1.27} & 1.07 \\
 & 1C & 2 $R_*$ & 3.61e+14 & 1.01 & \greent{1.14} & 1.01 & 1.68e+14 & 1.01 & \greent{1.24} & 1.02 & 1.27e+14 & 1.01 & 1.06 & \greent{1.13} \\
 & & 5 $R_*$ & 3.61e+14 & 1.01 & \greent{1.29} & 1.02 & 1.68e+14 & 1.02 & \greent{1.2} & 1.06 & 1.27e+14 & 1.01 & 1.07 & \greent{1.15} \\
\hline
\end{tabular}
\end{table*}

\begin{table*}
\caption{Column densities of selected parents and daughters in C-rich outflows. 
For each outflow density, the column density $[$cm$^{-2}]$ obtained including interstellar UV photons only (IS) is listed, followed by the ratios of the column density obtained including stellar and companion UV photons over that obtained including interstellar UV photons only, rounded to two significant figures. 
RD: red dwarf companion (\Tcomp\ = 4000 K, \Rcomp\ = $1.53\times 10^{10}$ cm), SL: solar-like companion (\Tcomp\ = 6000 K, \Rcomp\ = $8.14\times 10^{10}$ cm), WD: white dwarf companion (\Tcomp\ = 10 000 K, \Rcomp\ = $6.96 \times 10^{8}$ cm).
The different rows per molecule correspond to the smooth outflow (Sm.), two-component outflow (2C; \fic = 0.3, \fvol = 0.3, $l_* = 4 \times 10^{12}$ cm) and one-component outflow (1C; \fvol = 0.3, $l_* = 4 \times 10^{12}$ cm) for different onsets of dust extinction, \Rdust.
Increases/decreases larger than 10\% are marked in blue/red.
Changes larger than a factor five are marked in bold.
} 
\begin{tabular}{l l l c c c c c c c c c c c c}
\hline  
& & $R_\mathrm{dust}$ & \multicolumn{4}{c}{$\dot{M}= 10^{-5}\ \mathrm{M_\odot/yr},\ v_\infty = 15\ \mathrm{km/s}$} & \multicolumn{4}{c}{$\dot{M}= 10^{-6}\ \mathrm{M_\odot/yr},\ v_\infty = 5\ \mathrm{km/s}$} & \multicolumn{4}{c}{$\dot{M}= 10^{-7}\ \mathrm{M_\odot/yr},\ v_\infty = 5\ \mathrm{km/s}$}  \\  
\cline{3-3} \cline{4-7} \cline{8-11} \cline{12-15} 
\noalign{\smallskip}
& &  & IS & RD & SL & WD & IS & RD & SL & WD & IS & RD & SL & WD \\
 \cline{4-7} \cline{8-11} \cline{12-15} 
\noalign{\smallskip}
 & Sm. & 2 $R_*$ & 3.64e+17 & 1.0 & 1.0 & 1.0 & 1.06e+17 & 1.0 & 1.0 & 1.0 & 9.19e+15 & 1.0 & \red{0.89} & \red{0.81} \\
 & & 5 $R_*$ & 3.64e+17 & 1.01 & 1.01 & 1.01 & 1.06e+17 & 1.01 & 1.01 & 1.0 & 9.19e+15 & 1.01 & 1.04 & \red{0.83} \\
\ce{N2} & 2C & 2 $R_*$ & 3.64e+17 & 1.0 & 1.0 & 1.0 & 1.06e+17 & 1.0 & 0.99 & 0.98 & 9.33e+15 & 1.0 & \red{0.71} & \red{0.49} \\
 & & 5 $R_*$ & 3.64e+17 & 1.01 & 1.01 & 1.01 & 1.06e+17 & 1.01 & 0.97 & 0.94 & 9.33e+15 & 1.01 & 1.09 & \red{0.82} \\
 & 1C & 2 $R_*$ & 3.65e+17 & 1.0 & \red{0.49} & \red{0.34} & 1.07e+17 & 1.0 & \red{\textbf{0.16}} & \red{\textbf{0.08}} & 9.49e+15 & 1.0 & \greent{1.24} & \red{0.29} \\
 & & 5 $R_*$ & 3.65e+17 & 1.01 & \red{0.81} & \red{0.69} & 1.07e+17 & 1.01 & \red{0.59} & \red{0.35} & 9.49e+15 & 1.01 & \red{0.74} & 1.02 \\
\noalign{\smallskip}
 & Sm. & 2 $R_*$ & 3.12e+16 & 1.0 & 1.0 & 1.0 & 8.58e+15 & 1.0 & \greent{3.54} & 1.06 & 4.55e+14 & 1.01 & \red{\textbf{0.0}} & \greent{\textbf{11.98}} \\
 & & 5 $R_*$ & 3.12e+16 & 1.01 & 1.01 & 1.01 & 8.58e+15 & 1.01 & \red{\textbf{0.17}} & \red{0.87} & 4.55e+14 & 1.01 & \red{\textbf{0.0}} & \red{\textbf{0.0}} \\
\ce{CH4} & 2C & 2 $R_*$ & 3.12e+16 & 1.0 & \greent{1.48} & 1.02 & 8.53e+15 & 1.0 & \greent{\textbf{23.87}} & \greent{4.2} & 4.40e+14 & 1.05 & \red{\textbf{0.0}} & \greent{1.24} \\
 & & 5 $R_*$ & 3.12e+16 & 1.01 & \red{0.23} & 0.98 & 8.53e+15 & 1.01 & \red{\textbf{0.0}} & \red{\textbf{0.19}} & 4.40e+14 & 1.01 & \red{\textbf{0.0}} & \red{\textbf{0.0}} \\
 & 1C & 2 $R_*$ & 3.11e+16 & 0.99 & \red{\textbf{0.0}} & \red{\textbf{0.03}} & 8.46e+15 & 0.98 & \red{\textbf{0.0}} & \red{\textbf{0.0}} & 4.25e+14 & \greent{1.14} & \red{\textbf{0.0}} & \red{\textbf{0.0}} \\
 & & 5 $R_*$ & 3.11e+16 & 1.01 & \red{\textbf{0.0}} & \red{\textbf{0.0}} & 8.46e+15 & 1.0 & \red{\textbf{0.0}} & \red{\textbf{0.0}} & 4.25e+14 & 1.0 & \red{\textbf{0.0}} & \red{\textbf{0.0}} \\
\noalign{\smallskip}
 & Sm. & 2 $R_*$ & 2.28e+16 & 1.0 & 1.0 & 1.0 & 6.34e+15 & 1.0 & \greent{1.86} & \greent{1.13} & 3.95e+14 & 1.0 & \red{\textbf{0.04}} & \greent{2.84} \\
 & & 5 $R_*$ & 2.28e+16 & 1.01 & \greent{1.1} & 1.02 & 6.34e+15 & 1.01 & \red{0.28} & \greent{2.12} & 3.95e+14 & 1.02 & \red{\textbf{0.0}} & \red{\textbf{0.0}} \\
\ce{H2O} & 2C & 2 $R_*$ & 2.28e+16 & 1.0 & \greent{1.37} & 1.04 & 6.31e+15 & 1.0 & \greent{2.52} & \greent{2.84} & 3.90e+14 & 1.01 & \red{\textbf{0.0}} & 1.05 \\
 & & 5 $R_*$ & 2.28e+16 & 1.01 & \red{0.84} & \greent{1.48} & 6.31e+15 & 1.01 & \red{\textbf{0.01}} & \greent{1.85} & 3.90e+14 & 1.01 & \red{\textbf{0.0}} & \red{\textbf{0.0}} \\
 & 1C & 2 $R_*$ & 2.28e+16 & 1.01 & \red{\textbf{0.0}} & \red{\textbf{0.05}} & 6.28e+15 & 1.02 & \red{\textbf{0.0}} & \red{\textbf{0.0}} & 3.87e+14 & 0.99 & \red{\textbf{0.0}} & \red{\textbf{0.0}} \\
 & & 5 $R_*$ & 2.28e+16 & 1.01 & \red{\textbf{0.0}} & \red{\textbf{0.03}} & 6.28e+15 & 1.01 & \red{\textbf{0.0}} & \red{\textbf{0.0}} & 3.87e+14 & 1.0 & \red{\textbf{0.0}} & \red{\textbf{0.0}} \\
\noalign{\smallskip}
 & Sm. & 2 $R_*$ & 1.69e+17 & 1.0 & 1.0 & 1.0 & 4.85e+16 & 1.0 & \greent{1.12} & 1.0 & 4.04e+15 & 0.98 & \red{0.31} & \red{0.29} \\
 & & 5 $R_*$ & 1.69e+17 & 1.01 & 0.99 & 1.01 & 4.85e+16 & 1.01 & \red{\textbf{0.15}} & 1.07 & 4.04e+15 & 0.98 & \red{\textbf{0.0}} & \red{\textbf{0.03}} \\
\ce{SiC2} & 2C & 2 $R_*$ & 1.69e+17 & 1.0 & 1.04 & 1.0 & 4.84e+16 & 0.99 & \red{\textbf{0.04}} & \greent{1.2} & 4.07e+15 & 0.97 & \red{0.34} & \red{\textbf{0.14}} \\
 & & 5 $R_*$ & 1.69e+17 & 1.01 & \red{0.5} & 1.01 & 4.84e+16 & 0.99 & \red{0.86} & 0.93 & 4.07e+15 & 0.97 & \red{\textbf{0.0}} & \red{\textbf{0.08}} \\
 & 1C & 2 $R_*$ & 1.69e+17 & 0.93 & \red{\textbf{0.07}} & \red{0.56} & 4.83e+16 & 0.97 & \red{\textbf{0.11}} & \red{0.59} & 4.12e+15 & 0.94 & \red{\textbf{0.0}} & \red{0.37} \\
 & & 5 $R_*$ & 1.69e+17 & 0.96 & \red{\textbf{0.09}} & \greent{1.15} & 4.83e+16 & 0.94 & \red{\textbf{0.18}} & \greent{1.12} & 4.12e+15 & 0.96 & \red{\textbf{0.0}} & \red{\textbf{0.09}} \\
\noalign{\smallskip}
 & Sm. & 2 $R_*$ & 9.56e+16 & 1.0 & 1.0 & 1.0 & 2.71e+16 & 1.0 & \red{0.88} & 0.98 & 2.04e+15 & 1.0 & \greent{1.53} & \greent{1.22} \\
 & & 5 $R_*$ & 9.56e+16 & 1.01 & 1.0 & 1.01 & 2.71e+16 & 1.01 & \greent{1.57} & \red{0.82} & 2.04e+15 & 1.01 & \red{\textbf{0.01}} & \red{0.23} \\
\ce{CS} & 2C & 2 $R_*$ & 9.55e+16 & 1.0 & 0.94 & 0.99 & 2.69e+16 & 1.0 & \greent{1.37} & \red{0.39} & 2.02e+15 & 1.01 & \red{0.56} & \red{0.87} \\
 & & 5 $R_*$ & 9.55e+16 & 1.01 & \greent{1.46} & 0.9 & 2.69e+16 & 1.01 & \greent{1.32} & \greent{1.12} & 2.02e+15 & 1.02 & \red{\textbf{0.0}} & \red{0.31} \\
 & 1C & 2 $R_*$ & 9.54e+16 & 1.0 & \red{0.5} & \red{0.22} & 2.68e+16 & 1.06 & \red{0.36} & \red{\textbf{0.15}} & 2.00e+15 & 1.08 & \red{\textbf{0.01}} & \red{0.29} \\
 & & 5 $R_*$ & 9.54e+16 & 1.01 & \red{0.51} & \red{0.45} & 2.68e+16 & 1.02 & \red{0.54} & \red{0.24} & 2.00e+15 & 1.03 & \red{\textbf{0.0}} & \red{0.44} \\
\noalign{\smallskip}
 & Sm. & 2 $R_*$ & 3.85e+17 & 1.0 & 1.0 & 1.0 & 1.01e+17 & 1.0 & \red{\textbf{0.04}} & 0.97 & 4.82e+15 & \red{0.89} & \red{\textbf{0.13}} & \red{\textbf{0.04}} \\
 & & 5 $R_*$ & 3.85e+17 & 1.01 & \red{0.89} & 1.01 & 1.01e+17 & 1.01 & \red{\textbf{0.0}} & \red{0.23} & 4.82e+15 & \red{0.87} & \red{\textbf{0.0}} & \red{\textbf{0.0}} \\
\ce{C2H2} & 2C & 2 $R_*$ & 3.83e+17 & 1.0 & \red{0.55} & 0.99 & 9.96e+16 & 0.98 & \red{\textbf{0.01}} & \red{0.23} & 4.49e+15 & \red{0.62} & \red{\textbf{0.03}} & \red{\textbf{0.02}} \\
 & & 5 $R_*$ & 3.83e+17 & 1.01 & \red{\textbf{0.0}} & \red{0.83} & 9.96e+16 & 0.98 & \red{\textbf{0.0}} & \red{\textbf{0.0}} & 4.49e+15 & \red{0.72} & \red{\textbf{0.0}} & \red{\textbf{0.0}} \\
 & 1C & 2 $R_*$ & 3.81e+17 & \red{0.69} & \red{\textbf{0.0}} & \red{\textbf{0.0}} & 9.78e+16 & \red{\textbf{0.12}} & \red{\textbf{0.0}} & \red{\textbf{0.0}} & 4.17e+15 & \red{\textbf{0.05}} & \red{\textbf{0.0}} & \red{\textbf{0.01}} \\
 & & 5 $R_*$ & 3.81e+17 & \red{0.9} & \red{\textbf{0.0}} & \red{\textbf{0.0}} & 9.78e+16 & \red{0.67} & \red{\textbf{0.0}} & \red{\textbf{0.0}} & 4.17e+15 & \red{0.53} & \red{\textbf{0.0}} & \red{\textbf{0.0}} \\
\noalign{\smallskip}
 & Sm. & 2 $R_*$ & 3.64e+17 & 1.0 & 1.0 & 1.0 & 9.93e+16 & 1.0 & \red{0.62} & 1.0 & 5.07e+15 & 0.98 & \red{0.83} & 1.09 \\
 & & 5 $R_*$ & 3.64e+17 & 1.01 & 0.99 & 1.01 & 9.93e+16 & 1.01 & \red{0.39} & \red{0.84} & 5.07e+15 & 1.0 & \red{0.44} & 1.09 \\
\ce{HCN} & 2C & 2 $R_*$ & 3.63e+17 & 1.0 & 0.93 & 1.0 & 9.85e+16 & 1.0 & 0.95 & \red{0.83} & 4.90e+15 & 0.95 & 1.06 & \greent{1.91} \\
 & & 5 $R_*$ & 3.63e+17 & 1.01 & \red{\textbf{0.11}} & 0.99 & 9.85e+16 & 1.01 & \red{0.72} & \red{0.54} & 4.90e+15 & 0.98 & \red{\textbf{0.01}} & 0.93 \\
 & 1C & 2 $R_*$ & 3.62e+17 & 0.96 & \greent{1.89} & \greent{2.24} & 9.77e+16 & \red{0.85} & \greent{2.47} & \greent{2.84} & 4.76e+15 & \red{0.81} & \red{\textbf{0.1}} & \greent{2.41} \\
 & & 5 $R_*$ & 3.62e+17 & 0.99 & \greent{1.24} & \greent{1.45} & 9.77e+16 & 0.97 & \greent{1.24} & \greent{2.22} & 4.76e+15 & 0.96 & \red{\textbf{0.0}} & \red{0.49} \\
\noalign{\smallskip}
 & Sm. & 2 $R_*$ & 5.40e+16 & 1.0 & 1.0 & 1.0 & 1.53e+16 & 1.0 & \red{0.61} & 0.99 & 1.17e+15 & 0.96 & \red{\textbf{0.0}} & \red{\textbf{0.16}} \\
 & & 5 $R_*$ & 5.40e+16 & 1.01 & 0.97 & 1.01 & 1.53e+16 & 1.01 & \red{\textbf{0.01}} & 0.94 & 1.17e+15 & 0.96 & \red{\textbf{0.0}} & \red{\textbf{0.0}} \\
\ce{SiS} & 2C & 2 $R_*$ & 5.40e+16 & 1.0 & \red{0.85} & 1.0 & 1.53e+16 & 0.99 & \red{\textbf{0.0}} & \red{0.77} & 1.16e+15 & \red{0.87} & \red{\textbf{0.0}} & \red{\textbf{0.0}} \\
 & & 5 $R_*$ & 5.40e+16 & 1.01 & \red{0.21} & 0.99 & 1.53e+16 & 1.0 & \red{\textbf{0.0}} & \red{0.52} & 1.16e+15 & 0.91 & \red{\textbf{0.0}} & \red{\textbf{0.0}} \\
 & 1C & 2 $R_*$ & 5.40e+16 & \red{0.89} & \red{\textbf{0.0}} & \red{\textbf{0.0}} & 1.53e+16 & \red{0.69} & \red{\textbf{0.0}} & \red{\textbf{0.0}} & 1.15e+15 & \red{0.66} & \red{\textbf{0.0}} & \red{\textbf{0.0}} \\
 & & 5 $R_*$ & 5.40e+16 & 0.97 & \red{\textbf{0.0}} & \red{\textbf{0.0}} & 1.53e+16 & \red{0.89} & \red{\textbf{0.0}} & \red{\textbf{0.0}} & 1.15e+15 & \red{0.85} & \red{\textbf{0.0}} & \red{\textbf{0.0}} \\
\noalign{\smallskip}
 & Sm. & 2 $R_*$ & 4.46e+16 & 1.0 & 1.0 & 1.0 & 1.22e+16 & 1.0 & \red{0.57} & 0.93 & 6.09e+14 & 1.0 & \red{0.58} & \red{\textbf{0.08}} \\
 & & 5 $R_*$ & 4.46e+16 & 1.01 & 0.96 & 1.0 & 1.22e+16 & 1.01 & \red{0.84} & \red{0.43} & 6.09e+14 & 1.02 & \red{\textbf{0.0}} & \red{\textbf{0.15}} \\
\ce{SiO} & 2C & 2 $R_*$ & 4.46e+16 & 1.0 & \red{0.81} & 0.98 & 1.21e+16 & 1.0 & \red{\textbf{0.17}} & \red{\textbf{0.07}} & 5.87e+14 & 1.0 & \red{\textbf{0.0}} & \red{0.71} \\
 & & 5 $R_*$ & 4.46e+16 & 1.01 & \red{0.89} & \red{0.77} & 1.21e+16 & 1.01 & \red{\textbf{0.05}} & \red{0.4} & 5.87e+14 & 1.02 & \red{\textbf{0.0}} & \red{\textbf{0.0}} \\
 & 1C & 2 $R_*$ & 4.45e+16 & 1.0 & \red{\textbf{0.0}} & \red{0.54} & 1.20e+16 & 0.99 & \red{\textbf{0.0}} & \red{\textbf{0.06}} & 5.68e+14 & 1.02 & \red{\textbf{0.0}} & \red{\textbf{0.0}} \\
 & & 5 $R_*$ & 4.45e+16 & 1.01 & \red{\textbf{0.0}} & \red{\textbf{0.08}} & 1.20e+16 & 1.01 & \red{\textbf{0.0}} & \red{\textbf{0.0}} & 5.68e+14 & 1.02 & \red{\textbf{0.0}} & \red{\textbf{0.0}} \\
\hline 
\end{tabular}
\label{table:app-coldens-crich-parents}    
\end{table*}

\begin{table*}
\label{tab:continued}
\begin{tabular}{l l l c c c c c c c c c c c c}
\hline  
& & $R_\mathrm{dust}$ & \multicolumn{4}{c}{$\dot{M}= 10^{-5}\ \mathrm{M_\odot/yr},\ v_\infty = 15\ \mathrm{km/s}$} & \multicolumn{4}{c}{$\dot{M}= 10^{-6}\ \mathrm{M_\odot/yr},\ v_\infty = 5\ \mathrm{km/s}$} & \multicolumn{4}{c}{$\dot{M}= 10^{-7}\ \mathrm{M_\odot/yr},\ v_\infty = 5\ \mathrm{km/s}$}  \\  
\cline{3-3} \cline{4-7} \cline{8-11} \cline{12-15} 
\noalign{\smallskip}
 & Sm. & 2 $R_*$ & 5.27e+14 & 1.0 & 1.0 & 1.0 & 1.42e+14 & 1.0 & \red{0.81} & 1.1 & 6.78e+12 & 1.0 & \red{\textbf{0.0}} & \greent{\textbf{10.39}} \\
 & & 5 $R_*$ & 5.27e+14 & 1.01 & 0.97 & 1.01 & 1.42e+14 & 1.01 & \red{\textbf{0.0}} & \red{0.71} & 6.78e+12 & \red{0.64} & \red{\textbf{0.0}} & \red{\textbf{0.0}} \\
\ce{NH3} & 2C & 2 $R_*$ & 5.25e+14 & 1.0 & 1.0 & 1.03 & 1.40e+14 & 1.0 & \red{\textbf{0.03}} & \greent{\textbf{22.27}} & 6.32e+12 & 0.94 & \red{\textbf{0.0}} & \red{\textbf{0.0}} \\
 & & 5 $R_*$ & 5.25e+14 & 1.01 & \red{\textbf{0.0}} & 0.97 & 1.40e+14 & 1.01 & \red{\textbf{0.0}} & \red{\textbf{0.0}} & 6.32e+12 & \red{0.29} & \red{\textbf{0.0}} & \red{\textbf{0.0}} \\
 & 1C & 2 $R_*$ & 5.22e+14 & 0.93 & \red{\textbf{0.0}} & \red{\textbf{0.0}} & 1.37e+14 & \red{0.75} & \red{\textbf{0.0}} & \red{\textbf{0.0}} & 5.88e+12 & \red{0.3} & \red{\textbf{0.0}} & \red{\textbf{0.0}} \\
 & & 5 $R_*$ & 5.22e+14 & \red{0.83} & \red{\textbf{0.0}} & \red{\textbf{0.0}} & 1.37e+14 & \red{0.53} & \red{\textbf{0.0}} & \red{\textbf{0.0}} & 5.88e+12 & \red{\textbf{0.12}} & \red{\textbf{0.0}} & \red{\textbf{0.0}} \\
\noalign{\smallskip}
 & Sm. & 2 $R_*$ & 3.69e+13 & 1.0 & 1.03 & 1.0 & 1.04e+13 & \greent{1.27} & \greent{\textbf{815.58}} & \greent{\textbf{61.54}} & 3.29e+11 & \greent{\textbf{19.46}} & \red{\textbf{0.0}} & \red{\textbf{0.03}} \\
 & & 5 $R_*$ & 3.69e+13 & 1.01 & \greent{\textbf{10.17}} & \greent{2.77} & 1.04e+13 & \greent{1.2} & \red{\textbf{0.0}} & \red{0.33} & 3.29e+11 & \red{\textbf{0.03}} & \red{\textbf{0.0}} & \red{\textbf{0.0}} \\
\ce{H2S} & 2C & 2 $R_*$ & 3.68e+13 & 1.09 & \greent{\textbf{340.9}} & \greent{\textbf{21.63}} & 1.03e+13 & \greent{\textbf{10.14}} & \greent{\textbf{491.33}} & \greent{\textbf{1453.88}} & 3.09e+11 & \red{\textbf{0.1}} & \red{\textbf{0.0}} & \red{\textbf{0.0}} \\
 & & 5 $R_*$ & 3.68e+13 & \greent{1.43} & \red{\textbf{0.04}} & \greent{\textbf{38.7}} & 1.03e+13 & \red{0.66} & \red{\textbf{0.0}} & \red{0.21} & 3.09e+11 & \red{\textbf{0.04}} & \red{\textbf{0.0}} & \red{\textbf{0.0}} \\
 & 1C & 2 $R_*$ & 3.68e+13 & \greent{1.21} & \red{\textbf{0.0}} & \red{\textbf{0.01}} & 1.02e+13 & \red{\textbf{0.09}} & \red{\textbf{0.0}} & \red{\textbf{0.0}} & 3.18e+11 & \red{\textbf{0.06}} & \red{\textbf{0.0}} & \red{\textbf{0.0}} \\
 & & 5 $R_*$ & 3.68e+13 & \red{\textbf{0.04}} & \red{\textbf{0.0}} & \red{\textbf{0.01}} & 1.02e+13 & \red{\textbf{0.07}} & \red{\textbf{0.0}} & \red{\textbf{0.01}} & 3.18e+11 & \red{\textbf{0.03}} & \red{\textbf{0.0}} & \red{\textbf{0.0}} \\
\noalign{\smallskip}
 & Sm. & 2 $R_*$ & 6.74e+13 & 1.0 & 1.0 & 1.0 & 6.02e+13 & 1.0 & \greent{3.44} & 1.05 & 3.31e+13 & 1.02 & \red{\textbf{0.03}} & \greent{\textbf{19.57}} \\
 & & 5 $R_*$ & 6.74e+13 & 1.0 & 1.1 & 1.0 & 6.02e+13 & 1.0 & \red{0.22} & 1.05 & 3.31e+13 & 0.94 & \red{\textbf{0.0}} & \red{\textbf{0.0}} \\
\ce{CH3} & 2C & 2 $R_*$ & 7.35e+13 & 1.0 & \greent{1.58} & 1.02 & 5.92e+13 & 1.0 & \greent{\textbf{35.0}} & \greent{3.54} & 3.14e+13 & 0.98 & \red{\textbf{0.01}} & \red{0.39} \\
 & & 5 $R_*$ & 7.35e+13 & 1.0 & \red{0.29} & \greent{1.11} & 5.92e+13 & 1.01 & \red{\textbf{0.15}} & \red{0.35} & 3.14e+13 & \red{0.85} & \red{\textbf{0.0}} & \red{\textbf{0.0}} \\
 & 1C & 2 $R_*$ & 7.78e+13 & 1.08 & \red{\textbf{0.08}} & \red{\textbf{0.14}} & 5.75e+13 & \red{0.86} & \red{\textbf{0.08}} & \red{\textbf{0.11}} & 2.88e+13 & \red{0.77} & \red{\textbf{0.0}} & \red{\textbf{0.01}} \\
 & & 5 $R_*$ & 7.78e+13 & 1.05 & \red{\textbf{0.09}} & \red{\textbf{0.15}} & 5.75e+13 & 0.91 & \red{\textbf{0.04}} & \red{\textbf{0.12}} & 2.88e+13 & \red{0.72} & \red{\textbf{0.0}} & \red{\textbf{0.0}} \\
\noalign{\smallskip}
 & Sm. & 2 $R_*$ & 4.99e+15 & 1.0 & 1.0 & 1.0 & 4.42e+15 & 1.0 & 1.09 & 1.0 & 2.52e+15 & 1.0 & \greent{1.73} & \greent{1.52} \\
 & & 5 $R_*$ & 4.99e+15 & 1.0 & 1.0 & 1.0 & 4.42e+15 & 1.0 & \greent{1.55} & 1.08 & 2.52e+15 & 1.0 & 0.93 & \greent{1.69} \\
\ce{CN} & 2C & 2 $R_*$ & 4.97e+15 & 1.0 & 1.02 & 1.0 & 4.44e+15 & 1.0 & \greent{1.39} & 1.09 & 2.35e+15 & 1.01 & \greent{2.44} & \greent{2.77} \\
 & & 5 $R_*$ & 4.97e+15 & 1.0 & \greent{1.42} & 1.01 & 4.44e+15 & 1.0 & \greent{1.6} & \greent{1.55} & 2.35e+15 & 1.01 & \red{\textbf{0.08}} & \greent{2.04} \\
 & 1C & 2 $R_*$ & 4.85e+15 & 1.02 & \greent{4.05} & \greent{2.83} & 4.31e+15 & \greent{1.24} & \greent{\textbf{7.88}} & \greent{4.37} & 2.12e+15 & \greent{1.19} & \red{0.8} & \greent{4.51} \\
 & & 5 $R_*$ & 4.85e+15 & 1.0 & \greent{\textbf{6.12}} & \greent{2.3} & 4.31e+15 & 1.04 & \greent{\textbf{13.51}} & \greent{4.35} & 2.12e+15 & 1.07 & \red{\textbf{0.0}} & \greent{2.3} \\
\noalign{\smallskip}
 & Sm. & 2 $R_*$ & 4.03e+15 & 1.0 & 1.0 & 1.0 & 2.83e+15 & 1.0 & \red{\textbf{0.19}} & 0.98 & 1.10e+15 & 0.91 & \red{0.42} & \red{\textbf{0.18}} \\
 & & 5 $R_*$ & 4.03e+15 & 1.0 & 0.9 & 1.0 & 2.83e+15 & 1.0 & \red{0.47} & \red{0.34} & 1.10e+15 & \red{0.86} & \red{\textbf{0.0}} & \red{\textbf{0.02}} \\
\ce{C2H} & 2C & 2 $R_*$ & 3.49e+15 & 1.0 & \red{0.65} & 0.99 & 2.36e+15 & 0.99 & \red{0.28} & \red{0.41} & 8.93e+14 & \red{0.69} & \red{\textbf{0.19}} & \red{\textbf{0.15}} \\
 & & 5 $R_*$ & 3.49e+15 & 1.0 & \red{0.63} & \red{0.87} & 2.36e+15 & 0.98 & \red{0.38} & \red{0.3} & 8.93e+14 & \red{0.75} & \red{\textbf{0.0}} & \red{\textbf{0.05}} \\
 & 1C & 2 $R_*$ & 2.80e+15 & \red{0.8} & \red{\textbf{0.07}} & \red{\textbf{0.09}} & 1.63e+15 & \red{0.31} & \red{\textbf{0.14}} & \red{\textbf{0.14}} & 5.80e+14 & \red{0.21} & \red{\textbf{0.0}} & \red{\textbf{0.15}} \\
 & & 5 $R_*$ & 2.80e+15 & 0.93 & \red{\textbf{0.09}} & \red{\textbf{0.18}} & 1.63e+15 & \red{0.83} & \red{0.22} & \red{0.24} & 5.80e+14 & \red{0.61} & \red{\textbf{0.0}} & \red{\textbf{0.08}} \\
\noalign{\smallskip}
 & Sm. & 2 $R_*$ & 1.42e+13 & 1.0 & 1.0 & 1.0 & 2.55e+13 & 1.0 & \greent{\textbf{12.8}} & \greent{1.65} & 1.42e+13 & \greent{1.34} & \red{\textbf{0.0}} & \red{\textbf{0.19}} \\
 & & 5 $R_*$ & 1.42e+13 & 1.03 & \greent{\textbf{19.39}} & \greent{\textbf{7.95}} & 2.55e+13 & \greent{1.67} & \red{\textbf{0.12}} & \greent{1.81} & 1.42e+13 & \red{0.69} & \red{\textbf{0.0}} & \red{\textbf{0.0}} \\
\ce{HS} & 2C & 2 $R_*$ & 2.31e+13 & 1.0 & \greent{\textbf{21.38}} & \greent{1.4} & 2.99e+13 & 1.08 & \greent{4.71} & \greent{\textbf{13.8}} & 1.67e+13 & 0.98 & \red{\textbf{0.0}} & \red{\textbf{0.04}} \\
 & & 5 $R_*$ & 2.31e+13 & \greent{1.57} & \red{0.43} & \greent{\textbf{21.75}} & 2.99e+13 & \greent{1.45} & \red{\textbf{0.01}} & 1.01 & 1.67e+13 & \red{0.28} & \red{\textbf{0.0}} & \red{\textbf{0.0}} \\
 & 1C & 2 $R_*$ & 3.21e+13 & 1.07 & \red{\textbf{0.0}} & \red{\textbf{0.05}} & 3.59e+13 & 1.02 & \red{\textbf{0.0}} & \red{\textbf{0.0}} & 1.98e+13 & \red{0.2} & \red{\textbf{0.0}} & \red{\textbf{0.0}} \\
 & & 5 $R_*$ & 3.21e+13 & 1.02 & \red{\textbf{0.0}} & \red{\textbf{0.05}} & 3.59e+13 & 1.02 & \red{\textbf{0.0}} & \red{\textbf{0.0}} & 1.98e+13 & \red{\textbf{0.13}} & \red{\textbf{0.0}} & \red{\textbf{0.0}} \\
\noalign{\smallskip}
 & Sm. & 2 $R_*$ & 4.93e+11 & 1.0 & 1.0 & 1.0 & 1.91e+12 & 1.0 & \greent{\textbf{5.5}} & \greent{1.35} & 1.02e+12 & \greent{1.26} & \red{\textbf{0.0}} & \greent{\textbf{29.09}} \\
 & & 5 $R_*$ & 4.93e+11 & 1.0 & \greent{\textbf{13.23}} & \greent{1.21} & 1.91e+12 & 1.03 & \red{\textbf{0.05}} & \greent{\textbf{14.73}} & 1.02e+12 & \red{0.82} & \red{\textbf{0.0}} & \red{\textbf{0.0}} \\
\ce{NS} & 2C & 2 $R_*$ & 1.40e+12 & 1.0 & \greent{3.32} & \greent{1.13} & 3.20e+12 & 1.03 & \greent{1.57} & \greent{\textbf{56.81}} & 2.00e+12 & 0.97 & \red{\textbf{0.0}} & \red{\textbf{0.03}} \\
 & & 5 $R_*$ & 1.40e+12 & 1.01 & \red{0.28} & \greent{\textbf{80.8}} & 3.20e+12 & 1.05 & \red{\textbf{0.01}} & \red{0.89} & 2.00e+12 & \red{0.5} & \red{\textbf{0.0}} & \red{\textbf{0.0}} \\
 & 1C & 2 $R_*$ & 2.31e+12 & \greent{\textbf{5.81}} & \red{\textbf{0.03}} & \greent{\textbf{88.83}} & 4.53e+12 & \greent{3.11} & \red{\textbf{0.01}} & \red{\textbf{0.07}} & 2.94e+12 & \red{0.44} & \red{\textbf{0.0}} & \red{\textbf{0.0}} \\
 & & 5 $R_*$ & 2.31e+12 & \greent{\textbf{8.65}} & \red{\textbf{0.04}} & \greent{1.13} & 4.53e+12 & \greent{4.9} & \red{\textbf{0.01}} & \red{\textbf{0.01}} & 2.94e+12 & \red{0.45} & \red{\textbf{0.0}} & \red{\textbf{0.0}} \\
\noalign{\smallskip}
 & Sm. & 2 $R_*$ & 4.81e+12 & 1.0 & 1.0 & 1.0 & 5.11e+12 & 1.0 & \greent{4.78} & 1.05 & 2.28e+12 & 1.04 & \red{0.5} & \greent{\textbf{12.89}} \\
 & & 5 $R_*$ & 4.81e+12 & 0.97 & 1.09 & 0.98 & 5.11e+12 & 0.98 & \greent{2.54} & \greent{\textbf{7.35}} & 2.28e+12 & \red{0.69} & \red{\textbf{0.01}} & \red{0.26} \\
\ce{CH3CN} & 2C & 2 $R_*$ & 5.49e+12 & 1.0 & \greent{1.65} & 1.02 & 4.69e+12 & 1.01 & \greent{\textbf{53.97}} & \greent{\textbf{6.51}} & 1.96e+12 & \red{0.82} & \red{\textbf{0.09}} & \greent{\textbf{7.01}} \\
 & & 5 $R_*$ & 5.49e+12 & 0.98 & \greent{3.09} & \greent{3.01} & 4.69e+12 & 0.99 & \greent{1.12} & \greent{\textbf{13.72}} & 1.96e+12 & \red{0.63} & \red{\textbf{0.0}} & \red{\textbf{0.03}} \\
 & 1C & 2 $R_*$ & 5.92e+12 & 1.1 & \red{\textbf{0.12}} & \greent{2.42} & 4.25e+12 & \greent{1.22} & \red{\textbf{0.06}} & \red{0.54} & 1.55e+12 & \red{0.74} & \red{\textbf{0.0}} & \red{\textbf{0.02}} \\
 & & 5 $R_*$ & 5.92e+12 & 1.01 & \red{\textbf{0.07}} & \greent{2.93} & 4.25e+12 & \red{0.89} & \red{\textbf{0.02}} & \red{0.25} & 1.55e+12 & \red{0.6} & \red{\textbf{0.0}} & \red{\textbf{0.02}} \\
\noalign{\smallskip}
 & Sm. & 2 $R_*$ & 3.64e+13 & 1.0 & 1.0 & 1.0 & 6.46e+13 & 1.0 & \greent{2.27} & 1.09 & 1.47e+13 & \greent{1.64} & \red{0.26} & \red{\textbf{0.16}} \\
 & & 5 $R_*$ & 3.64e+13 & 1.0 & \greent{3.42} & 1.02 & 6.46e+13 & 1.02 & \red{0.79} & \greent{1.2} & 1.47e+13 & \greent{1.66} & \red{\textbf{0.0}} & \red{\textbf{0.03}} \\
\ce{C6H-} & 2C & 2 $R_*$ & 5.32e+13 & 1.0 & \greent{4.27} & 1.09 & 6.75e+13 & 1.1 & \red{0.79} & \greent{1.8} & 2.09e+13 & \greent{1.78} & \red{0.26} & \red{\textbf{0.09}} \\
 & & 5 $R_*$ & 5.32e+13 & 1.01 & 0.94 & \greent{1.72} & 6.75e+13 & \greent{1.2} & 1.07 & \red{0.71} & 2.09e+13 & \greent{1.22} & \red{\textbf{0.0}} & \red{\textbf{0.04}} \\
 & 1C & 2 $R_*$ & 7.06e+13 & \greent{2.29} & \greent{2.78} & \greent{1.32} & 7.20e+13 & \greent{1.55} & \greent{1.5} & 0.98 & 2.68e+13 & \greent{1.28} & \red{\textbf{0.0}} & \red{\textbf{0.02}} \\
 & & 5 $R_*$ & 7.06e+13 & \greent{1.75} & \greent{3.26} & 1.09 & 7.20e+13 & \greent{1.44} & \greent{1.69} & 0.93 & 2.68e+13 & \red{0.83} & \red{\textbf{0.0}} & \red{\textbf{0.04}} \\
\hline
\end{tabular}
\end{table*}



\bsp	
\label{lastpage}
\end{document}